\documentclass[10pt,preprint]{aastex}
\usepackage{color}
\usepackage{graphicx}

\newcommand{\etal}{et~al.}

\newcommand{\ks}{$K_{\rm s}$}

\newcommand{\teff}{$T_{\rm eff}$}

\newcommand{\vsini}{$v \sin i$}
\newcommand{\mum}{$\mu$m}

\begin{document}

\title{On Infrared Excesses Associated With Li-Rich K Giants}

\slugcomment{Version from \today}

\author{Luisa~M.~Rebull\altaffilmark{1}, 
Joleen K.~Carlberg\altaffilmark{2,3},
John C. Gibbs\altaffilmark{4},
J. Elin Deeb\altaffilmark{5}, 
Estefania Larsen\altaffilmark{6}, 
David V. Black\altaffilmark{7}, 
Shailyn Altepeter\altaffilmark{6}, 
Ethan Bucksbee\altaffilmark{6}, 
Sarah Cashen\altaffilmark{4}, 
Matthew Clarke\altaffilmark{6},
Ashwin Datta\altaffilmark{4}, 
Emily Hodgson\altaffilmark{4}, 
Megan Lince\altaffilmark{4}}

\altaffiltext{1}{Spitzer Science Center (SSC) and Infrared Science
Archive (IRSA), Infrared Processing and Analysis Center (IPAC), 1200
E.\ California Blvd., California Institute of Technology, Pasadena, CA
91125 USA; rebull@ipac.caltech.edu}
\altaffiltext{2}{NASA Goddard Space Flight Center, Code 667, Greenbelt
MD 20771 USA}
\altaffiltext{3}{NASA Postdoctoral Program Fellow}
\altaffiltext{4}{Glencoe High School, 2700 NW Glencoe Rd., Hillsboro, OR 97124 USA}
\altaffiltext{5}{Bear Creek High School, 9800 W. Dartmouth Pl., Lakewood, CO 80227 USA}
\altaffiltext{6}{Millard South High School, 14905 Q St., Omaha, NE
68137 USA}
\altaffiltext{7}{Walden School of Liberal Arts, 4230 N. University
Ave., Provo, UT 84604 USA}

\begin{abstract}

Infrared (IR) excesses around K-type red giants (RGs) have previously
been discovered using Infrared Astronomy Satellite (IRAS) data, and
past studies have suggested a link between RGs with overabundant Li and IR
excesses, implying the ejection of circumstellar shells or disks.  We
revisit the question of IR excesses around RGs using higher spatial
resolution IR data, primarily from the Wide-field Infrared Survey
Explorer (WISE).    Our goal was to elucidate the link between 
three unusual RG properties: fast rotation, enriched Li, and IR
excess.  Our sample of RGs includes those with previous IR detections,
a sample with well-defined rotation and Li abundance measurements with
no previous IR measurements, and a large sample of RGs asserted to be
Li-rich in the literature; we have 316 targets thought to be K giants,
about 40\% of which we take to be Li-rich. In 24 cases with previous
detections of IR excess at low spatial resolution, we believe that
source confusion is playing a role, in that either (a) the source that
is bright in the optical is not responsible for the IR
flux, or (b) there is more than one source responsible for the IR flux
as measured in IRAS.  We looked for IR excesses in the remaining
sources, identifying 28 that have significant IR excesses by $\sim$20
\mum\ (with possible excesses for 2 additional sources).  There appears to
be an intriguing correlation in that the largest IR excesses are all
in Li-rich K giants, though very few Li-rich K giants have IR excesses
(large or small). These largest IR excesses also tend to be found in
the fastest rotators. There is no correlation of IR excess with the
carbon isotopic ratio, $^{12}$C/$^{13}$C. IR excesses by 20 \mum,
though relatively rare, are at least twice as common among our sample
of Li-rich K giants. If dust shell production is a common by-product
of Li enrichment mechanisms, these observations suggest that the IR
excess stage is very short-lived, which is supported by theoretical
calculations. Conversely, the Li-enrichment mechanism may only
occasionally produce dust, and an additional parameter (e.g.,
rotation) may control whether or not a shell is ejected. 

\end{abstract}

\keywords{stars: late-type; stars:evolution; infrared:stars}

\section{Introduction}
\label{sec:intro}

As stars evolve from the main-sequence (MS) to the red giant branch
(RGB), they exhibit several characteristic changes. As the outer layers
expand and cool, the star's rotation rate slows, the convection zone
deepens and a series of shell-burning and core-burning phases begin to
take place. A number of RGB K-type giants, however, exhibit
uncharacteristically rapid rotation rates that also seem to be
correlated with high lithium abundances, $A$(Li) (e.g., Carlberg
\etal\ 2012, hereafter C12). These higher rotation rates and $A$(Li)
are inconsistent with those predicted by standard stellar evolutionary
models. It has also been suggested that many of these high-Li RGB
stars have infrared (IR) excesses suggestive of a circumstellar shell
or disk (de la Reza \etal\ 1996, 1997, Drake \etal\ 2002, and
references therein). Various hypotheses have been proposed to explain
the combination of high Li, rapid rotation rates and IR excesses,
including the accretion of nearby giant planets equivalent to a few
Jupiter masses (e.g., Siess \& Livio 1999) or a newly triggered
nuclear fusion stage that could eject a dusty shell (e.g., de la Reza
\etal\ 2015 and references therein). 

The de la Reza \etal\ (1997; dlR97) study (and related studies) used
data from the Infrared Astronomy Satellite (IRAS; Neugebauer \etal\
1984), which surveyed at 12, 25, 60, and 100 \mum.  The spatial
resolution, however, was relatively low, up to a few arcminutes.
Despite the relatively low spatial resolution, many stars with IR
excesses were identified (e.g., Gillett 1986, Paresce \& Burrows
1987). In some regions with high source density, identifying the
optical counterpart to the infrared source can be difficult, but in
many cases, the counterparts are easily identifiable. dlR97, as well
as other studies, used these IRAS data to look for K giants with IR
excesses. 

Data are now available from the much higher spatial resolution
(6-12$\arcsec$) and much more sensitive Wide-field Infrared Survey
Explorer (WISE; Wright \etal\ 2010). WISE surveyed the whole sky at
3.4, 4.6, 12, and 22 \mum.  While WISE does not detect wavelengths
longer than IRAS channel 2 (25 \mum), it is much higher spatial
resolution, and can provide insight into how to interpret the IRAS
data at 60 and 100 \mum.  

Some investigations subsequent to dlR97 have begun to question the
connection between IR excesses and lithium abundance in RGs. Fekel \&
Watson (1998) found 6 giants with larger than typical lithium
abundances out of 39 giants with IR excess (as determined from IRAS),
which they point out is a similar fraction of stars with enhanced Li
as found in normal field giants. Jasniewicz \etal\ (1999) finds 8
Li-rich stars out of 29 stars with IR excesses (also as determined via
IRAS), finding no correlation between Li abundance and IR excess.
Lebzelter \etal\ (2012) report on 3 Li-rich giants (out of more than
400 studied), none of which have IR excesses suggestive of mass loss.
The IR excesses in their paper were identified based on WISE data, but
were limited to 12 \mum\ and shorter because few of their targets were
detected at 22 \mum. However, they also looked for evidence of gas
mass loss in their spectra, and found none. Kumar \etal\ (2015) report
on a search for IR excesses in 2000 K giants. None of their far-IR
excess sources are lithium-rich, and of their 40 Li-rich sources, they
identify 7 as having IR excess of any sort. These authors combined
IRAS and WISE data to look for IR excesses.

In the present paper, we have also combined IRAS and WISE data, as
well as data from several other IR surveys as discussed below.  Large
IR exceses are easily identified in the SEDs, and we used tools
developed in the context of the study of young stars to look for small
but significant excesses. We started with the dlR97 IRAS-selected
targets; these IRAS-selected targets are all quite bright in the IR.
We added to this the sample from C12, who assembled a set of K giants
consisting of rapid and slow rotators in which the relationship
between \vsini, lithium abundances, and carbon isotope ratios
($^{12}$C/$^{13}$C) could be explored with an intention of exploring
evidence for planetary accretion. The C12 sample was assembled without
regard to IR excess, so these objects are on average much fainter in
the IR than the dlR97 sample. Finally, many Li-rich K giants (and
candidates) have been reported in the literature, many of which are
not in the dlR97 or C12 samples. We have included such Li-rich K
giants in our sample. We looked for IR excesses among all of these
targets, performing visual inspection of multi-wavelength images and
assembling broadband spectral energy distributions (SEDs) for our
targets.

We now assemble the list of targets (Sec.~\ref{sec:assemblecatalog}),
and search various archives for images and photometry for these
targets (Sec.~\ref{sec:largedatasets}). Then we drop some stars from
our sample, some by necessity due to missing data, and some because
they are likely subject to source confusion 
(Sec.~\ref{sec:droppedsrcs}). From the remaining sample, we can
identify sources likely to have an IR excess (Sec.~\ref{sec:irx}).
Then we discuss some properties of the sample as a whole
(Sec.~\ref{sec:disc1}). We find few stars with IR excesses, so our
ability to find correlations with abundances is somewhat limited, but
we discuss this in Sec.~\ref{sec:abun} before summarizing in
Sec.~\ref{sec:concl}.

\section{Assembling the Target List}
\label{sec:assemblecatalog}

There are 82 targets published in dlR97. They obtained spectra of the
targets on their list, and reported the targets as Li-rich if the Li
line at $\lambda$6708 had an intensity comparable to or higher than
the \ion{Ca}{1} line at $\lambda$6718. If the Li abundance was
actually known, they took those with abundances larger than log
$\epsilon$(Li)=1.2 dex as Li-rich. Many of the targets have HD
numbers, and thus finding coordinates for an optical counterpart is
straightforward. However, many of their targets have only IRAS names,
and so they took spectra of objects they believed to be counterparts
of the sources listed; see discussion below. The target position we
used for these sources is that reported in the IRAS catalogs.

There are 86 K giants reported in C12. Comparing this to the targets
from dlR97, there is only one source in common between them, HD 31993.
C12 reports lithium abundances, \vsini, and $^{12}$C/$^{13}$C ratios,
among other things, for their targets. Not all of these targets are
Li-rich.

There is wide-ranging literature reporting on Li-rich K giants (and
candidates). We compiled 149 additional targets that have been
identified either consistently or at one time as confirmed or possible
Li-rich K giants, but not included in either the dlR97 or the C12
samples. They include targets from  Adam\'ow \etal\ (2014), 
Anthony-Twarog \etal\ (2013), Carney \etal\ (1998), Carlberg \etal\
(in prep),  Castilho \etal\ (2000), Drake \etal\ (2002), Fekel \&
Watson (1998), Hill \& Pasquini (1999), Jasniewicz \etal\ (1999), 
Kirby \etal\ (2012), Kraft \etal\ (1999), Kumar \etal\ (2011) and
references therein, Liu \etal\ (2014), Luck \& Heiter (2007), Martell
\etal\ (2013), Monaco \etal\ (2014), Pilachowski \etal\ (2003), Ruchti
\etal\ (2011), Silva Aguirre \etal\ (2014), Smith \etal\ (1999), and
Torres \etal\ (2000).  Kumar \etal\ (2015) appeared as we were
finishing our analysis, and it has similar goals as the present paper.
It uses as a starting point the list of 2000 low-mass K giants from
Kumar \etal\ (2011), so all of the Li-rich sources from that sample
are already in our sample. All of the literature sources are
identified simply as `literature sources' in
Table~\ref{tab:bigdatatable}; the Appendix identifies the paper of
origin for any given target. 

Our complete list of 316 targets appears in
Table~\ref{tab:bigdatatable}. We obtained RA, Dec positions for these
targets, most of which are quite bright in the optical, primarily from
SIMBAD, though literature was consulted for fainter sources as
required. The positions we used are in Table~\ref{tab:bigdatatable},
which also includes all the bandmerged brightness measurements
discussed below. In the Appendix, Table~\ref{tab:specialnotes}
collects a few special notes about any special circumstances attached
to that star, e.g., information about typos in previously published
tables, or necessary tweaks to the positions. There are copious notes
in the main text about targets called out as special (e.g., those with
IR excesses).

\begin{deluxetable}{cccp{8cm}}
\tabletypesize{\scriptsize}
\tablecaption{Contents of Online Catalog\tablenotemark{a}\label{tab:bigdatatable}}
\tablewidth{0pt}
\tablehead{ \colhead{Format}
&\colhead{Units} & \colhead{Label} & \colhead{Explanations}}
\startdata
A26& ---& name & name of source\\
F11.6 & dec & RA & right ascension of source (J2000) \\
F11.6 & dec & Dec  & declination of source (J2000) \\
A6    & --- & dlR97 & flag to indicate source is part of the dlR97 sample\\
A4    & --- & C12 & flag to indicate star is part of the C12 sample\\
A5    & --- & lit & flag to indicate star is selected from the literature as a Li-rich giant (but not C12 or dlR97) \\
A4   & --- & LirichLit & value of `yes' means source is identified in the literature as a Li-rich giant\\
A4   & --- & LirichHere & value of `yes' means source is selected here as a Li-rich giant\\
F7.2 & --- & ALiNLTE & $A$(Li) under the assumption of NLTE from the literature \\
A2   & ---  & ALiLTEl & limit flag for $A$(Li) under the assumption of LTE from the literature\\
F7.2 & --- & ALiLTE & $A$(Li) under the assumption of LTE from the literature \\
A2 & --- & Cratiol & limit flag on $^{12}$C/$^{13}$C ratio \\
F5.1 & --- & Cratio & $^{12}$C/$^{13}$C ratio \\
F5.1 & km s$^{-1}$ & vsini & projected rotational velocity  \vsini\ in km s$^{-1}$  from the literature\\
 I5 & K & Teff & Effective temperature from the literature\\
 F5.1 & --- & logg & log $g$ from the literature\\
 F6.2 & mag & Umag  & Vega-based magnitude in $U$ band\\
 F6.2 & mag & Umerr & Vega-based magnitude error in $U$ band; taken to be 20\% unless specified\\
 F6.2 & mag & Bmag  & Vega-based magnitude in $B$ band\\
 F6.2 & mag & Bmerr & Vega-based magnitude error in $B$ band; taken to be 20\% unless specified\\
 F6.2 & mag & Vmag  & Vega-based magnitude in $V$ band\\
 F6.2 & mag & Vmerr & Vega-based magnitude error in $V$ band; taken to be 20\% unless specified\\
 F6.2 & mag & Rmag  & Vega-based magnitude in $R$ band\\
 F6.2 & mag & Rmerr & Vega-based magnitude error in $R$ band; taken to be 20\% unless specified\\
 F6.2 & mag & umag  & AB SDSS magnitude in $u$ band\\
 F6.2 & mag & umerr & AB SDSS magnitude error in $u$ band\\
 F6.2 & mag & gmag  & AB SDSS magnitude in $g$ band\\
 F6.2 & mag & gmerr & AB SDSS magnitude error in $g$ band\\
 F6.2 & mag & rmag  & AB SDSS magnitude in $r$ band\\
 F6.2 & mag & rmerr & AB SDSS magnitude error in $r$ band\\
 F6.2 & mag & imag  & AB SDSS magnitude in $i$ band\\
 F6.2 & mag & imerr & AB SDSS magnitude error in $i$ band\\
 F6.2 & mag & zmag  & AB SDSS magnitude in $z$ band\\
 F6.2 & mag & zmerr & AB SDSS magnitude error in $z$ band\\
 A22  & --- & 2Mname & Name from 2MASS or 2MASX catalog\\
 A2   & --- & Jlim  & limit flag for 2MASS $J$ band \\
 F6.2 & mag & Jmag  & Vega-based magnitude in 2MASS $J$ band\\
 F6.2 & mag & Jmerr & Vega-based magnitude error in 2MASS $J$ band\\
 A2   & --- & Jqual & 2MASS data quality flag for $J$ (A=best)\\
 A2   & --- & Hlim  & limit flag for 2MASS $H$ band \\
 F6.2 & mag & Hmag  & Vega-based magnitude in 2MASS $H$ band\\
 F6.2 & mag & Hmerr & Vega-based magnitude error in 2MASS $H$ band\\
 A2   & --- & Hqual & 2MASS data quality flag for $H$ (A=best)\\
 A2   & --- & Klim  & limit flag for 2MASS $K_s$ band \\
 F6.2 & mag & Kmag  & Vega-based magnitude in 2MASS $K_s$ band\\
 F6.2 & mag & Kmerr & Vega-based magnitude error in 2MASS $K_s$ band\\
 A2   & --- & Kqual & 2MASS data quality flag for $K_s$ (A=best)\\
 A18  & --- & DENISname & Name from DENIS catalog\\
 F6.2 & mag & Imag  & Vega-based DENIS magnitude in $I$ band\\
 F6.2 & mag & Imerr & Vega-based DENIS magnitude error in $I$ band\\
 F6.2 & mag & Jmag  & Vega-based DENIS magnitude in $J$ band\\
 F6.2 & mag & Jmerr & Vega-based DENIS magnitude error in $J$ band\\
 F6.2 & mag & Kmag  & Vega-based DENIS magnitude in $K$ band\\
 F6.2 & mag & Kmerr & Vega-based DENIS magnitude error in $K$ band\\
 A26  & --- & WISEname & Name from WISE (AllWISE) catalog or reject catalog\\
 A2   & --- & W1lim  & limit flag for WISE-1 ([3.4])  \\
 F6.2 & mag & W1mag  & Vega-based magnitude in WISE-1 ([3.4]) \\
 F6.2 & mag & W1merr & Vega-based magnitude error in WISE-1 ([3.4]); value of $-9$ for those measures that are limits\\
 A2   & mag & W1qual & WISE-1 ([3.4]) data quality flag (A=best)\\
 A2   & --- & W2lim  & limit flag for WISE-2 ([4.6])  \\
 F6.2 & mag & W2mag  & Vega-based magnitude in WISE-2 ([4.6]) \\
 F6.2 & mag & W2merr & Vega-based magnitude error in WISE-2 ([4.6]); value of $-9$ for those measures that are limits\\
 A2   & mag & W2qual & WISE-2 ([4.6]) data quality flag (A=best)\\
 A2   & --- & W3lim  & limit flag for WISE-3 ([12])  \\
 F6.2 & mag & W3mag  & Vega-based magnitude in WISE-3 ([12]) \\
 F6.2 & mag & W3merr & Vega-based magnitude error in WISE-3 ([12]); value of $-9$ for those measures that are limits\\
 A2   & mag & W3qual & WISE-3 ([12]) data quality flag (A=best)\\
 A2   & --- & W4lim  & limit flag for WISE-4 ([22])  \\
 F6.2 & mag & W4mag  & Vega-based magnitude in WISE-4 ([22]) \\
 F6.2 & mag & W4merr & Vega-based magnitude error in WISE-4 ([22]); value of $-9$ for those measures that are limits\\
 A2   & mag & W4qual & WISE-4 ([22]) data quality flag (A=best)\\
 A27  & --- & SEIPname & Name from SEIP source list\\
 F6.2 & mag & I1mag  & Vega-based magnitude in IRAC-1 ([3.6]) \\
 F6.2 & mag & I1merr & Vega-based magnitude error in IRAC-1 ([3.6]) \\
 F6.2 & mag & I2mag  & Vega-based magnitude in IRAC-2 ([4.5]) \\
 F6.2 & mag & I2merr & Vega-based magnitude error in IRAC-2 ([4.5]) \\
 F6.2 & mag & I3mag  & Vega-based magnitude in IRAC-3 ([5.8]) \\
 F6.2 & mag & I3merr & Vega-based magnitude error in IRAC-3 ([5.8]) \\
 F6.2 & mag & I4mag  & Vega-based magnitude in IRAC-4 ([8]) \\
 F6.2 & mag & I4merr & Vega-based magnitude error in IRAC-4 ([8]) \\
 F6.2 & mag & M1mag  & Vega-based magnitude in MIPS-1 ([24]) \\
 F6.2 & mag & M1merr & Vega-based magnitude error in MIPS-1 ([24]) \\
 A14  & --- & IRASPSCname & Name from IRAS Point Source Catalog (PSC)\\
 F6.2 & mag & IRAS1PSCmag & Vega-based magnitude in IRAS-1 ([12]) from PSC; errors taken to be 0.22 mags \\
 F9.2 & Jy & IRAS1PSCfd  & flux density in Jy in IRAS-1 from PSC \\
 A2   & mag & IRAS1PSCqual & Data quality flag for IRAS-1 from PSC (3=best, 1=limit)\\
 F6.2 & mag & IRA21PSCmag & Vega-based magnitude in IRAS-2 ([25]) from PSC; errors taken to be 0.22 mags \\
 F9.2 & Jy & IRAS2PSCfd  & flux density in Jy in IRAS-2 from PSC \\
 A2   & mag & IRAS2PSCqual & Data quality flag for IRAS-2 from PSC (3=best, 1=limit)\\
 F6.2 & mag & IRAS3PSCmag & Vega-based magnitude in IRAS-3 ([60]) from PSC; errors taken to be 0.22 mags \\
 F9.2 & Jy & IRAS3PSCfd  & flux density in Jy in IRAS-3 from PSC \\
 A2   & mag & IRAS3PSCqual & Data quality flag for IRAS-3 from PSC (3=best, 1=limit)\\
 F6.2 & mag & IRAS4PSCmag & Vega-based magnitude in IRAS-4 ([100]) from PSC; errors taken to be 0.22 mags \\
 F9.2 & Jy & IRAS4PSCfd  & flux density in Jy in IRAS-4 from PSC \\
 A2   & mag & IRAS4PSCqual & Data quality flag for IRAS-4 from PSC (3=best, 1=limit)\\
 A14  & --- & IRASFSCname & Name from IRAS Faint Source Catalog (FSC)\\
 F6.2 & mag & IRAS1FSCmag & Vega-based magnitude in IRAS-1 ([12]) from FSC; errors taken to be 0.22 mags \\
 F9.2 & Jy & IRAS1FSCfd  & flux density in Jy in IRAS-1 from FSC \\
 A2   & mag & IRAS1FSCqual & Data quality flag for IRAS-1 from FSC (3=best, 1=limit)\\
 F6.2 & mag & IRA21FSCmag & Vega-based magnitude in IRAS-2 ([25]) from FSC; errors taken to be 0.22 mags \\
 F9.2 & Jy & IRAS2FSCfd  & flux density in Jy in IRAS-2 from FSC \\
 A2   & mag & IRAS2FSCqual & Data quality flag for IRAS-2 from FSC (3=best, 1=limit)\\
 F6.2 & mag & IRAS3FSCmag & Vega-based magnitude in IRAS-3 ([60]) from FSC; errors taken to be 0.22 mags \\
 F9.2 & Jy & IRAS3FSCfd  & flux density in Jy in IRAS-3 from FSC \\
 A2   & mag & IRAS3FSCqual & Data quality flag for IRAS-3 from FSC (3=best, 1=limit)\\
 F6.2 & mag & IRAS4FSCmag & Vega-based magnitude in IRAS-4 ([100]) from FSC; errors taken to be 0.22 mags \\
 F9.2 & Jy & IRAS4FSCfd  & flux density in Jy in IRAS-4 from FSC \\
 A2   & mag & IRAS4FSCqual & Data quality flag for IRAS-4 from FSC (3=best, 1=limit)\\
 A28  & --- & AKARIIRCname & Name from AKARI IRC catalog, v1\\
 F9.2 & Jy & AKARI9fd  & flux density in Jy in 9 microns from AKARI IRC \\
 F9.2 & Jy & AKARI9fderr  & error in flux density in Jy in 9 microns from AKARI IRC \\
 F9.2 & Jy & AKARI18fd  & flux density in Jy in 18 microns from AKARI IRC \\
 F9.2 & Jy & AKARI18fderr  & error in flux density in Jy in 18 microns from AKARI IRC \\
 A28  & --- & AKARIFISname & Name from AKARI FIS catalog, v1\\
 F9.2 & Jy & AKARI65fd  & flux density in Jy in 65 microns from AKARI FIS \\
 F9.2 & Jy & AKARI65fderr  & error in flux density in Jy in 65 microns from AKARI FIS \\
 F9.2 & Jy & AKARI90fd  & flux density in Jy in 90 microns from AKARI FIS \\
 F9.2 & Jy & AKARI90fderr  & error in flux density in Jy in 90 microns from AKARI FIS \\
 F9.2 & Jy & AKARI140fd  & flux density in Jy in 140 microns from AKARI FIS \\
 F9.2 & Jy & AKARI140fderr  & error in flux density in Jy in 140 microns from AKARI FIS \\
 F9.2 & Jy & AKARI160fd  & flux density in Jy in 160 microns from AKARI FIS \\
 F9.2 & Jy & AKARI160fderr  & error in flux density in Jy in 160 microns from AKARI FIS \\
 A18  & --- & MSXname & Name from MSX catalog \\
 F9.2 & Jy & MsxAfd  & flux density in Jy in MSX Band A (7.76 \mum) \\
 F9.2 & Jy & MsxAfderr & error in flux density in Jy in MSX Band A (7.76 \mum) -- as reported, may be very large \\
 F9.2 & Jy & MsxB1fd & flux density in Jy in MSX Band B1 (4.29 \mum) \\
 F9.2 & Jy & MsxB1fderr& error in flux density in Jy in MSX Band B1 (4.29 \mum) -- as reported, may be very large \\
 F9.2 & Jy & MsxB2fd & flux density in Jy in MSX Band B2 (4.35 \mum) \\
 F9.2 & Jy & MsxB2fderr& error in flux density in Jy in MSX Band B2 (4.35 \mum) -- as reported, may be very large \\
 F9.2 & Jy & MsxCfd  & flux density in Jy in MSX Band C (11.99 \mum) \\
 F9.2 & Jy & MsxCfderr & error in flux density in Jy in MSX Band C (11.99 \mum) -- as reported, may be very large \\
 F9.2 & Jy & MsxDfd  & flux density in Jy in MSX Band D (14.55 \mum) \\
 F9.2 & Jy & MsxDfderr & error in flux density in Jy in MSX Band D (14.55 \mum) -- as reported, may be very large \\
 F9.2 & Jy & MsxEfd  & flux density in Jy in MSX Band E (20.68 \mum) \\
 F9.2 & Jy & MsxEfderr & error in flux density in Jy in MSX Band E (20.68 \mum) -- as reported, may be very large \\
\enddata
\tablenotetext{a}{This catalog is available in its entirety in the
online version of Table 1.}
\end{deluxetable}

\begin{deluxetable}{cccp{2.5cm}p{5cm}}
\tabletypesize{\scriptsize}
\rotate
\tablecaption{Overview of Photometric Studies and Data Included\label{tab:allstudies}}
\tablewidth{0pt}
\tablehead{\colhead{Dataset} & \colhead{Band(s) used} & \colhead{Search
radius\tablenotemark{a} ($\arcsec$)} & \colhead{Fraction with match} & \colhead{notes}}
\startdata
2MASS &  $JHK_s$ (1.2-2.2 \mum) & 1\tablenotemark{b}& 304/316=96\% & primary catalog; many saturated\\
WISE & 3.4, 4.5, 12, 22 \mum\ & 1 & 311/316=98\% & primary catalog; many saturated in part. ($\sim$20\% of sample required $>$1$\arcsec$ counterpart match)\\
IRAS & 12, 25, 60, 100 \mum & $<$20 & PSC: 159/316=50\%; FSC: 121/316=38\% & data used for original de la Reza studies\\
AKARI & 9, 18, 65, 90, 140, 160 \mum\ & 1 & IRC: 221/316=70\%; FIS: 36/316=11\% & supplementary data ($\sim$20\% of IRC sources that had matches 
required $>$1$\arcsec$ counterpart match, most of the FIS required $>$1$\arcsec$.)\\
MSX & 4.3, 4.4, 7.8, 12.0, 14.6, 20.6 \mum\ & 20 & 54/316=17\%& supplementary data\\
SEIP & 3.6, 4.5, 5.8, 8, 24 \mum\ & 1 & 39/316=12\%& very low fractional coverage\\
DENIS & 0.82, 1.25, 2.15 \mum\ & 1 & 83/316=26\%& provides deeper $K_s$ than 2MASS; only 58 have $K_s$, and none of those are lacking in 2MASS $K_s$ \\
SDSS & $ugriz$ (0.29-0.91 \mum) & 2 & 94/316=30\% & supplementary data \\
NOMAD,C12 & $UBVR$ (0.36-0.7 \mum) & 1 & 277/316=88\% & literature data from NOMAD or C12 \\
\enddata
\tablenotetext{a}{Characteristic distance to source match; some sources with
poor positions required a much larger search radius.}
\tablenotetext{b}{Most sources found a match within 1$\arcsec$ (histogram
of distances peaks strongly below 0.2$\arcsec$), but $\sim$10\% required
larger (by-eye) matches.}
\end{deluxetable}

\clearpage

\section{Archival Data}
\label{sec:largedatasets}
\label{sec:archivaldata}

In this section, we discuss the catalogs we searched for
detections of our sources; they are summarized in
Table~\ref{tab:allstudies}, along with the fraction of the sample
having matches in each of these catalogs. Table~\ref{tab:bigdatatable}
includes all the crossmatched sources (names and reported
brightnesses) discussed below.

\subsection{Overall Approach to Archival Data}

All of the large-area catalogs described here were merged by position with
a catalog-dependent search radius to the position we obtained for our
targets as described above. Typically, the closest source by
position was taken to be the match, and often the best match was within
1$\arcsec$. However, each source was investigated in the images from the
Palomar Observatory Digital Sky Survey, the Sloan Digital Sky Survey, the
Two-Micron All Sky Survey, and WISE. Nebulosity, source confusion, or
extended sources were all noted. Spectral energy distributions (SEDs) were
constructed (using zero points if necessary as provided in the
corresponding survey documentation) as an additional check on the source
matching -- obvious discontinuities in the SED suggested problems with
source matching, and a better match was sought (but not always found).  If
a source other than the closest source by position was determined to be a
better match, then the match was forced to be the better match, even if it
was $>1\arcsec$ away. We primarily used the Infrared Science Archive (IRSA)
tool
FinderChart\footnote{http://irsa.ipac.caltech.edu/applications/finderchart/}
for this process, along with the one-to-one catalog matching feature in the
IRSA catalog search tool. Significant issues with images and SEDs will be
discussed in the next section (\S\ref{sec:droppedsrcs}).

\subsection{Primary Catalogs}
\label{sec:primarycatalogs}

The Two-Micron All Sky Survey (2MASS; Skrutskie \etal\ 2006) obtained
data over the whole sky at $JHK_s$ bands.  We found matches to most of
our sources well within 1$\arcsec$ -- a histogram of distances peaks
strongly below 0.2$\arcsec$. However, $\sim$10\% of the targets
(largely those still having original coordinates from IRAS) required
larger (by-eye) matches, up to 15$\arcsec$ away.  Many of our targets
are quite bright and are therefore saturated in the 2MASS catalog. For
most of these, we can obtain at least estimates of $K_s$ from the
Naval Observatory Merged Astrometric Dataset (NOMAD; Zacharias \etal\
2005), though empirically we have found that the errors as reported
there are likely significantly underestimated, perhaps representing
statistical errors only (not including systematics). For one source,
the $JHK_s$ brightnesses had to be retrieved from the extended source
catalog (rather than the point source catalog). For many of our bright
targets, the formal photometric quality as reported in 2MASS may be
poor, but the points are in good agreement with the rest of the SED
assembled here. We thus retained 2MASS measurements even if the
photometric quality was deemed poor by the 2MASS pipeline. (About 60\%
of the sources with 2MASS counterparts have $K_s$ photometric quality
`A'; $\sim$35\% have nominal photometric quality `D' or worse.) Limits
reported in the catalog were retained as limits here.

2MASS provides the coordinate system to which other catalogs including
WISE are anchored, so, given the very close positional matches for
2MASS, we expected (and found) comparable high-quality matches with
those other catalogs.  

IRAS surveyed the sky in 1983 in four bands, 12, 25, 60, and 100 \mum.
As the first all-sky infrared survey, it is relatively low spatial
resolution and relatively shallow. The Point Source Catalog (PSC;
Beichman \etal\ 1988) reports on sources smaller than 0.5-2$\arcmin$
in the in-scan direction (where the native survey pixels are
rectangular).  The typical full width at half max (FWHM) of sources in
the IRAS Sky Survey Atlas (ISSA) data products (which is what appears
in FinderChart) is
3.4$\arcmin$-4.7$\arcmin$\footnote{http://irsa.ipac.caltech.edu/IRASdocs/issa.exp.sup/ch1/C.html}. 
The IRAS Faint Source Catalog (FSC; Moshir \etal\ 1992) is a
reprocessing of the IRAS data that obtains, among other things, better
positional accuracy and reaches fainter flux densities. We searched
both the IRAS PSC and FSC for counterparts to our sources.  Since the
dlR97 sources were selected based on IRAS properties, all of them have
detections in the PSC, and 36 are also detected in the FSC.  Nearly
half (103/235) of our remaining sources have IRAS detections in either
the PSC or FSC in any band, though only about a third of the C12
sources have an IRAS detection (in any band).  Overall, 181 have
sources in either the PSC or FSC, and 99 have counterparts in both the
PSC and FSC.

The IRAS data (where available) for our targets appear in
Table~\ref{tab:bigdatatable}. We used the non-color corrected flux
density at the various bands as reported in the catalogs, and used the
Vega zero points as reported in the online IRAS
documentation\footnote{http://irsa.ipac.caltech.edu/IRASdocs/exp.sup/ch6/C2a.html}
to convert the flux densities to magnitudes, namely 28.3, 6.73, 1.19,
and 0.43 Jy for the four bands, respectively. No errors are reported,
so we took a flat 20\% flux density uncertainty, which is 0.22 mag. 
We merged catalogs without regard to flux quality, though the quality
is noted in our catalog. Nearly all of the detected sources are the
highest quality (qual=3) in both the PSC and FSC at 12 \mum, but
$>$80\% of these sources are the {\em lowest} quality in the PSC and
FSC at 100 \mum\ (qual=1). The lowest quality measurements are,
according to the documentation, meant to be limits. In some cases,
even the nominal limits are in good agreement with detections from
other instruments. We retained the measurements and the flux quality
flags in our database. 

WISE surveyed the whole sky at 3.4, 4.6, 12, and 22 \mum; all of the
available WISE data taken between 2010 Jan and 2011 Feb were
incorporated into the AllWISE catalog (Cutri \etal\ 2014), which we
used for this work.  In three cases, brightnesses for a target had to
be retrieved from the AllWISE catalog of rejects. Since the dlR97
sources were selected based on the relatively shallow IRAS data, many
were saturated in at least one WISE band; many fewer of the C12
sources were saturated.  Of the WISE detections, only those with data
quality flags `A', `B', or `C' were retained, but most detections were
`A' or `B'; the fraction of detections with data quality flag `C' is
10\%, 5\%, 1\%, and 3\%  for the four WISE channels, respectively. 
For many of our very bright targets, the formal photometric quality as
reported in WISE may be poor, but the points are in good agreement
with the rest of the SED; limits from the catalog were retained in our
database.

Given the relatively low spatial resolution of IRAS compared to 2MASS
or WISE, we did not necessarily expect to find very close positional
matches to sources with solely IRAS positions. However, many sources
whose coordinates were the original IRAS positions found very
close matches in 2MASS and/or WISE, demonstrating the high quality of
those original IRAS positions. The places where IRAS did not match
well were largely those where source confusion pulled the photocenter
position off from the brightest source in WISE; see additional
discussion on source confusion issues below.

All of the abundances and associated information (\teff, log $g$) were
most often taken from the papers reporting the star as Li-rich -- see
the Appendix table for the specific literature reference. We allowed
$A$(Li) from the non-local-thermodynamic-equilibrium (NLTE) estimates
to take precedence over LTE estimates, but in some casese, only LTE
abundances were available.  In a few cases, only Li equivalent widths
were available in the literature, in which case we did not even
attempt to estimate an $A$(Li), and thus the sources are effectively
dropped from analysis requiring $A$(Li). In $\sim$20 cases, McDonald
\etal\ (2012) provided a \teff\ estimate when no other was available
from the literature. For the 14 stars with super-solar metallicities
in C12, we provide corrected NLTE abundances here.

\subsection{Secondary Catalogs}

The AKARI mission (Murakami \etal\ 2007) surveyed the sky in
2006-2007, in wavelengths between 1.8 and 180 \mum\ using two
instruments, the Infrared Camera (IRC) and the Far-Infrared Surveyor
(FIS). We searched the IRC catalog for counterparts at 9 and 18 \mum,
and the FIS catalog for counterparts at 65, 90, 140, and 160 \mum. A
large fraction ($\sim$70\%) of our sources have a counterpart in at
least one IRC band; relatively few ($\sim$11\%) of our sources have a
counterpart in at least one FIS band. The AKARI data cover the whole
sky, and can provide valuable data to help populate the SEDs of our
targets. For the majority of targets where we have AKARI counterparts,
the AKARI data are consistent with WISE and/or IRAS measurements.  We
did not sort the AKARI matches by photometric quality; because our
sources are bright, even those with low photometric quality flags
matched the existing SED quite well. For the FIS sources where the
photometric quality is the lowest, no errors are given, so we adopted
a conservative uncertainty of 50\% for those sources. 

The Midcourse Space Experiment (MSX; Egan \etal\ 2003) surveyed the
Galactic Plane in 1996-1997 at several bands between 8 and 21 \mum\ --
Band A=7.76 \mum, B1=4.29 \mum, B2=4.35 \mum, C=11.99 \mum, D=14.55
\mum, and E=20.68 \mum. The spatial resolution of these images range
from 20$\arcsec$ to 72$\arcsec$. Relatively few (just 54) of our
targets have MSX counterparts, in no small part because of sky
coverage, but the lower sensitivity of the MSX instruments also plays
a role.  The errors we report in Table~\ref{tab:bigdatatable} are the
errors reported in the MSX catalog, and as such may be very large.

The Deep Near Infrared Survey of the Southern Sky (DENIS) conducted a
survey of the southern sky at $I$, $J$, and $K$ bands.  It is deeper than
2MASS. In several cases, because the search radius for a counterpart had to
be large, it was clear upon construction of the SEDs that the nearest
source by position was not the best match (often because a fainter source
was closer to the given position than the real target), and so the match
was rejected. In the end, only 58 of our targets have a $K_s$ magnitude
from DENIS, and none of those are lacking a 2MASS $K_s$. Therefore, the
DENIS measurements do not play a role in identification of IR
excesses, but were retained in those few cases to better define the SED. 


The entire archive of photometric 3-24 \mum\ cryogenic-era Spitzer
Space Telescope (Werner \etal\ 2004) data has been reprocessed and
images and source lists released as part of the Enhanced Imaging
Products (SEIP). Spitzer is generally more sensitive (and has higher
spatial resolution) than WISE or AKARI. However, the SEIP source list
is limited to those with signal-to-noise ratios greater than 10.
Because Spitzer is a pointed mission, the entire sky is not covered,
and only 12\% of our sources have counterparts in the SEIP source
list. These measurements were retained in those few cases specifically
because they are higher spatial resolution, and can provide valuable
insight into the reliability of the flux densities provided by IRAS,
AKARI, and WISE. We have found that the errors as reported in the
SEIP are likely statistical and probably do not include a calibration
uncertainty floor. We have added 4\% errors in quadrature to the
reported errors.

The Sloan Digital Sky Survey (SDSS; see, e.g., Ahn \etal\ 2014 and
references therein) has surveyed a significant fraction of the sky at
$ugriz$ (optical) bands. These data, where available, help define the
Wien side of our objects' SEDs; they do not aid in calculation of IR
excesses, but they `guide the eye' to identify the photosphere. About
30\% of our targets have SDSS counterparts in at least one band.
Several more of our targets appear in SDSS images, but are far too
bright for reliable photometry. We used images and photometry as
retrieved via IRSA's FinderChart. 

The Digitized Sky Survey (DSS) is a digitization of the photographic
sky survey plates from the Palomar (the Palomar Observatory Sky
Survey, POSS) and UK Schmidt telescopes.  We used images from the DSS
retrieved via IRSA's FinderChart to check on source confusion and
multiplicity.

NOMAD reports broadband optical photometry for most of our targets; 
C12 reports optical photometry for their targets. Those values were
included in our database, as for the SDSS optical data above, to define the
short-wavelength side of the SED. 


\section{Dropped Sources}
\label{sec:droppedsrcs}

In this section, we describe the set of targets that we have to drop
from our dataset because they are not detected at sufficient bands
(Sec.~\ref{sec:verysparseseds}, Sec.~\ref{sec:sparseseds}), or that
are likely subject to source confusion where the IRAS detection is
likely composed of more than one source, or where the bright source in
POSS is not responsible for the IR flux (Sec.~\ref{sec:srcconf}).  The
24  sources we identify as subject to source confusion come from
studies with sources first identified in the low spatial resolution
IRAS data and followed up in the optical.

\subsection{Sources with Very Sparse SEDs}
\label{sec:verysparseseds}

There are 10 sources that do not appear in many of the catalogs we
used here, such that they have very sparsely populated SEDs. These
SEDs are sparse enough that they cannot be handled in the same way as
the other sources in the set -- they have no $K_s$ or [22] measures,
and sometimes no WISE data at all.  These sources are listed in
Table~\ref{tab:sparsesrcs}.  All of these very sparse SED stars are
from Kirby \etal\ (2012), and are in dwarf spheroidal galaxies. They
are just too far away to be detected in enough bands in the surveys we
used.    Comments on these objects appear in
Table~\ref{tab:sparsesrcs}. For each source, we inspected the SED and
the photometry, looking for any evidence of IR excess at the available
bands, and found none.  

These 10 sparse SED sources are, of necessity, frequently dropped from
subsequent figures and discussion here. We reiterate, however, that
they are all from the `addtional literature' sample; none are from C12
or dlR97.

\subsection{Sources with Relatively Sparse SEDs beyond 10 \mum}
\label{sec:sparseseds}

There are 36 sources that have relatively well-populated SEDs, but are
missing detections past 10 or 20 \mum. We cannot treat those sources
in exactly the same way as the rest of the stars in the set, but at
least we can constrain whether or not there is an IR excess, more so
than for the stars in the previous section. These sources are listed
in Table~\ref{tab:sparsesrcs}. There are two sources for which an IR
excess cannot be ruled out given the available detections. The star
known as ``For 90067'' could be consistent with an IR excess at 8
\mum, given the available IRAC data, but the error on the [8] point is
large. (Following the approach below in Sec.~\ref{sec:smirx}, but
customized to this star, $\chi_{K,[8]}$ is 3.1.) SDSS J0632+2604 has a
much more convincing excess, with a [3.4]$-$[12]=1.45. 


These sources do not often appear in the subsequent figures and
discussion, because they are, for example, missing [22], and thus
cannot appear in a figure plotting [3.4]$-$[22]. However, the two
possible excess sources here are sometimes included in the counts of
sources with IR excesses, and where we do so, we note it explicitly.
Nearly all of these sources (26/36) are objects we take to be Li-rich.
Of these, SDSS J0632+2604 has the largest IR excess and also the
largest Li abundance, with $A$(Li)$_{\rm LTE}=4.2$~dex. 

None of these relatively sparse sources are from the dlR97 sample.
Seven of them are from C12, and the remainder are from the `addtional
literature' sample.

\begin{deluxetable}{lccp{12cm}}
\tabletypesize{\scriptsize}
\tablecaption{Objects with Sparse SEDs\label{tab:sparsesrcs}}
\tablewidth{0pt}
\rotate
\tablehead{\colhead{name} &  \colhead{very sparse SED?\tablenotemark{a}}
&\colhead{IR excess?\tablenotemark{b}} &\colhead{notes}}
\startdata
Scl 1004838	&	x	&	\nodata	&	Distant object; not well-populated SED. No evidence for IR excess.	\\
Scl 1004861	&	x	&	\nodata	&	Distant object; not well-populated SED. No evidence for IR excess.	\\
For 55609	&	\nodata	&	\nodata	&	No W3W4 but 3 IRAC bands; no evidence for excess.	\\
For 60521	&	\nodata	&	\nodata	&	No W3W4 but 3 IRAC bands; no evidence for excess.	\\
For 90067	&	\nodata	&	x	&	No W3W4, but
all 4 IRAC bands. Excess possible at 8 \mum\ despite large error.
Following the approach in Sec.~\ref{sec:smirx} below, but customized to this star,
$\chi_{K,[8]}$ is 3.1. \\
For 100650	&	x	&	\nodata	&	Distant object; not well-populated SED. No evidence for IR excess.	\\
G0300+00.29	&	\nodata	&	\nodata	&	No W4; no evidence for excess	\\
SDSS J0304+3823	&	\nodata	&	\nodata	&	No W3W4; no evidence for excess	\\
RAVEJ043154.1-063210	&	\nodata	&	\nodata	&	No W4; no evidence for excess	\\
G0453+00.90	&	\nodata	&	\nodata	&	No W4; no evidence for excess	\\
SDSS J0535+0514	&	\nodata	&	\nodata	&	No W4; no evidence for excess	\\
Be 21 T50 	&	\nodata	&	\nodata	&	No W4; no evidence for excess	\\
SDSS J0632+2604	&	\nodata	&	x	&	No W4; W3
suggests most likely has excess ([3.4]$-$[12]=1.45).	\\
Tr5 3416	&	\nodata	&	\nodata	&	No W4; colors
suggest could have small excess at [12] ([3.4]$-$[12]=0.41);
following the approach in Sec.~\ref{sec:smirx} below, but customized to this star,
$\chi_{[3.4],[12]}$=1.97, not significant.	\\
SDSS J0654+4200	&	\nodata	&	\nodata	&	No W4; no evidence for excess	\\
G0653+16.552	&	\nodata	&	\nodata	&	No W4; no evidence for excess	\\
G0654+16.235	&	\nodata	&	\nodata	&	No W4; no evidence for excess	\\
SDSS J0720+3036	&	\nodata	&	\nodata	&	No W3W4; no evidence for excess	\\
SDSS J0808-0815	&	\nodata	&	\nodata	&	No W4; no evidence for excess	\\
SDSS J0831+5402	&	\nodata	&	\nodata	&	No W3W4; no evidence for excess	\\
SDSS J0936+2935	&	\nodata	&	\nodata	&	No W3W4; no evidence for excess	\\
G0935-05.152	&	\nodata	&	\nodata	&	No W4; no evidence for excess	\\
G0946+00.48	&	\nodata	&	\nodata	&	No W4; no evidence for excess	\\
LeoI 71032	&	x	&	\nodata	&	Distant object; not well-populated SED. No evidence for IR excess.	\\
LeoI 60727	&	x	&	\nodata	&	Distant object; not well-populated SED. No evidence for IR excess.	\\
LeoI 32266	&	x	&	\nodata	&	Distant object; not well-populated SED. No evidence for IR excess.	\\
LeoI 21617	&	x	&	\nodata	&	Distant object; not well-populated SED. No evidence for IR excess.	\\
C1012254-203007	&	\nodata	&	\nodata	&	No W4; no evidence for excess	\\
SDSS J1105+2850	&	\nodata	&	\nodata	&	No W3W4; no evidence for excess	\\
LeoII C-7-174	&	x	&	\nodata	&	Distant object; not well-populated SED. No evidence for IR excess.	\\
LeoII C-3-146	&	x	&	\nodata	&	Distant object; not well-populated SED. No evidence for IR excess.	\\
G1127-11.60	&	\nodata	&	\nodata	&	No W4; no evidence for excess	\\
M68-A96=Cl* NGC 4590 HAR 1257	&	\nodata	&	\nodata	&	No W4, but all 4 IRAC bands. No evidence for excess.	\\
SDSS J1310-0012	&	\nodata	&	\nodata	&	No W3W4; no evidence for excess	\\
CVnI 195\_195	&	\nodata	&	\nodata	&	No W3W4, but 3 IRAC bands. No evidence for excess.	\\
CVnI 196\_129	&	x	&	\nodata	&	Distant object; not well-populated SED. No evidence for IR excess.	\\
M3-IV101=Cl* NGC 5272 SK 557	&	\nodata	&	\nodata	&	No W4, but all 4 IRAC bands. No evidence for excess.	\\
SDSS J1432+0814	&	\nodata	&	\nodata	&	No W3W4; no evidence for excess	\\
SDSS J1522+0655	&	\nodata	&	\nodata	&	No W3W4; no evidence for excess	\\
SDSS J1607+0447	&	\nodata	&	\nodata	&	No W3W4; no evidence for excess	\\
SDSS J1901+3808	&	\nodata	&	\nodata	&	No W4; no evidence for excess	\\
SDSS J1909+3837	&	\nodata	&	\nodata	&	No W4; no evidence for excess	\\
KIC 4937011	&	\nodata	&	\nodata	&	No W4, but all 4 IRAC bands. No evidence for excess.	\\
SDSS J2019+6012	&	\nodata	&	\nodata	&	No W4; no evidence for excess	\\
SDSS J2200+4559	&	\nodata	&	\nodata	&	No W4; no evidence for excess	\\
SDSS J2206+4531	&	\nodata	&	\nodata	&	No W4; no evidence for excess	\\
\enddata
\tablenotetext{a}{This column is populated if the SED is very sparse
(Sec.~\ref{sec:verysparseseds}).}
\tablenotetext{b}{This column is populated if the object could have an
IR excess by 8 to 12 \mum\ (Sec.~\ref{sec:sparseseds}).}
\end{deluxetable}

\clearpage
\subsection{Source Confusion}
\label{sec:srcconf}

As mentioned above, it is often the case that a bright source in IRAS
is also a bright source in the optical, but sometimes this is not a
good assumption. Now that 2MASS and WISE images are available, it is
far easier than it was in the 1990s to identify sources over 3 orders
of magnitude in wavelength (0.5 to 100 \mum).  In 24 of our targets,
we believe that there are likely issues of source confusion. In these
cases, at least one of the following things can be seen in the images:
(a) the IR excess seen in the IRAS images is due to more than one WISE
(or 2MASS) source, likely unresolved in the IRAS catalog; or (b) the
nearest bright source in POSS, which is most likely the source for
which an optical spectrum to measure lithium was obtained, is not
responsible for most of the IR flux density detetected via IRAS. 

\begin{figure}[h]
\epsscale{0.4}
\plotone{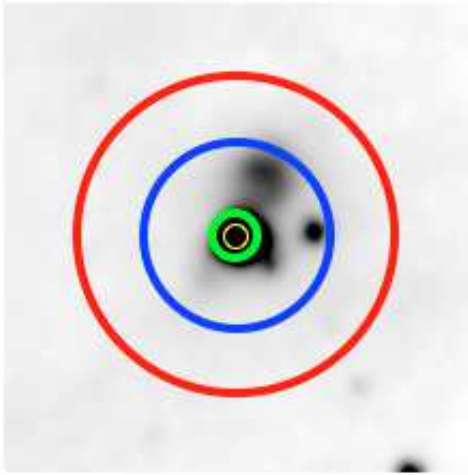}
\caption{Image of IRAS06365+0223 in [12], reverse greyscale,
300$\arcsec$ on a side as most of the subsequent image postage stamps
are. The circles represent the range of IRAS spatial resolutions; see
text for more discussion. The red circle is 3.4$\arcmin$ in diameter,
representative of the typical FWHM of shorter-wavelengths sources in
the ISSA images. The blue circle and the green circle are
2 and 0.5$\arcmin$ in diameter, and represent the in-scan direction
resolution of the IRAS PSC.  The small yellow circle here is
15$\arcsec$ in diameter, representative of the `target' blue circle in
subsequent 300$\arcsec$ images. (It is yellow instead of blue just for
enhanced visibility in this reverse greyscale image.)
\label{fig:beamsizes}}
\end{figure}

For the first kind of source confusion, where the IR excess seen in
IRAS is due to more than one WISE source, understanding the original
IRAS resolution is important, because these issues of IRAS spatial
resolution can be subtle. Figure~\ref{fig:beamsizes} shows a WISE 12
\mum\ image of one of our targets, IRAS06365+0223, with circles
overlaid to represent the various possible IRAS resolutions. The image
is 300$\arcsec$=5$\arcmin$ on a side. The original IRAS PSC was
derived from images that had pixels that were substantially
rectangular, but included sources believed to be point sources with
sizes less than $\sim$0.5-2.0$\arcmin$ {\em in the in-scan
direction}\footnote{http://irsa.ipac.caltech.edu/IRASdocs/exp.sup/ch5/A2.html}.
The small green circle in Fig.~\ref{fig:beamsizes} is representative
of this 12 \mum\ 0.5$\arcmin$ resolution, and the large blue circle is
representative of this 100 \mum\ 2$\arcmin$ resolution. Note that
these are shown in the figure as circles, whereas in reality, this
resolution is only obtained in the in-scan direction, which is roughly
along lines of ecliptic longitude; the resolution is considerably
worse in the cross-scan direction. (For this target, lines of ecliptic
longitude are locally about 5$\arcdeg$ east of North, but this angle
varies over the sky.) The typical FWHM of sources in the
ISSA data products (ISSA images are shown in FinderChart) is
3.4$\arcmin$-4.7$\arcmin$\footnote{http://irsa.ipac.caltech.edu/IRASdocs/issa.exp.sup/ch1/C.html},
which is represented by the large red circle in
Fig.~\ref{fig:beamsizes} at 3.4$\arcmin$ (diameter). The multiple
sources seen in this WISE image, for example, are likely convolved
together for at least some of the IRAS measurements. This is an
important factor in several of the IRAS sources we discuss in this
section, meaning that the IR flux attributed to a single optical
source may not be correct, and the IR excess previously measured for a
given optical source may be significantly overestimated. 

The second kind of source confusion, where the bright source in POSS
is not responsible for the IR flux, is source confusion of a different
nature. In many cases, it is a good assumption that the bright source
in IRAS is also the bright source in POSS. However, in a significant
number of cases, all the sources in the region are of comparable
brightness in POSS, or the optically bright source is not the source
of most of the IR light. Now that 2MASS and WISE are  available, we
can trace the source across wavelengths to securely identify the
optical counterpart in the POSS images. In these cases, however, the
spectrum obtained to assess lithium may very well have been of the
bright POSS source, and not of the origin of the IR flux at all,
especially in those cases where no optical counterpart can be found in
the POSS images. 

\begin{figure}[h]
\epsscale{0.9}
\plotone{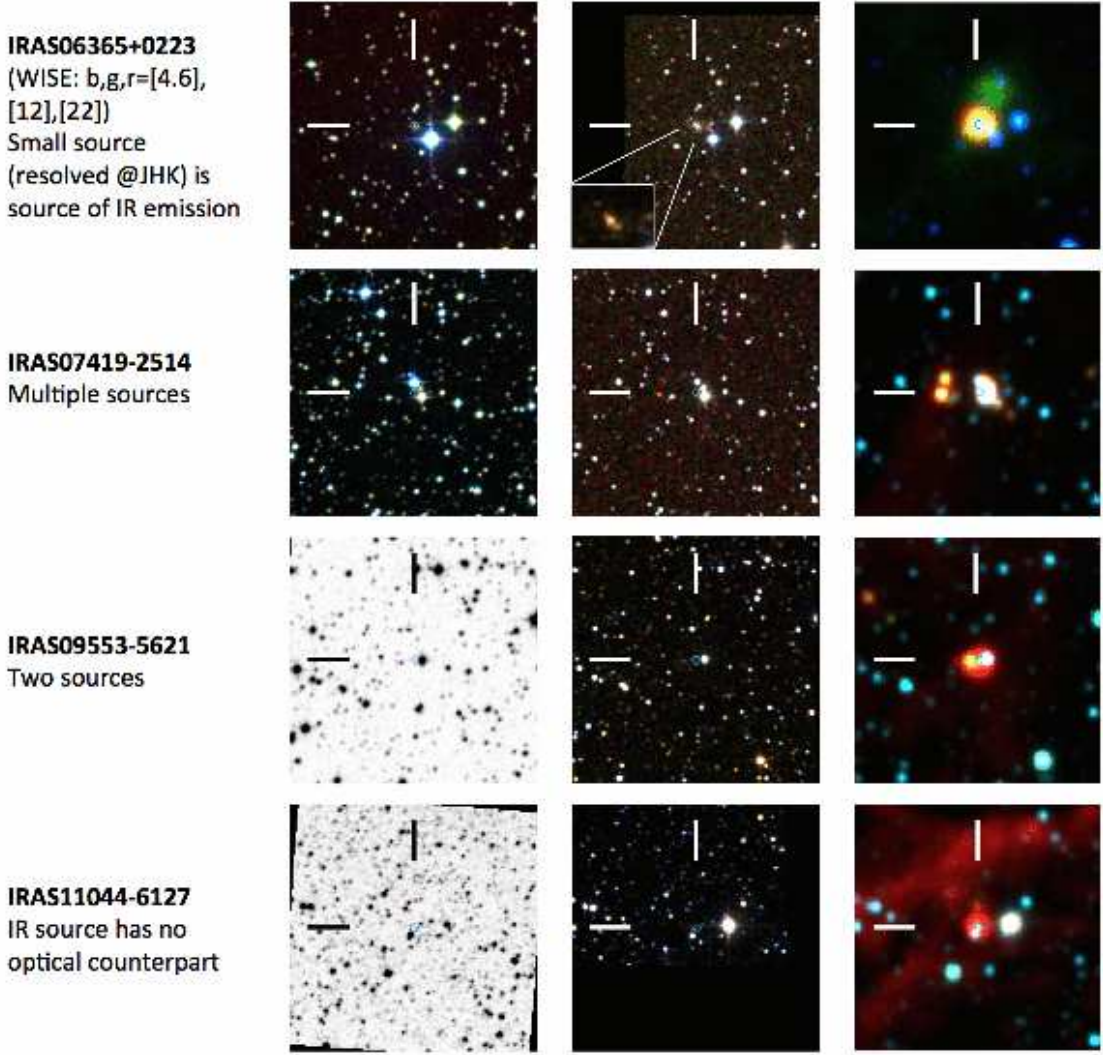}
\caption{First image column: POSS 3-color (DSS2 Blue/Red/IR for b/g/r
planes), or it is a single-band reverse greyscale DSS2 Red. Second
image column: 2MASS $JHK_s$ color image. Third image column: WISE
[3.4], [4.6], [12] for b/g/r, respectively, unless specified.  Images
are all 300$\arcsec$ on a side unless specified. North-up. Small blue
circle centered on position used for the target, and white/black hash
marks above and to the left help guide the eye to this position. 
Four rows are: 
(1) IRAS06365+0223, where the 2MASS image has an additional inset with
an enlargement of the source of the IR flux, and WISE has
b,g,r=[4.6],[12],[22]. The target position is correctly the source of
most of the long-wavelength flux, but is not a match to either of the
optically bright sources. 
(2) IRAS07419-2514. The target position is in amongst a small cluster
of sources, and corresponds to the photocenter of the aggregate source
seen by IRAS. 
(3) IRAS09553-5621. The target position is in between two sources
bright at WISE bands, but only one source appears at $K_s$. That
source that appears at $K_s$ also appears in POSS images. The sources
are not separable at [22], and are likely both contributing to the
measured IRAS flux density. 
(4) IRAS11044-6127. There is no easily visible source at the target
position in POSS; the bright source seen as white immediately to the
West of the target is the brightest source in the field in the DSS-IR
and 2MASS, but it is comparably bright to the other stars in DSS.
\label{fig:imagefig1}}
\end{figure}

\begin{figure}[h]
\epsscale{0.9}
\plotone{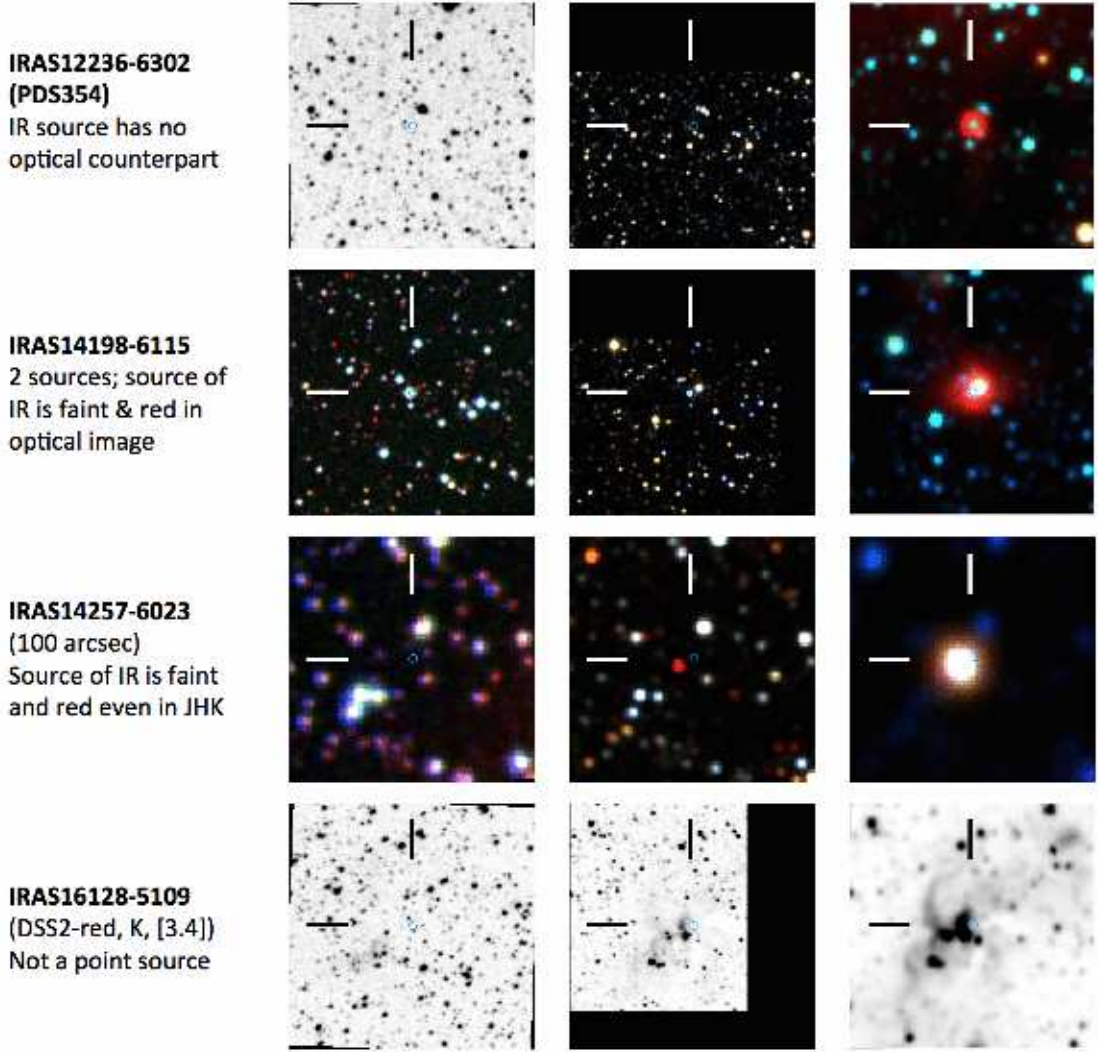}
\caption{(Notation is as in Fig.~\ref{fig:imagefig1}.)
Four rows are: 
(1) IRAS12236-6302 (PDS354). The blue arc immediately above the
target in WISE can be seen to be 3 distinct sources in 2MASS and, combined,
are the brightest thing in the POSS 300$\arcsec$ images. These sources
are not responsible for most of the IR flux measured at the target
position.
(2) IRAS14198-6115.  There are two sources here that are distinct in
2MASS and marginally resolved in [3.4], but are indistinguishable by
[22].  
(3) IRAS14257-6023.  Images are 100$\arcsec$ on a side to better show
the source that is very red in 2MASS and is dominating the measured
flux by WISE bands. The brightest source in POSS is the source
appearing as white in 2MASS to the North and slightly West of the target
position, but it is not responsible for the IR flux. 
(4) IRAS16128-5109. This is not a point source in the IR, and is very
bright (saturated in WISE) by [22].  Single bands are shown in the
optical, NIR, and MIR to better show the nebulosity.
\label{fig:imagefig2}}
\end{figure}

\begin{figure}[h]
\epsscale{0.9}
\plotone{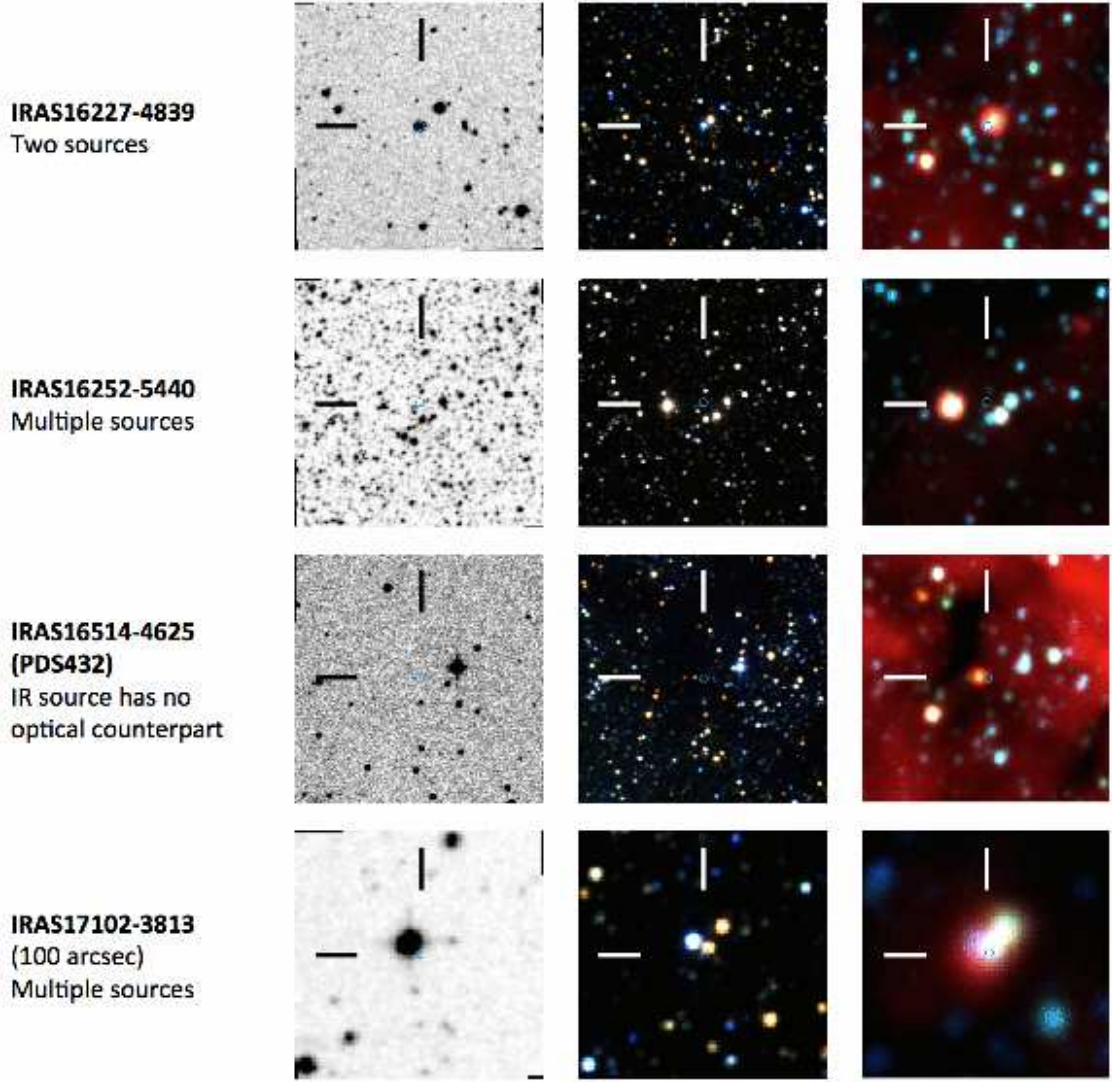}
\caption{(Notation is as in Fig.~\ref{fig:imagefig1}.)
Four rows are: 
(1) IRAS16227-4839. There are two sources here that are resolved in 2MASS and
marginally resolved at [3.4], but merged by [22], with the more northerly
source dominating the IR flux. The more southerly source is brighter in
POSS. 
(2) IRAS16252-5440. There is a cluster of sources that is responsible for the
IR flux, with the target position at the photocenter. The field has no
clearly dominant source in POSS.
(3) IRAS16514-4625(PDS432). A dark cloud is apparent in 2MASS and WISE. 
The brightest source in
the optical is the center of the source seen in WISE as blue but not a point
source because it aggregates 3 sources seen in 2MASS. The source that
dominates by [22] is faint at [3.4].
(4) IRAS17102-3813. Images are 100$\arcsec$ on a side to better show
the trio of sources seen at $JHK_s$ that become a smear dominated by
the two sources to the southeast by [22]. The optically brightest
source may contribute some but not all of the 22 \mum\ flux density.
\label{fig:imagefig3}}
\end{figure}

\begin{figure}[h]
\epsscale{0.9}
\plotone{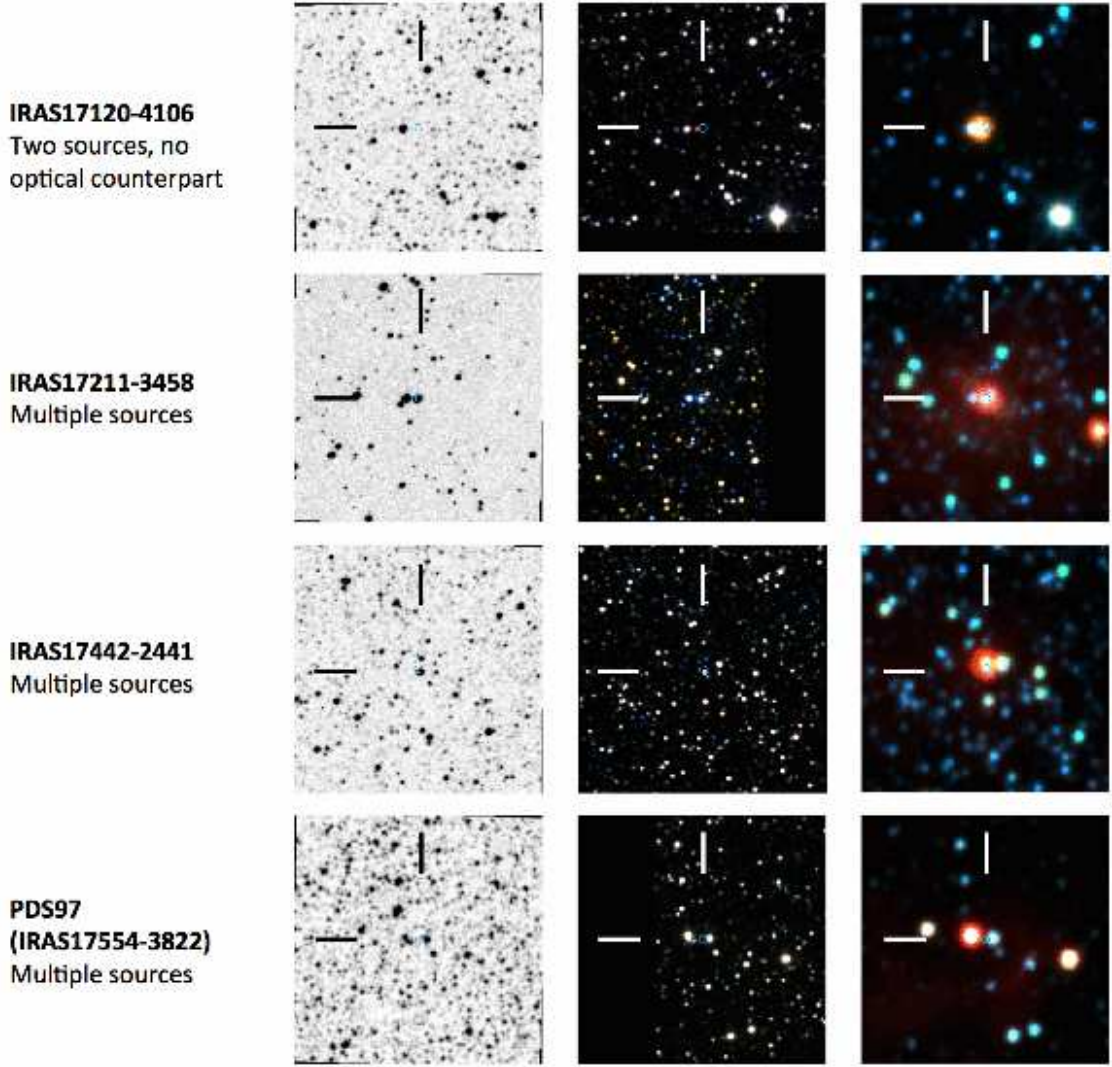}
\caption{(Notation is as in Fig.~\ref{fig:imagefig1}.)
Four rows are: 
(1) IRAS17120-4106. Two sources are barely resolved at [3.4] that
merge by [12] and [22]. There is no optical counterpart at the source
position.
(2) IRAS17211-3458. There are multiple sources that are barely
resolved at [3.4], which merge at longer wavelengths. This is a
complicated source; see the text.
(3) IRAS17442-2441. Two sources at the source position are barely resolved at [3.4] that merge
by [12] and [22] with each other and with other sources in the region. 
(4) PDS97(IRAS17554-3822). Source position is in between two sources
that are of comparable brightness at [3.4] (and, for that matter, POSS
and 2MASS),  with the easterly source dominating by [22]. 
\label{fig:imagefig4}}
\end{figure}

\begin{figure}[h]
\epsscale{0.9}
\plotone{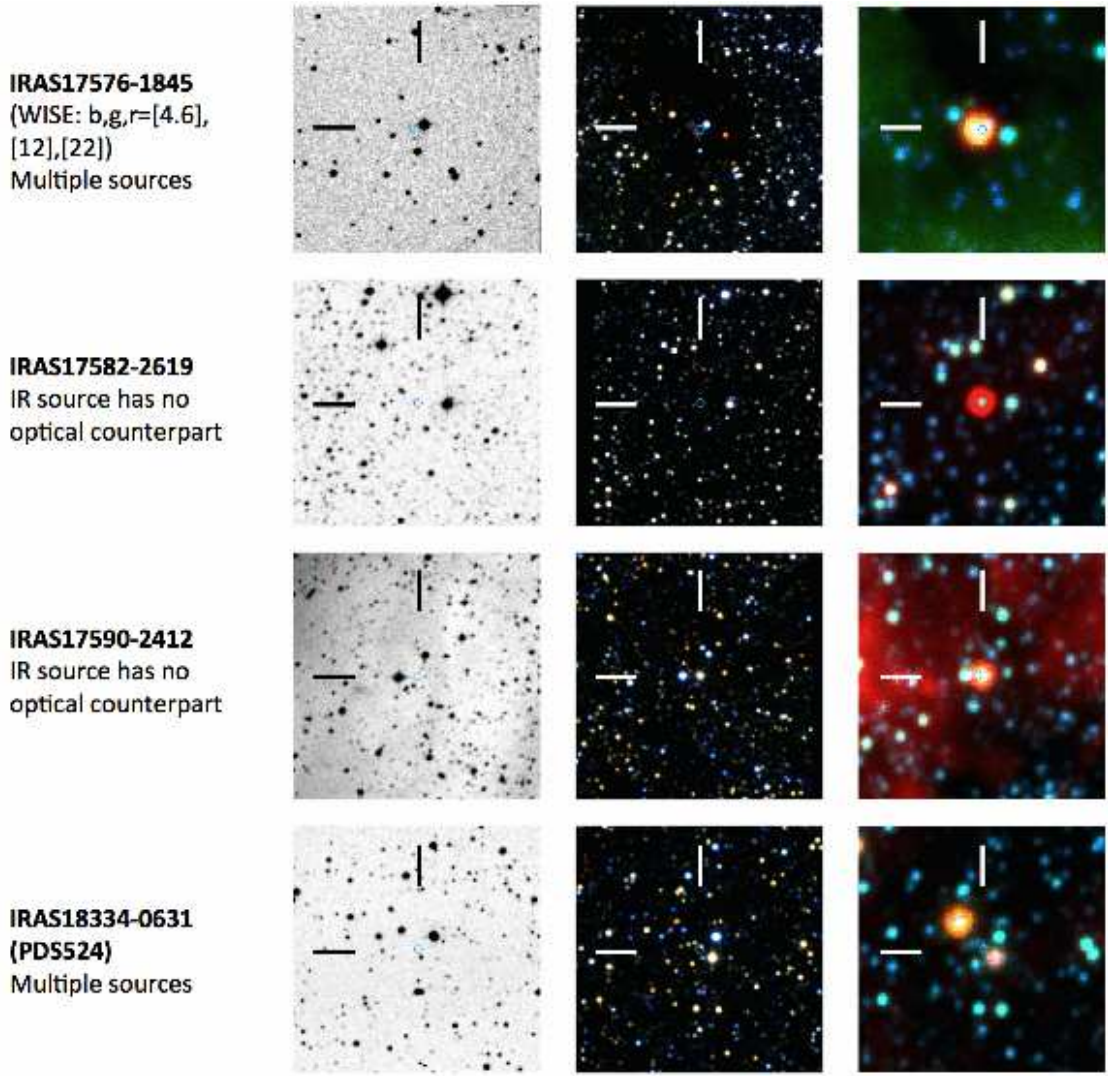}
\caption{(Notation is as in Fig.~\ref{fig:imagefig1}.)
Four rows are: 
(1) IRAS17576-1845. The source dominating at [22] (and presumably IRAS
bands) is smeary at [3.4]. The brightest source at POSS bands is not
the source of the IR.
(2) IRAS17582-2619. The brightest source in the IR has no optical
counterpart, and is strongly dominated by the longest wavelengths.
(3) IRAS17590-2412.  The brightest source in
the IR has no  optical counterpart, and is strongly dominated by the
longest wavelengths. Diffuse emission can also be seen.
(4) IRAS18334-0631(PDS524). There is no optical source at the target
position; the target's IRAS flux has contributions from several
sources in this viscinity, including the arc of blue in WISE precisely at the
target position. The brightest POSS source is blue in WISE.
\label{fig:imagefig5}}
\end{figure}

\begin{figure}[h]
\epsscale{0.9}
\plotone{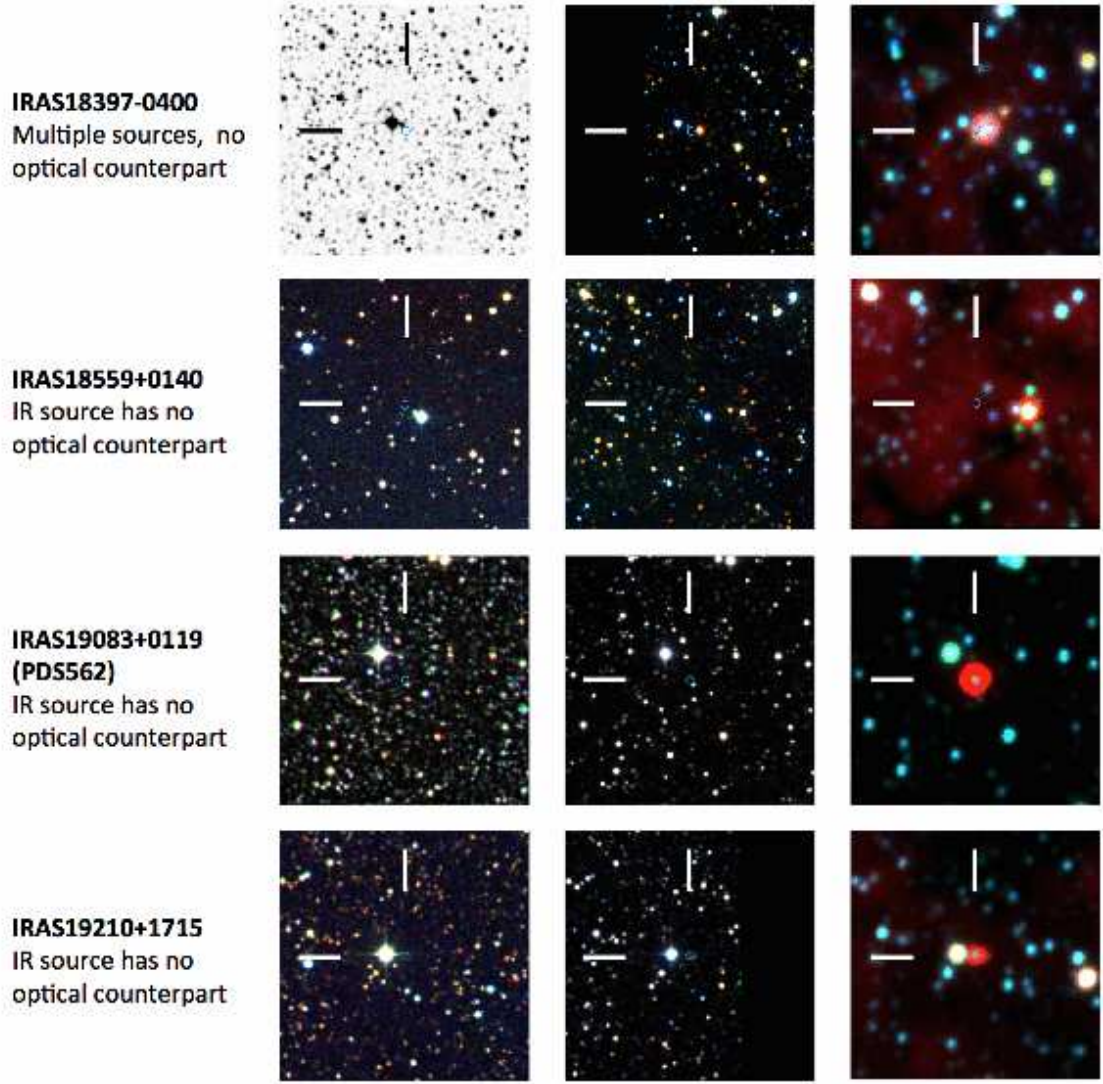}
\caption{(Notation is as in Fig.~\ref{fig:imagefig1}.)
Four rows are: 
(1) IRAS18397-0400. The brightest source in the IR has no  optical
counterpart, and is dominated by the longest wavelengths. The
brightest POSS source is blue in WISE.
(2) IRAS18559+0140. There is no optical source at the target position;
the bright POSS source to the southwest is not bright at all in the IR
and is blue in the WISE data; the target's IRAS flux has contributions
from several sources, most likely including the flux from the bright
source to the upper left, pulling the photocenter off of the bright
clump of sources to the right.
(3) IRAS19083+0119(PDS562). The brightest source in the IR has no
optical counterpart, and is strongly dominated by the longest
wavelengths. The brightest POSS source is blue in WISE.
(4) IRAS19210+1715.  There is no optical source at the target
position; the bright POSS source to the east can be seen in the IR,
but the target position matches the very red source to the west, which
most likely dominates the IRAS flux.
\label{fig:imagefig6}}
\end{figure}

To demonstrate these issues of source confusion, we provide POSS,
2MASS, and WISE images for each of the 24 targets in
Figures~\ref{fig:imagefig1}--\ref{fig:imagefig6}, specifically to
allow readers to follow the same sources across wavelengths. (The FITS
images can be interactively explored via IRSA's FinderChart.)  For
each of these figures, either a multi-color or single band image
appears for each of POSS, 2MASS, and WISE, and the images are
300$\arcsec$ on a side unless specified.  For POSS, it is either DSS2
Blue/Red/IR for the blue, green, and red bands, respectively, or it is
a single-band DSS2 Red. For 2MASS, $JHK_s$ corresponds to
blue/green/red, respectively, and for WISE, [3.4], [4.5], and [12]
correspond to blue/green/red unless specified. The source position is
indicated with a small blue circle; white or black hash marks above
and to the left help guide the eye to this position.  A brief
discussion of each of these sources appears in
Table~\ref{tab:droppedsrcs}; a briefer still summary appears in the
Figure itself and the caption.

\begin{figure}[h]
\epsscale{0.9}
\plotone{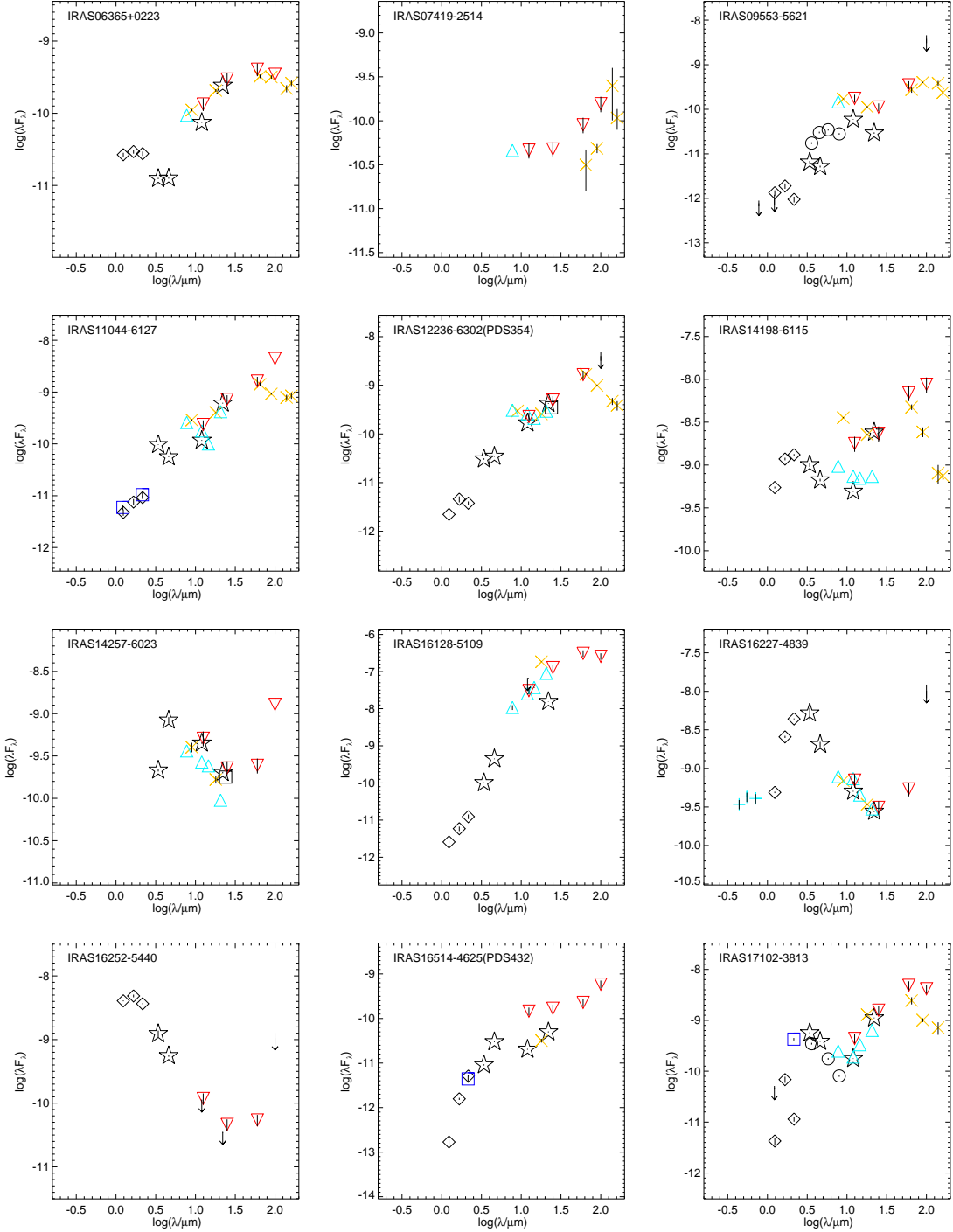}
\caption{SEDs for things that are likely subject to source confusion,
part 1.  Notation for all SEDs in this paper is as follows. The axes
are log $\lambda F_{\lambda}$ in cgs units (ergs s$^{-1}$ cm$^{-2}$)
and log $\lambda$ in microns. Symbols: cyan $+$ are literature
$UBRI_c$;  black $+$ are SDSS $ugriz$;  black diamonds are 2MASS
$JHK_s$; blue squares are Denis $IJK$;  black circles are from
Spitzer/IRAC; black stars are WISE; yellow $\times$ are AKARI; cyan
triangles are MSX; black squares are Spitzer/MIPS (24 \mum); red
downward pointing triangles are IRAS PSC and FSC. Any arrows are
limits at the corresponding wavelength. Error bars are indicated as
vertical black bars at the center of each point. See text and
Table~\ref{tab:droppedsrcs} for discussion of individual objects.
\label{fig:srcconfseds1}}
\end{figure}

\begin{figure}[h]
\epsscale{1.0}
\plotone{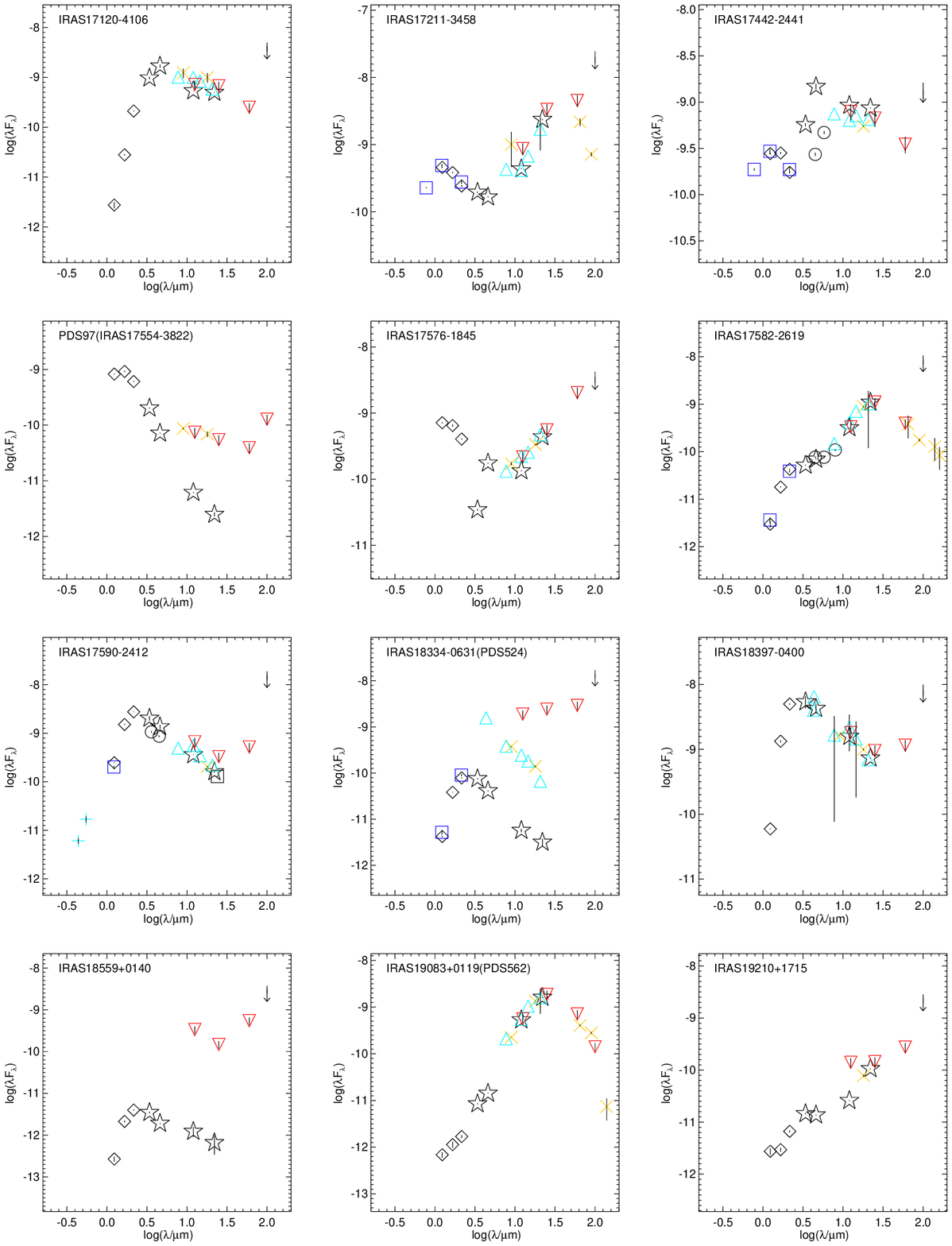}
\caption{SEDs for things that are likely subject to source confusion,
part 2; notation is as described in Fig.~\ref{fig:srcconfseds1}.
See text and Table~\ref{tab:droppedsrcs} for discussion of individual objects.
\label{fig:srcconfseds2}}
\end{figure}

SEDs for these 24 sources appear in Figures~\ref{fig:srcconfseds1} and
\ref{fig:srcconfseds2}.  These SEDs correspond in most cases to the
source position measured in IRAS, but may or may not be the origin of
all of the IRAS flux density, or correspond to the optically bright
source. Because we are matching sources largely by position to the
sources in the catalogs, in several of the cases illustrated in
Figure~\ref{fig:imagefig1}-\ref{fig:imagefig6} where the true
counterpart is is impossible to match, the counterparts across
catalogs are not the same source, and the SED clearly betrays this
conglomeration of sources (see notes in Table~\ref{tab:droppedsrcs}).
IRAS18334-0631(PDS524) (Fig.~\ref{fig:srcconfseds2}, 3rd row center)
is the clearest example of this, where AKARI and MSX are seeing the
same source, but it is a different source than the source that
2MASS, DENIS, and WISE identify, both of which are inconsistent with
the flux densities measured in IRAS.

Eight of these sources have SEDs that do not resemble isolated stars.
They rise steadily from 1 to 20 or even 100 \mum\ and beyond. These
would be SEDs consistent with extragalactic sources, or very heavily
obscured stars of any age. These sources are noted in
Table~\ref{tab:droppedsrcs}; IRAS16128-5109 is one of the best
examples (Fig.~\ref{fig:srcconfseds1}, 3rd row center). This source
has measurements from 2MASS, WISE, MSX, AKARI, and IRAS, with the
energy density at $\sim$20 \mum\ being $\sim$5 orders of magnitude
larger than the energy density at $\sim$1 \mum. The IRAS data suggest
a rollover in this SED near 60 \mum. Most of these steep SEDs turn over
at long wavelengths, but two
of these eight sources have a peak in the SED at $\sim$22-25 \mum,
shorter than the others. These two sources, IRAS17582-2619 and
IRAS19083+0119(PDS562), could be of a different nature than the
others.

\begin{figure}[h]
\epsscale{0.3}
\plotone{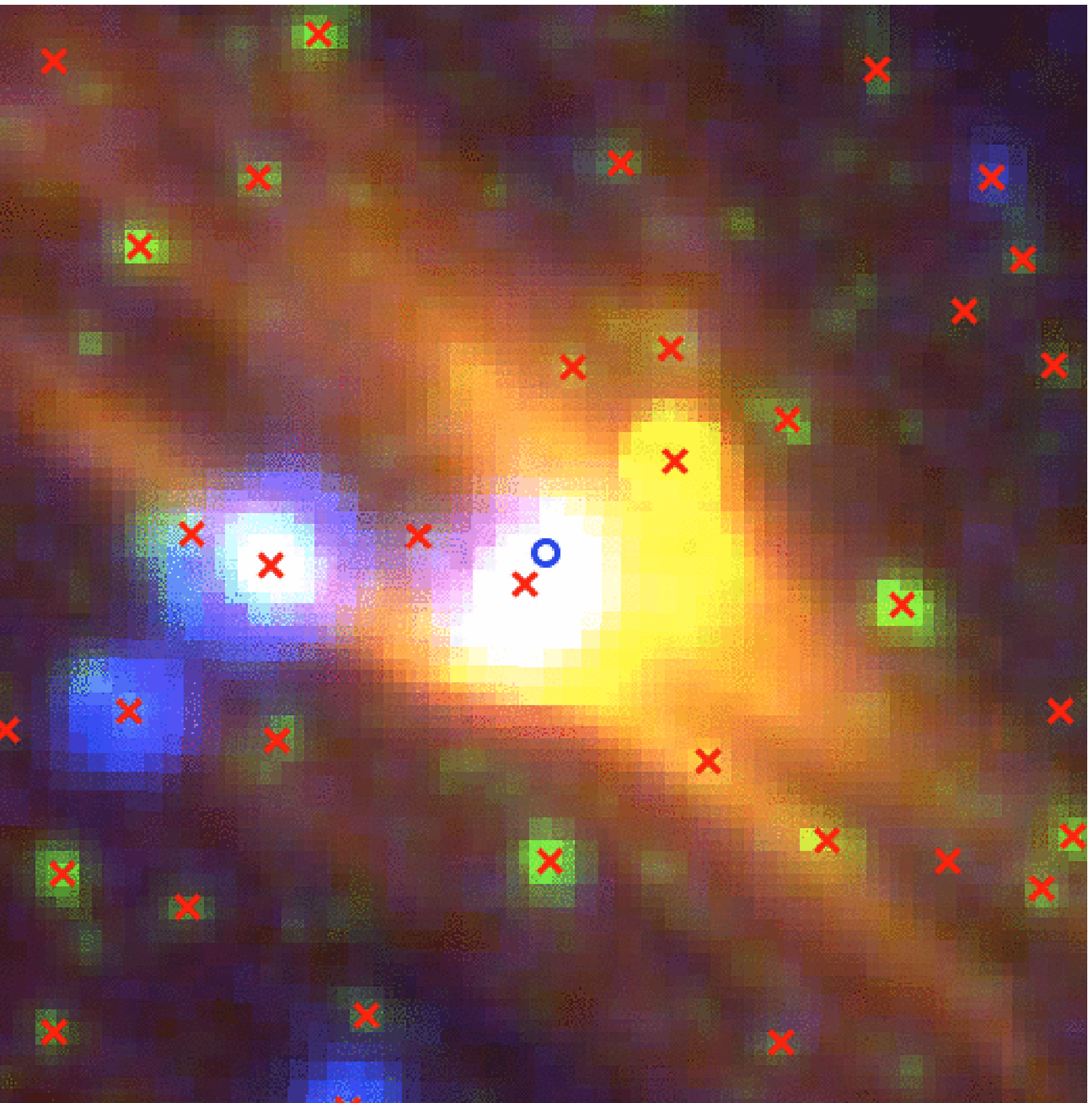}
\caption{3-color image of IRAS17211-3458, about an arcminute on a
side, North-up. Blue plane is DSS 2 red plates, green plane is IRAC-2 (4.5
\mum), and red plane is IRAC-4 (8 \mum). Blue circle is target
position, and red $\times$ are sources from the 2MASS (point source)
catalog. The target position is $\sim$2$\arcsec$ from the source taken as
the 2MASS match. There is complex long-wavelength emission here.
Photometric measurements $>$10 \mum\ combine flux densities from more
than one source, plus nebulosity, and so we have left this source in
the set of sources likely subject to source confusion.
\label{fig:IRAS17211-3458}}
\end{figure}

One of these sources, IRAS17211-3458, is worthy of a few additional
comments here because it is the most borderline of these source
confusion cases. The images shown in Fig.~\ref{fig:imagefig4} (second
row) show a small concave arc of sources in the optical near the
target location; by $H$ (middle panel), it is a convex arc of sources,
where only two sources (including our target) are bright in the
POSS+2MASS images -- they appear white in the 2MASS image in
Fig.~\ref{fig:imagefig4}. We identified source confusion here in part
because these two sources are of comparable brightness at POSS, but
also because WISE has trouble resolving the three close, IR-bright
sources. WISE identifies some faint nebulosity here; in the WISE image
in Fig.~\ref{fig:imagefig4}, the surface brightness is barely
visible.  However, there are Spitzer data here too, which resolves
complex striated nebulosity; see Figure~\ref{fig:IRAS17211-3458}. The
IRAS target position is $\sim$2$\arcsec$ from the nearest 2MASS
source, which is larger than typical uncertainties over the whole
catalog. The Spitzer source corresponding to this object does not
appear in the SEIP, perhaps because it either is or appears to be
slightly resolved because of the high surface brightness nebulosity
surrounding it. The SED (Fig.~\ref{fig:srcconfseds2}, top center) is
the source closest to the target position, where available. On the
face of it, its shape would be consistent with a photosphere with
large excess. That, plus the fact that there is a POSS source exactly
at the location of the IRAS source (and is presumably the object for
which an optical spectrum to assess lithium was obtained), could
conceivably place it in the set of objects with large excesses
identified and discussed below. However, based on the images, bands
longer than 12 \mum\ (WISE, MSX, AKARI, IRAS) certainly are measuring
net flux density from more than one point source, plus nebulosity near
(in projection) to the source. This is also the point at which the SED
starts to significantly diverge from the apparent photosphere. Because
of this ambiguity, we have left it in the set of confused sources. It
is not tagged as a Li-rich source in dlR97, so even if we were able to
add it to the analysis we perform below, it is unlikely to have
contributed significantly.

Because of the ambiguity about the sources and their counterparts,
these 24 sources have to be dropped from our sample. IRAS07419-2514
was identified as a possible K giant in Torres \etal\ (2000), and
PDS97 (IRAS17554-3822) is from de la Reza, Drake, \& da Silva (1996). 
The remaining 22 sources compose $\sim$30\% of the original dlR97
sample. We drop these sources from the bulk of our analysis, but show
them in certain plots where relevant below.

\clearpage

\begin{deluxetable}{lccp{12cm}}
\tabletypesize{\scriptsize}
\tablecaption{Objects that are dropped from the main sample\label{tab:droppedsrcs}}
\tablewidth{0pt}
\rotate
\tablehead{\colhead{name} &  \colhead{src conf?\tablenotemark{a}}
&\colhead{steep SED?\tablenotemark{b}} &\colhead{notes}}
\startdata
IRAS06365+0223	&	x	&	\nodata	&	Target
position on an optically faint source within a grouping of optically
bright sources, and corresponds to a very red, resolved source seen as
resolved in $K_s$ (fluxes retrieved from 2MASS extended source
catalog). This extended source is the origin of most of the
long-wavelength flux. dlR97 source.	\\
IRAS07419-2514	&	x	&	\nodata	&	Target
position in amongst a grouping of optically bright sources, with somewhat different sources bright in the IR. Position corresponds to the photocenter of the aggregate source seen by IRAS.  Because target center does not correspond to an optically detected source, there are no short-wavelength measurements in the SED (so it appears to have a sparse SED). Source identified in Torres \etal\ (2000), and notes there say that IRAS flux may come from CO cloud WB 1046. IRAS flux probably attributable to $\leq$ 5 IR-bright sources seen in WISE.	\\
IRAS09553-5621	&	x (m?)	&	x	&	Target position in between two sources that are comparably bright at 4.6 \mum. Only one source is apparent at optical through $K_s$, and by 22 \mum, the sources have merged into one apparent source, likely responsible for the measured IRAS flux. SED assembled from closest source by position and thus may represent fluxes from different sources. Steep SED. AKARI consistent with IRAS at longest bands. dlR97 source.	\\
IRAS11044-6127	&	x	&	x	&	High surface
density of comparably bright sources in the optical; no easily visible
source exactly at the target position in the optical. There is a faint
source that starts to appear in 2MASS images, and a source to the West
that becomes bright. The source exactly at the target position is
still faint and blended with a source to the south at 3.4 \mum. The
target source is rising fast and dominates by 12 \mum, dominating all
the sources in the field by 22 \mum\ (and therefore probably dominates the measured IRAS flux). Steep SED.  dlR97 source.	\\
IRAS12236-6302(PDS354)	&	x	&	x	&	Another crowded field with comparably bright sources in POSS; there is an optically brighter source above and to the West of the target position that can be seen to be three distinct sources in 2MASS. In WISE, these sources are not resolved and form a blue arc immediately above the target. A source at the target location strongly dominates the field at 12 and 22 \mum, and is presumably responsible for most of the IRAS flux (and, for that matter, that from MSX and AKARI). Steep SED.  dlR97 source. Torres \etal\ (2000) mention that their optical spectrum of the source they took to be the  counterpart of the IR emission has strong H$\alpha$ emission and could be an \ion{H}{2} region. 	\\
IRAS14198-6115	&	x (m?)	&	\nodata	&	There are two
sources here that are distinct in 2MASS and marginally resolved in
[3.4] (though AllWISE catalog identifies only one), but are indistinguishable by [22]. There are several sources in this viscinity in the optical. SED assembled from closest source by position and thus likely represents fluxes from different sources. dlR97 source. 	\\
IRAS14257-6023	&	x (m?)	&	\nodata	&	The brightest source in POSS is to the North and slightly West of the target position, but it is not responsible for the IR flux.  There is a source to the southeast that is very red in 2MASS and is dominating the measured flux by WISE bands, and is likely responsible for the IRAS flux. SED assembled from closest source by position and thus may represent fluxes from different sources.  dlR97 source.	\\
IRAS16128-5109	&	x	&	x	&	This is not a point source in the IR, and is very bright (saturated) by [22]. It appears in SIMBAD as an \ion{H}{2} region; the morphology of the image suggests a dense clump of sources from which emanate long streamers of extended emission.  Steep SED. dlR97 source.	\\
IRAS16227-4839	&	x	&		&	There are two
sources that are resolved in 2MASS and marginally resolved at [3.4],
but merged by [22], with the more northerly source dominating the IR
flux. The more southerly source is brighter in POSS.  Some extended
emission visible in WISE.  dlR97 source.\\
IRAS16252-5440	&	x	&		&	This is also
PDS 146. In POSS, there is a high surface density of comparably bright
sources. By 2MASS, an aggregate of at least 4 IR-bright sources is
apparent. In WISE, it can be seen that the target position corresponds
roughly to the photocenter of the aggregate of sources seen at [12]. 
The brightest source by [22] is to the East of the target position and
is probably responsible for most of the IRAS flux. The SED we have
assembled for this source corresponds to the source closest by
position to the target position, and as such does not represent
correctly the source of the longest wavelength flux. dlR97 source.
Torres \etal\ (1995) lists this as ``other" and ``probably not
young", and later implies it may be a normal MS star. 	\\
\tablebreak
IRAS16514-4625(PDS432)	&	x	&	x	&	A dark
cloud is apparent near this source in WISE images. The brightest
source in the optical is to the northwest of the target position,
which is resolved into at least 3 sources in 2MASS. The source that
dominates the IR by [22] (and probably also IRAS) is faint at [3.4].  Steep SED.  dlR97 source.  Torres \etal\ (2000) list it as a confirmed giant but the source that was measured may not be responsible for the IR flux.	\\
IRAS17102-3813	&	x (m?)	&	\nodata	&	A trio of sources seen at 2MASS become a smear dominated by the two sources to the southeast by [22]. The optically brightest source may contribute some but not all of the [22] (and IRAS) flux. SED assembled from closest source by position and thus may represent fluxes from different sources. dlR97 source.	\\
IRAS17120-4106	&	x	&		&	Nothing is at
the target position in POSS, though there is a source straight East of
the target position. That easterly source persists through [3.4]. 
There is a faint source closer to the target position appearing by
$K_s$. The two sources are barely resolved at [3.4], and merge by [12] and [22].  Given the [22] photocenter, some flux is likely attributable to each source, even in WISE. dlR97 source.	\\
IRAS17211-3458	&	x (m?)	&	\nodata	&	Two comparably
bright sources are resolved in POSS, and in 2MASS. The two sources are
barely resolved at [3.4] and merge by [12] and [22]. SED assembled
from closest source by position, so may represent fluxes
from different sources.  dlR97 source. See text for additional
discussion.\\
IRAS17442-2441	&	x (m?)	&	\nodata	&	The DSS images
are dense with sources of comparable brightness, though there is a
source close to the target position. Similarly, in 2MASS, no source
dominates, though there is a source close to the target location.
There is a source at the target position and another comparably bright
one to the West just resolved at [3.4]; they merge by [12] and [22]. SED assembled from closest source by position and thus may represent fluxes from different sources. dlR97 source.	\\
PDS97(IRAS17554-3822)	&	x (m?)	&	\nodata	&	Target
position in between two sources of comparable brightness at POSS and
2MASS. It is also between two comparably bright sources at [3.4] and
[4.6]; by [12] the easterly source dominates, and by [22], all the
flux is likely from the easterly source. Source from
la Reza, Drake, \& da Silva (1996).  Gregorio-Hetem
\etal\ (1992) list it as a high velocity giant with strong Li, and
``incorrect identification in PSC" because KL Cra is 78$\arcsec$ to east,
outside error ellipse.  SIMBAD lists it as a T~Tauri. 	\\
IRAS17576-1845	&	x	&		&	The multi-wavelength images suggest extinction in this field. The source position is reasonably close to a bright POSS source, but the target position can be seen to have several sources in 2MASS and WISE.  The source dominating at [22] (and presumably IRAS bands) is `smeary' at [3.4].  dlR97 source. Coadella \etal\ (1995) list it as a candidate to be related to high-mass star forming regions with an ultracompact \ion{H}{2} region, though it remained undetected in their survey. 	\\
IRAS17582-2619	&	x	&	x	&	The brightest
source in the IR has no optical counterpart, and the IR is strongly
dominated by the longest wavelengths. The optically brightest source
is to the west of the target source, and has faded substantially by
[12] and [22] \mum. Steep SED to 20 \mum, well-defined, with data from
multiple surveys in good agreement with each other. Turnover from
steep SED happens abruptly at $\sim$20 \mum, shorter wavelengths than
most of  the other steep SEDs identified here (the other one like this
is IRAS19083+0119(PDS562)).  dlR97 source. SIMBAD
lists this as an OH/IR star. Yoon \etal\ (2014) and references therein
identify it as a post-AGB star (OH4.02-1.68). It appears in
Ramos-Larios \etal\ (2012) and as a heavily obscured post-AGB star or
PN candidate, and Garc\'ia-Lario \etal\ (1997)  as a PN
candidate. It appears in de la Reza \etal\ (2015) as an early AGB
star. \\
IRAS17590-2412	&	x	&		&	The brightest source in the IR has no optical counterpart, and is strongly dominated by the longest wavelengths. Diffuse emission can also be seen in the field in various bands. dlR97 source. Messineo \etal\ (2004) identify a SiO emitter in this region but suggest that it may not be associated with the source from which an optical spectrum had been obtained by dlR97. They note that this IRAS source is the only mid-infrared source within their 86 GHz beam.	\\
\tablebreak
IRAS18334-0631(PDS524)	&	x (m?)	&	\nodata	&	There
is no optical source at the target position; the target's IRAS flux
likely has contributions from several sources in this viscinity,
including a source resolved as an arc in WISE [3.4] \& [4.6] precisely at the target position. The brightest POSS source is to the northwest of the target, and the brightest source by $K_s$ is to the southwest. The brightest source by [22] is to the northeast. SED assembled from closest source by position and thus likely represents fluxes from different sources. dlR97 source. 	\\
IRAS18397-0400	&	x	&		&	The brightest source in the IR has no  optical counterpart, and is dominated by the longest wavelengths. The brightest POSS source is to the northeast, though it is no longer the brightest source by 2MASS. The source responsible for most of the IR flux is slightly to the west of the target position, and is visible in 2MASS. MSX measurements have very large errors but are consistent with the rest of the SED. dlR97 source.	\\
IRAS18559+0140	&	x (m?)	&	\nodata	&	There is no
optical source at the target position; a bright POSS source is to the
southwest and is not bright at all in the IR. The target's IRAS flux
has contributions from several sources, including a small clump to the
east; the photocenter is pulled to the west, off of the bright clump
of sources, and this offset of the IRAS position is the farthest off
from the WISE source in  the entire dataset. The assembled SED
corresponds to the closest source by position and as such does not
represent the brightest IR source here, and likely represents more
than one source. dlR97 source.	\\
IRAS19083+0119(PDS562)	&	x	&	x	&	The
brightest source in the IR has no optical counterpart, and is strongly
dominated by the longest wavelengths. There is a faint source at the
target position by $K_s$ and it rises quickly through the WISE bands.
Steep SED, with data from several surveys in good agreement with each
other. Turnover from steep SED happens near $\sim$20 \mum, shorter
wavelengths than most of the other steep SEDs identified here (the
other one like this is IRAS17582-2619).  dlR97 source. SIMBAD lists it
as a possible planetary nebula. Yoon \etal\ (2014) and references
therein identify it as a post-AGB star.	\\
IRAS19210+1715	&	x	&	x	&	There is no optical source at the target position; though there is a bright POSS source to the east, which can be seen in the IR. However, the target position matches a very red source, marginally visible in the 2MASS images, and rising quickly through the WISE bands, which most likely dominates the IRAS flux. There may be a contribution at [22] from a nearby source bright at [12]. Steep SED. dlR97 source.	\\
\enddata
\tablenotetext{a}{This column is populated if the object is likely
subject to source confusion of either of the sorts described in the
text. ``x (m?)'' indicates that there may be multiple sources
represented in the SED, e.g., the object shown in the SED at 2 \mum\
may not be the same object as that shown at 22 \mum.}
\tablenotetext{b}{This column is populated if the object's SED rises
steadily from 2 to at least 20 \mum, as described in the text.}
\end{deluxetable}

\clearpage

\begin{deluxetable}{lccllccc}
\tabletypesize{\scriptsize}
\tablecaption{Sources with IR excesses\tablenotemark{a}\label{tab:irxsrcs}}
\tablewidth{0pt}
\rotate
\tablehead{\colhead{name} &  \colhead{[3.4]$-$[22]} &\colhead{$\chi_{[3.4],[22]}$}
&\colhead{Sample}&\colhead{data codes\tablenotemark{b}} &
\colhead{drop?\tablenotemark{c}} & \colhead{lit?\tablenotemark{d}} &
\colhead{start\tablenotemark{e}}}
\startdata

IRAS00483-7347	&	4.71	&	(large)	&	Castilho \etal\ 1998	&	W	M	A	I	S	&	D?	&	\nodata	&	$<$2	\\
NGC 362 V2	&	1.13	&	9.7	&	Smith \etal\ 1999	&	W				S	&	\nodata	&	\nodata	&	5	\\
HD19745	&	2.32	&	(large)	&	dlR97, Kumar \etal\ 2011	&	W		A	I		&	\nodata	&	IRx	&	10	\\
IRAS03520-3857	&	8.80	&	(large)	&	dlR97	&	W		A	I		&	\nodata	&	\nodata	&	2?	\\
IRASF04376-3238	&	3.83	&	(large)	&	Torres \etal\ 2000	&	W		A	I		&	\nodata	&	\nodata	&	3?	\\
IRAS07227-1320(PDS132)	&	6.64	&	(large)	&	dlR97	&	W	M	A	I		&	\nodata	&	\nodata	&	5?	\\
IRAS07456-4722(PDS135)	&	5.40	&	(large)	&	dlR97	&	W		A	I		&	\nodata	&	\nodata	&	3?	\\
HD65750	&	\nodata	&	\nodata	&	dlR97, Castilho \etal\ (2000)	&	(W)		A	I		&	\nodata	&	IRx	&	2?	\\
IRAS07577-2806(PDS260)	&	8.75	&	(large)	&	dlR97	&	W	M	A	I		&	\nodata	&	\nodata	&	3?	\\
HD233517	&	5.53	&	(large)	&	dlR97, Kumar \etal\ 2011	&	W		A	I		&	\nodata	&	IRx	&	10	\\
IRASF08359-1644	&	6.00	&	(large)	&	Torres \etal\ 2000	&	W		A	I		&	\nodata	&	\nodata	&	3?	\\
G0928+73.2600	&	0.63	&	6.4	&	C12	&	W					&	\nodata	&	\nodata	&	22	\\
HD96195	&	\nodata	&	\nodata	&	Castilho \etal\ 2000 &	(W)	M	A	I		&	D?	& IRx	&	10	\\
Tyc0276-00327-1	&	0.36	&	3.4	&	C12	&	W		A			&	\nodata	&	\nodata	&	22	\\
IRAS12327-6523(PDS355)	&	3.56	&	(large)	&	dlR97	&	W	M	A	I		&	\nodata	&	\nodata	&	2	\\
HD111830	&	0.63	&	3.7	&	dlR97	&	W		A	I	S	&	\nodata	&	\nodata	&	22	\\
PDS365(IRAS13313-5838)	&	7.37	&	(large)	&	dlR97, Kumar \etal\ 2011	&	W	M	A	I		&	\nodata	&	IRx	&	2?	\\
PDS68(IRAS13539-4153)	&	6.04	&	(large)	&	dlR97, Kumar \etal\ 2011	&	W		A	I		&	\nodata	&	IRx	&	10	\\
IRAS16086-5255(PDS410)	&	9.75	&	(large)	&	dlR97	&	W	M	A	I		&	\nodata	&	\nodata	&	3?	\\
HD146834	& 	1.07	&	2.6	&	dlR97	&W		A		S	&	\nodata	&	IRx&	20?	\\
IRAS17578-1700	&	\nodata	&	\nodata	&	dlR97	&	(W)	M	A	I		&	D	&	\nodata	&	$<$2	\\
IRAS17596-3952(PDS485)	&	5.15	&	(large)	&	dlR97, Kumar \etal\ 2011	&	W		A	I		&	\nodata	&	IRx	&	3?	\\
V385 Sct	&	\nodata	&	\nodata	&	Castilho \etal\ 2000	&	(W)	M	A	I		&	D	&	\nodata	&	3?	\\
IRAS19012-0747	&	1.39	&	7.2	&	dlR97	&	W	M	A	I		&	\nodata	&	\nodata	&	10?	\\
IRAS19038-0026	&	1.45	&	4.9	&	Castilho \etal\ 2000	&	W		A	I		&	D?	&	\nodata	&	10?	\\
PDS100	&	5.86	&	(large)	&	dlR97, Kumar \etal\ 2011	&	W		A	I		&	\nodata	&	IRx	&	4?	\\
Tyc9112-00430-1	&	0.86	&	3.1	&	Ruchti \etal\ 2011	&	W					&	\nodata	&	\nodata	&	22	\\
HD219025	&	3.74	&	9.6	&	dlR97, Kumar \etal\ 2011	&	W		A	I		&	\nodata	&	IRx	&	3	\\
\enddata
\tablenotetext{a}{Note that up to two more stars with IR excess
could be identified from the relatively sparse SEDs in
Sec.~\ref{sec:sparseseds}; these are For 90067 and SDSS J0632+2604,
with SDSS J0632+2604 being more compelling.}
\tablenotetext{b}{W=WISE data in SED, with `(W)' meaning some bands
are missing; M=MSX data in SED; A=AKARI data in SED; I=IRAS data in
SED; S=Spitzer data in SED.}
\tablenotetext{c}{This column is populated if there is a reasonable
likelihood that the star isn't a first ascent K giant, e,g., of the sort
appropriate for this study. `D' means we are fairly confident it
should be dropped, and `D?' means there is some doubt as to whether it
should be dropped; see text.}
\tablenotetext{d}{This column is populated if recent literature
has already identified this source as having an IR excess.}
\tablenotetext{e}{This column contains the approximate wavelength, in
microns, of the start of the IR excess.}
\end{deluxetable}

\clearpage

\section{Sources With IR Excesses By $\sim$25 \mum}
\label{sec:irx}

\subsection{Overview of Approach}

Having omitted the objects for which we have substantial difficulty
making matches across catalogs above, we have now a subset of objects
for which we have established reliable multi-wavelength matches across
catalogs. All of these objects have no or little ambiguity in the
images to which we have access, e.g., they appear as clean point
sources. We can now inspect the resultant assembled SEDs for evidence
of an infrared excess.

In assessing whether or not a star has an IR excess, one needs to
compare a measure of brightness at relatively short wavelengths
(expected to be dominated by the stellar photosphere) with that at
relatively long wavelengths, where dust emission is likely to be
present. We take a two-pronged approach to identifying excesses. There
are some objects for which an IR excess is immediately apparent upon
inspection of the SED; no detailed analysis is required. These objects
are summarized in Sec.~\ref{sec:lgirx} and detailed in
Sec.~\ref{sec:lgirxnotes}. There are other objects for which the IR
excess is more subtle. For these latter objects, we employ an approach
developed in the context of finding small IR excesses around young
stars, which is described in Sec.~\ref{sec:smirx}. Details of objects
found to have these more subtle IR excesses can be found in
Sec.~\ref{sec:smirxnotes}.  Table~\ref{tab:irxsrcs} summarizes the
stars we identify as having either a large or small IR excess. .

\subsection{Overview: Sources with Very Large Excesses}
\label{sec:lgirx}

There are 19 stars whose SEDs immediately reveal significant IR
excesses at wavelengths $<$20 \mum, and maintain large excesses out to
at least 25 \mum. These sources often have detections in more surveys
than just WISE; they often have data from MSX, AKARI, IRAS, and even
Spitzer. Over the entire sample, including these sources with large
excesses, the data from these various surveys are in reasonably good
agreement, though some objects have more scatter than others. (The
scatter could be due to complex backgrounds and variable beamsize
across the surveys, or even intrinsic variability in the source.) If
there is disagreement, however, it is typically IRAS that
overestimates the flux density from the object, which makes sense
since IRAS is the lowest spatial resolution of all the surveys used
here.

\begin{figure}[h]
\epsscale{1.0}
\plotone{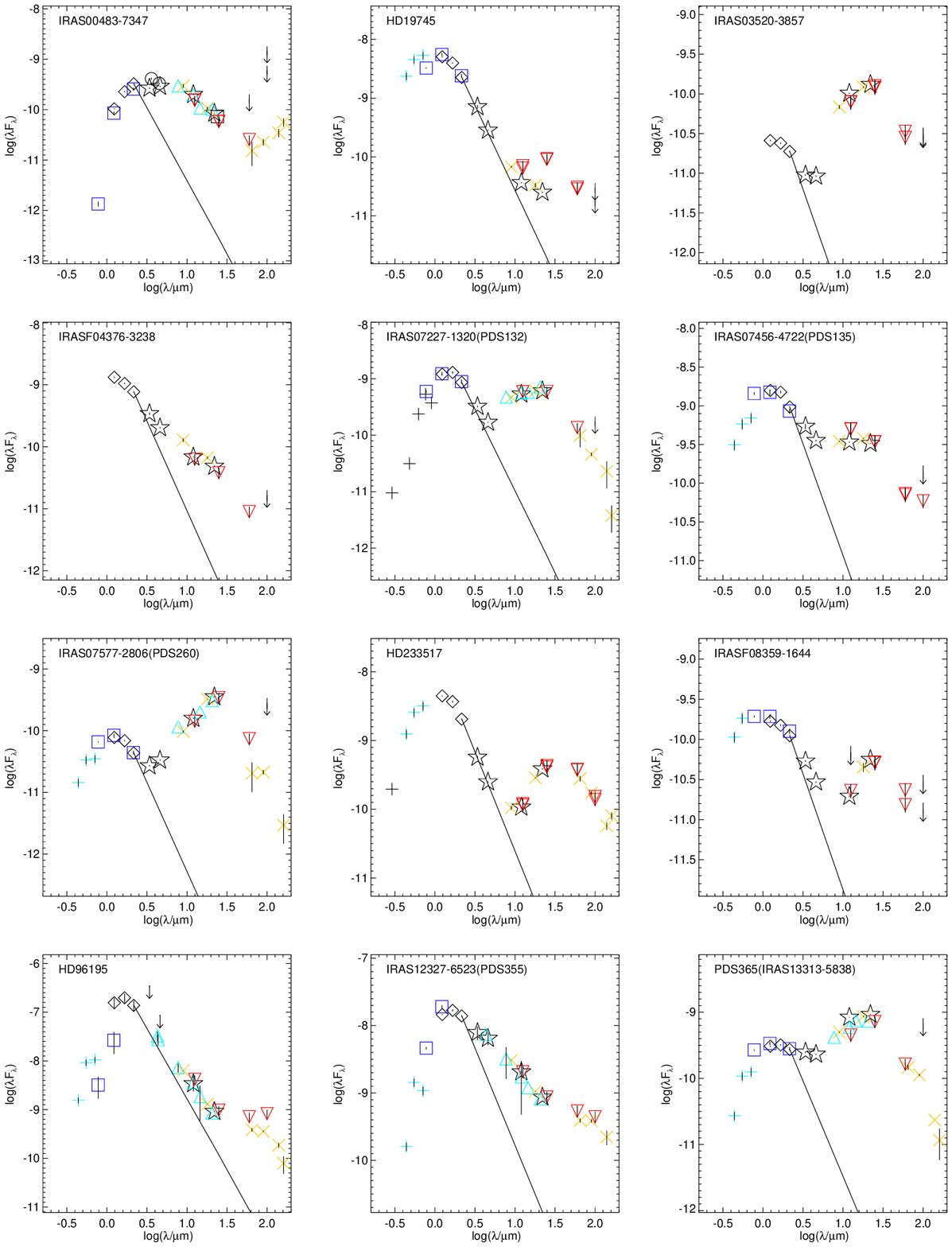}
\caption{SEDs for sources with large IR excesses, part 1; notation
is as described in Fig.~\ref{fig:srcconfseds1}, with an additional line
with a Rayleigh-Jeans slope extended from $K_s$ (e.g., if $K_s$ is on
the photosphere, the photosphere longward of $K_s$ should fall on this
line). All of these objects have a significant excess above this line.
See text and table for discussion of individual objects.
\label{fig:lgirxseds1}}
\end{figure}

\begin{figure}[h]
\epsscale{1.0}
\plotone{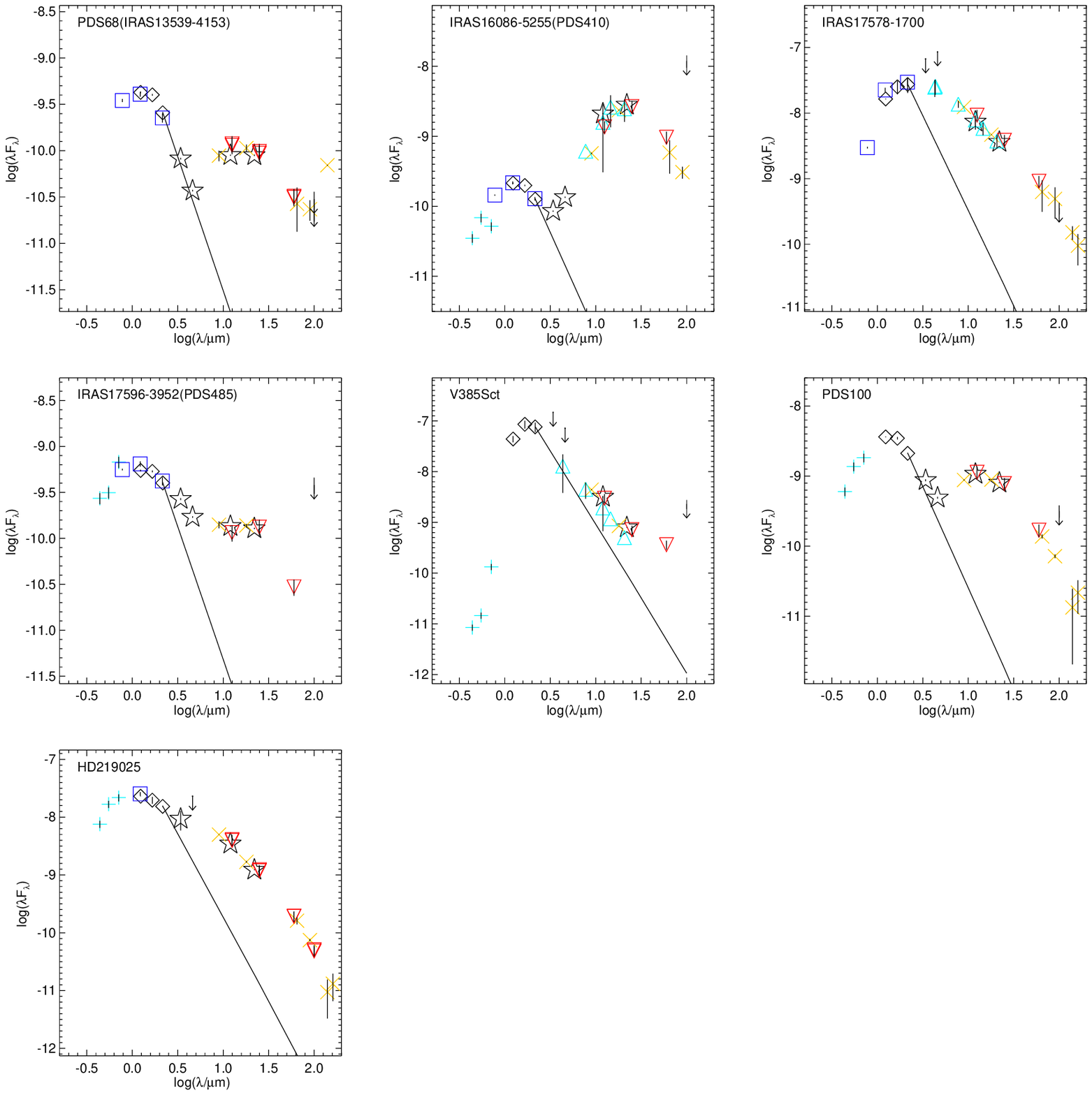}
\caption{SEDs for sources with large IR excesses,
part 2; notation is as in Fig.~\ref{fig:lgirxseds1}.
See text and table for discussion of individual objects.
\label{fig:lgirxseds2}}
\end{figure}

The SEDs for these objects with unambiguous, large excesses appear in
Figures \ref{fig:lgirxseds1} and \ref{fig:lgirxseds2}. For stars
without circumstellar dust (at least those warm enough to have the
peak of their photospheric SED be at $<$1 \mum), measurements in the
infrared ($\geq$2 \mum) should fall on a line consistent with a
Rayleigh-Jeans (R-J) slope. To guide the eye, in
Figs.~\ref{fig:lgirxseds1} and \ref{fig:lgirxseds2}, we have added an
R-J line extended from 2 \mum.  All of the sources with very large
excesses can be seen to deviate from this line.

Out of these 19 stars with large IR excesses, 14 (73\%) are from the
dlR97 sample, five are from the literature sample, and none are from
C12.  However, four of the 19 may not be K giants -- one of the dlR97
stars, IRAS17578-1700, is a carbon star, one of the stars from the
literature sample, V385 Sct, is a very cool S-type star, and the other
two may be too cool to be first ascent K giants. Some of these stars
have detections indicating that the SEDs are rising beyond 90 \mum,
suggesting that sub-mm observations are needed to constrain the outer
extent of the IR excess.  Some of these objects have been identified in
recent literature (e.g., Kumar \etal\ 2015) as having an IR excess,
but for others, this is the first confirmation that the objects have
an IR excess using data more recently obtained than IRAS.

\subsection{Notes on Sources with Very Large Excesses}
\label{sec:lgirxnotes}

{\bf 	IRAS00483-7347	}	This is an extremely well-populated
SED, with data from WISE, MSX, AKARI, IRAS, and even Spitzer. However,
the SED is wide compared to other SEDs in this study. The SED
suggests that $K_s$ is not on the R-J side of the SED, perhaps because
the star is significantly cooler than a K giant. A R-J line extended
from 2 \mum\ as shown, or even a R-J line extended from $\sim$5 \mum,
suggests a substantial IR excess around this star. Data from multiple
sources are in good agreement with each other, and AKARI suggests that
there is a significant long-wavelength component to the IR excess, with the
SED rising again at the longest wavelengths. This source is identified
in Castilho \etal\ (1998) as a Li-rich K giant, though no \teff\
estimates are available in the literature.  This star may be
too cool to be a K giant, though it clearly has a large IR excess.
Additional spectroscopy of this source would be helpful for
a better understanding of the \teff\ and where the excess starts.

{\bf 	HD19745	}	The IR excess for this source starts to appear
past 10 \mum; WISE and AKARI are in good agreement. IRAS overestimated
the IR flux density from this star, but it still has a clear excess.
This source is known to be a Li-rich K giant (Reddy \& Lambert 2005),
and is incorrectly identified in SIMBAD as a T~Tauri. Reddy \& Lambert
(2005) identify it as a red clump star.  This star was identified in
Kumar \etal\ (2015) as having an IR excess. 

{\bf 	IRAS03520-3857	}	This object is a dlR97 source, and it
had to be offset from the nominal IRAS position by 11$\arcsec$ to pick
up the counterparts, which is very large in the context of the other
positional shifts needed in the rest of our sample. However, the field
is relatively clean (consisting of one bright source) and is not
suggestive of source confusion. There is good agreement between WISE,
AKARI, and IRAS for 10-20 \mum. There is about an order of magnitude
more energy density emerging at 10-20 \mum\ than at 3-4 \mum. 
However, compared to other sources here, there are few detections
blueward of $\sim$2 \mum, and it would be nice to see the SED turn
over to define the Wien side of the SED.   This object is identified
in two papers as a possible galaxy based on IRAS colors (Saunders
\etal\ 2000, Wang \& Rowan-Robinson 2009) but is not identified as a
confirmed galaxy in either paper. SIMBAD identifies it as a ``peculiar
star." This source has no \teff\ in the literature.

{\bf 	IRASF04376-3238	}	This object's IR excess starts at
least by 5 \mum, if not actually at 3 \mum. WISE, AKARI, and IRAS are
in good agreement. However, compared to other sources here, there are
few detections blueward of $\sim$2 \mum, and it would be nice to see
the SED turn over to define the Wien side of the SED. This object is
identified as a K giant in Torres \etal\ (2000), but as a candidate
T~Tauri in Magnani \etal\ (1995). It does have an SED  consistent with
SEDs found in young stars (see, e.g., Rebull \etal\ 2011). SIMBAD
lists it as a ``peculiar star.'' Spectroscopy would be useful to
distinguish a young star with low gravity from an old star with low
gravity, and assess its Li abundance. (An uncertain equivalent width
for Li is given for it in Torres \etal\ 2000, but no abundance.) This
source has no \teff\ in the literature.

{\bf 	IRAS07227-1320(PDS132)	}	This star has a substantial IR
excess that evidently starts abruptly between 4.6 and 7.8 \mum. AKARI,
MSX, WISE, and IRAS are consistent with each other. This object is
part of the dlR97 sample, and Torres \etal\ (2000) list it as a
confirmed giant. However, Torres et al. (1995) lists this as ``other"
and ``probably not young" in their paper on young stars; it is implied
to be a normal MS star.  Garc\'ia-Lario \etal\ (1997) identify it as a
possible PN (planetary nebula) based on its IR excess. It is
identified as a post-AGB star and an M 1 I giant in Su\'arez \etal\
(2006).  Szczerba \etal\ (2007) identify it as `not a post-AGB
(asymptotic giant branch) star', but Yoon \etal\ (2014) identify it as
a post-AGB star of type M3 IV. This source has no \teff\ in the
literature. Additional data are needed to clarify the status of this
object.

{\bf 	IRAS07456-4722(PDS135)	}	This star's IR excess is small
at 3.4 and 4.6 \mum, but becomes substantial by $\sim$10 \mum. There
is scatter but generalized agreement among IRAS, AKARI, and WISE. This
object is part of the dlR97 sample.  Torres \etal\ (2000) list it as a
confirmed giant, though Torres \etal\ (1995) categorize it as a
``Probable post-FU Ori star.''  SIMBAD lists it as a T~Tauri,
apparently on the basis of Torres \etal\ (1995). As with other sources
here, it has an SED  consistent with SEDs found in young stars (see,
e.g., Rebull \etal\ 2011), and spectroscopy would be useful to
distinguish a young star with low gravity from an old star with low
gravity. This source has no \teff\ in the literature.

{\bf 	IRAS07577-2806(PDS260)	}	There is good agreement
between WISE, AKARI, MSX, and IRAS here, and there is more energy
density near 20 \mum\ (from dust) than near 1 \mum\ (from the
photosphere). AKARI provides detections out past 100 \mum. The IR
excess may start at 3.4 \mum. This source is from the dlR97 set, and
Torres \etal\ (2000) list it as a confirmed giant.  SIMBAD identifies
it as a post-AGB star, and reference Garc\'ia-Lario \etal\ (1997), but
this object does not appear in the paper. Szczerba \etal\ (2007)
retain it as a candidate AGB star. It does not have a \teff\ in the
literature.

{\bf 	HD233517	}	This star's SED suggests an excess
that starts past 5 \mum. This is an original dlR97 source, and this
star was identified in Kumar \etal\ (2015) as having an IR excess, as
well as in Fekel \& Watson (1998), Jasniewicz \etal\ (1999), and Drake
\etal\ (2002), among others.

{\bf 	IRASF08359-1644	}	This SED suggests an IR excess that
most likely starts at $\sim$10 \mum. WISE and IRAS (and the single
AKARI point) are in good agreement with each other. Torres \etal\
(2000) identify this source as a Li-rich K giant. No other information
about this source is apparently available (including \teff);
additional data would be useful.

{\bf 	HD96195	}	This SED is more complicated than ones above.
Reliable WISE points at the shortest two bands do not exist; MSX
provides a link between $K_s$ and [12]. MSX, AKARI, WISE, and IRAS are
all in rough agreement between 10 and 20 \mum, though there is some
scatter (and large error bars in one case). Assuming that $K_s$ is on
the photosphere (and on the R-J side of the SED, that is, assuming
that this source is hot enough), there is a significant IR excess by
10 \mum: $K_s-[12]$=1.14 mag. We have placed this object in this
section with the other large excesses; we might have included it in
the set of objects with more subtle excesses (following the method
laid out in Sec.~\ref{sec:smirx} below, $\chi_{K,[22]}$=6.8), but
close inspection of the SED shows that points $>$10 \mum\ are probably
above the photosphere. Moreover, the AKARI points, if detecting flux
density truly associated with this source, identify a substantial IR
excess at the bands $>$50 \mum. This source appears in Castilho \etal\
(2000) and Pereyra \etal\ (2006) as a Li-rich giant, but it may 
be too cool to be a K giant. Its reported \teff\ in the literature
(Castilho \etal\ 2000) is $\sim$3400-3600 K.  It is also identified in
McDonald \etal\ (2012) as having an IR excess (their $E_{IR}$=1.932)
and having \teff=3400.

{\bf 	IRAS12327-6523(PDS355)	}	The IR excess for this source
appears by 3.4 \mum, and increases from there. WISE, MSX, AKARI, and
IRAS are all in good agreement with each other (despite large errors
on the MSX points). This is part of the dlR97 sample. Torres \etal\
(2000) list it as a confirmed giant, but also mention that there is
some reddening here that likely comes from the Coalsack, and the star
may be only 200 pc away, so the IR may be contaminated. Reddy \&
Lambert (2005) confirm the evolved nature of the star. de la Reza
\etal\ (2015) categorizes this source as an early AGB star. 

{\bf 	PDS365(IRAS13313-5838)	}	This star's IR excess likely
starts between 2 and 3 \mum; there is more energy density near 20
\mum\ (from dust) than near 1 \mum\ (from the photosphere). There is
good agreement among WISE, MSX, AKARI, and IRAS. This star is part of
the dlR97 sample and was identified in Kumar \etal\ (2015) as having
an IR excess. While it is a confirmed Li-rich K giant (e.g., Kumar
\etal\ 2011, Drake \etal\ 2002, among others), SIMBAD 
lists it as a `Post-AGB Star (proto-PN).'

{\bf 	PDS68(IRAS13539-4153)	}	The IR excess here starts
abruptly between 5 and 8 \mum; 3.4 and 4.6 \mum\ are on the
photosphere if $K_s$ is as well. WISE, AKARI, and IRAS are in good
agreement. If the longest wavelength AKARI bands are detecting flux
attributable solely to this source, it suggests that the SED is rising
again at the longest bands. This star was identified in Kumar \etal\
(2015) as having an IR excess. While it is listed as a confirmed
Li-rich K giant in several studies (e.g., Kumar \etal\ 2011, Kumar \&
Reddy 2009, among others), it also appears in Valenti \etal\ (2003) as
a T~Tauri candidate, though no useable spectra are reported of this
object in that paper.   SIMBAD has adopted the `T Tauri'
categorization. The $^{12}$C/$^{13}$C ratio is only a limit  ($>$20;
Reddy \& Lambert 2005), and cannot conclusively determine whether
early RGB first dredge-up mixing (which lowers $^{12}$C/$^{13}$C from
the main sequence value) has occurred.  

{\bf 	IRAS16086-5255(PDS410)	}	This star's IR excess likely
starts between 2 and 3 \mum; there is more energy density near 20
\mum\ (from dust) than near 1 \mum\ (from the photosphere). There is
good agreement among WISE, MSX, AKARI, and IRAS. This star is part of
the dlR97 sample, and is listed in Torres \etal\ (2000) as a confirmed
giant. However, SIMBAD categorizes it as `post-AGB, proto-PN,' and
Szczerba \etal\ (2007) retain it as a candidate AGB. No \teff\ is
available. 

{\bf 	IRAS17578-1700	}	This star has a substantial IR excess
that may start at $K_s$; it may also be too cool for $K_s$ to be on
the R-J side of the SED. There is good agreement between MSX, AKARI,
WISE, and IRAS, though there are no viable WISE points at the shortest
two bands. This object is part of the dlR97 sample. Chen, Yang, \&
Zhang (2007), and references therein including Lloyd Evans (1991) 
identify it as a J-type carbon star based on optical spectra and IR
excesses, and that categorization is inherited by SIMBAD. This
star is likely too cool to be a K giant, and moreover is likely to be
a carbon star.

{\bf 	IRAS17596-3952(PDS485)	}	This object is part of the
dlR97 sample, and its position had to be slightly adjusted from the
nominal IRAS position to pick up the counterparts, but the field is
relatively clean and is not suggestive of source confusion. This
star's IR excess likely starts between 2 and 3 \mum; there is a larger
excess by $\sim$10 \mum, and WISE, AKARI, and IRAS are in good
agreement where the points exist. This star was identified in Kumar
\etal\ (2015) as having an IR excess. It is a confirmed Li-rich K
giant (e.g., Kumar \etal\ 2011, Reddy \& Lambert 2005).

{\bf 	V385 Sct	}	This is a very bright star, and the
$K_s$ mag may not be on the R-J side of the SED; it may be too cool to
be a K giant. There are no WISE data for the shortest two bands, and
MSX fills the gap between 2 and 10 \mum. There is scatter among the
MSX, WISE, AKARI, and IRAS data, likely because it is so bright. It
does seem to have an obvious IR  excess, however. It appears in
Castilho \etal\ (2000) and Pereyra \etal\ (2005) as a Li-rich K
giant.  However, it appears as GCSS 557 in Stephenson (1976) and is
identified as star of type S in there and subsequent literature (such
as Stephenson 1984, where it is CSS 1043). Stars of type S have ZrO
bands as well as TiO bands, and other abundance anomalies; they are
thermally pulsing AGBs that experience substantial dredge-up. It
seems  unlikely to be a first-ascent Li rich K giant. The \teff\ value
for it from the literature (Castilho \etal\ 2000) is $\sim$3300 K, so it
is also too cool to be a first-ascent K giant.

{\bf 	PDS100	}	There may be a small excess at 3.4 and 4.6
\mum, but there is a clear excess by $\sim$10 \mum, as seen by WISE,
AKARI, and IRAS. This star is part of the dlR97 set, and is also known
as V859 Aql. SIMBAD indicates it as a T~Tauri, but the literature is
clear that it is instead a Li-rich K giant (Reddy \etal\ 2002, among
others). This star was identified in Kumar \etal\ (2015) as having an
IR excess.

{\bf 	HD219025	}	This star is also known as BI Ind, and
its IR excess starts between 2 and 3 \mum. There is a viable
WISE detection at 3.4 \mum, but not at 4.6 \mum; there is good
agreement past $\sim$10 \mum\ among WISE, AKARI, and IRAS. It has a
substantial IR excess. This star is part of the dlR97 set, and is
identified in many places in the literature as a Li-rich K giant (e.g,
Kumar \etal\ 2011). SIMBAD has this as an RS CVn, but it is unlikely
to be such an object. This star was also identified in Kumar \etal\
(2015) as having an IR excess. Jasniewicz \etal\ (1999) and Whitelock
\etal\ (1991) have previously found a NIR excess for this star. They
verify using Hipparcos parallax that this is a red giant and not a
young star. It is also a rapid rotator.

\subsection{Overview: Sources with Small Excesses}
\label{sec:smirx}

Identifying smaller excesses around the remaining stars requires more
analysis than simple SED inspection. We can quantitively compare the
brightness at 2-3 \mum\ (which should be dominated by the stellar
photosphere in these cases where there is not much circumstellar dust
to create an IR excess) with that at relatively long wavelengths,
where dust emission may be present. In considering significant IR
excesses, the IR excess is larger than any uncertainties in
calibration. However, for small excesses, well-defined errors are
important, uncertainties in calibration should be taken into account,
and it becomes more important to use data that are uniformly obtained,
calibrated, and processed  so as to minimize systematics.
Additionally, data obtained not over decades but obtained close in
time minimize influence from intrinsic stellar variation.  

WISE data meet these criteria, as the data are uniformly obtained (at
nearly the same time), reduced, and calibrated. Therefore,
[3.4]$-$[22] is essentially an ideal metric with which to identify IR
excesses.

There have been several approaches in the literature used to determine
with confidence whether or not a star has an IR excess.  For
example, Mizusawa \etal\ (2012) tested several methods of finding IR
excesses in F stars.  To identify sources with small IR excesses, we
adopt here the following approach (as in Mizusawa \etal\ 2012, or
Trilling \etal\ 2008). We calculate $\chi$: \begin{equation} \chi_{\rm
[3.4],[22]} = \frac{([3.4]-[22])_{\rm observed} - ([3.4]-[22])_{\rm
predicted})}{\sigma_{([3.4]-[22])}}\end{equation} and take as a
significant excess those stars for which $\chi>3$. For K giants,
$[3.4]-[22]_{\rm predicted}$ is 0, but for cooler objects (such as M
giants), the predicted value is not 0 (see, e.g., Gautier \etal\
2007).  Mizusawa \etal\ (2012) were able to combine $\chi$
calculations for two independent measures of IR excess for most of the
targets in their sample, using $K_s - [24]$ and [3.4]$-$[22]. We do
not have such uniform independent measures of IR excess, so for the
most part, we use $\chi_{\rm [3.4],[22]}$. In order for the $\chi$
calculation to be successful, however, one needs good estimates of the
star's brightness at the relevant bands, as well as good estimates of
the error on that measurement. Many of our targets are very bright,
and saturated in $K_s$ and [3.4], which limits our ability to
correctly estimate brightnesses (and errors) and therefore $\chi$.
However, we can use the existing measurements and reported errors to
identify objects that are likely to have a small IR excess.  

Table~\ref{tab:irxsrcs} includes [3.4]$-$[22] and $\chi_{\rm
[3.4],[22]}$ for the sources with IR excesses. (Note that the $\chi$
values as calculated for the objects with large and obvious IR
excesses above in Sec.~\ref{sec:lgirx} are often $>$100, despite there
being, in some cases, an IR excess even at 3.4 \mum.)

There are nine stars we identify as having small but significant
excesses. Two of the stars with small IR excesses are from the C12
sample, three are from the dlR97 sample, and the remaining four come
from the literature sample of Li-rich giants. One may be too cool to
be a K giant. Two have been recently identified in the literature as
having an IR excess.  Comments on each of these excess objects follow
in the next section.

\begin{figure}[h]
\epsscale{1.0}
\plotone{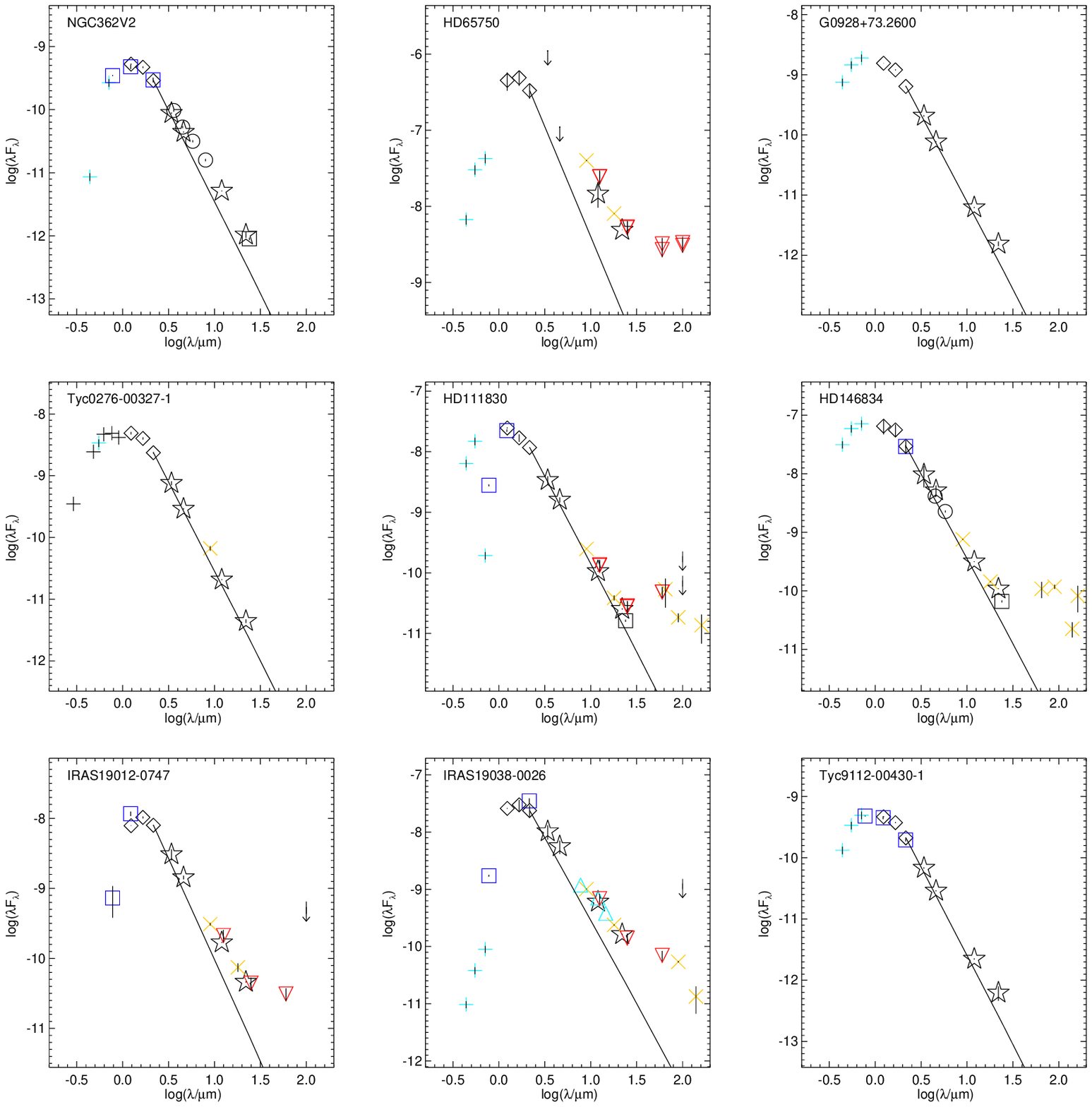}
\caption{SEDs for sources with smaller IR excesses; notation
is as in Fig.~\ref{fig:lgirxseds1}.
See text and table for discussion of individual objects.
\label{fig:smirxseds}}
\end{figure}

\subsection{Notes on Sources with Small Excesses}
\label{sec:smirxnotes}

{\bf 	NGC 362 V2	}	There are Spitzer data for this
source, and Spitzer agrees well with WISE. There is a small IR excess
here. By inspection of the SED, the excess probably starts relatively
early in the SED, $\sim$5 \mum, for a small excess. (The specific
values are [3.4]$-$[22]=1.13 mag, $\chi_{[3.4],[22]}$=9.7). The star
is identified as a Li-rich K giant in Smith \etal\ (1999).	

{\bf 	HD65750	}	This star is very bright, and as such is
missing 3.4 and 4.6 \mum\ data (and the $K_s$ data come from NOMAD).
There are, however, WISE, AKARI, and IRAS, all of which are in good
agreement with each other. This source can be seen in the SED to have
an excess; it has no [3.4], but $\chi_{K,[22]}$=11. It is highly
nebulous at POSS, though not at longer bands, so the measured IR flux
is not likely contaminated by extended emission.   This source comes
from the dlR97 sample; it appears in Castilho \etal\ (2000) as a
Li-rich giant. SIMBAD says this is V341 Car, a pulsating variable
star of type M0III. It has a \teff\ of 3600, so it is most likely a
borderline case for being a first ascent K giant. McDonald \etal\
(2012) identified it as having an IR excess in their study of IR
excesses around Hipparcos stars, calling it out (HIP 38834) in their
Table 3 of luminous giant stars with detected circumstellar emission
(their $E_{IR}$=5.88).	

{\bf 	G0928+73.2600	}	This star does not have a very
well-populated SED, or an immediately obvious IR excess based on the
SED, but the $\chi$ calculation supports there being a small but
significant IR excess here ([3.4]$-$[22]=0.63 mag,
$\chi_{[3.4],[22]}$=6.4). This star is identified as a particularly
interesting source in C12 and Carlberg \etal\ (2010) because it has
particularly high Li ($A$(Li)$_{\rm NLTE}$=3.30 dex), relatively rapid
rotation (8.4 km s$^{-1}$), and high $^{12}$C/$^{13}$C (28). We have
more discussion of this source in Sec.~\ref{sec:c12} below.	

{\bf 	Tyc0276-00327-1	}	This source does not have an
immediately obvious IR excess from the SED (which is also not terribly
well populated), but the $\chi$ calculation supports there being a
very small but significant IR excess here ([3.4]$-$[22]=0.36 mag,
$\chi_{[3.4],[22]}$=3.4). In the images, it appears within the
extended halo of emission associated with an extended source (possibly
a galaxy) that is bright at [12] and [22]. It is possible that this
may affect the IR excess, though the AllWISE catalog does profile
fitting photometry that should take into account this higher
background. Additional data to secure the association of the IR excess
with this object would be helpful. This object is from the C12 sample,
but is unremarkable in that study, showing low Li, slow rotation, and
average $^{12}$C/$^{13}$C.

{\bf 	HD111830	}	On first glance at the SED, this
object seems to have good agreement with photospheric measurements
from $K_s$ to 25 \mum, with WISE, AKARI, IRAS, and even Spitzer/MIPS
falling on the R-J line to 25 \mum. However, IRAS and AKARI, if
detecting flux density truly associated with this source, identify a
substantial IR excess at  bands longer than 50 \mum. There turns out
to be a  small but significant excess at 22 \mum\ ([3.4]$-$[22]=0.63
mag, $\chi_{[3.4],[22]}$=3.7). $K_s-[24]$=0.4 mag, consistent with a
small excess. This star is part of the dlR97 sample; it appears in
Jasniewicz \etal\ (1999) as Li-rich but is listed in their table with an
upper limit on the Li abundance.	

{\bf 	HD146834	}	This source is also HR 6076. In POSS,
there is nebulosity (and SIMBAD categorizes it as ``star in nebula").
It is very bright in 2MASS and WISE, with no strong nebular emission,
though there is some faint extended emission in the background at 12
and 22 \mum. It has WISE, Spitzer (IRAC and MIPS), and AKARI data, all
of which are in fairly good agreement. However, near 20 \mum, it is
confusing. AKARI measures 860 ($\pm$15) mJy at 18 \mum, WISE measures
797 ($\pm$12) mJy at 22 \mum, and MIPS measures significantly less,
528 mJy, at 24 \mum.   The original error at 24 \mum\ reported in the
SEIP is $\pm$16 mJy, and is likely underestimated, so we have added a
4\% error floor, as described above; the net error is 21 mJy, still
not enough to bring the measures into alignment within 1$\sigma$. The
$K_s$ mag is bright, and its error is also likely underestimated.
Following the measurements and errors as reported, however,
[3.4]$-$[22]=1.07, $\chi_{[3.4],[22]}$= 2.6, and $K_s-[24]$= 1.19,
$\chi_{K,[24]}$=3.2. The errors on [3.4] and \ks\ are both large, which
lowers the $\chi$ values. It is difficult to decide if the excess
here is real and significant. We have opted to call the excess
significant because there are three measures of the brightness near 20
\mum, and it does seem to be significantly brighter near 20 \mum\ than
the photospheric expectations based on the brightness near 2-3 \mum.
It also has some long-wavelength AKARI detections; if AKARI is
measuring flux associated solely with this star, then it has a
significant long-wavelength excess. It appears in McDonald \etal\
(2012) as having a small IR excess in their study of IR excesses
around Hipparcos stars (their $E_{IR}$=1.532). It is from the dlR97
sample.

{\bf 	IRAS19012-0747	}	There are two sources here in close
proximity in POSS, but the target position matches one of the two
optically bright sources. Both sources can be seen in $J$ through
[12], though the target source is clearly dominating, and overwhelms
any flux from the apparent companion by [22] (and presumably in
IRAS).  While source confusion is possible, it's reasonably likely
that a spectrum was obtained of the source of most of the IR flux.
This source has WISE, AKARI, and MSX data, all of which are in good
agreement with each other. IRAS and AKARI, if detecting flux density
truly associated with this source, identify a substantial IR excess
at  bands longer than 25 \mum. At $\sim$20 \mum, it has a weaker
excess -- [3.4]$-$[22]=1.39, $\chi_{[3.4],[22]}$=7.2. This star is
part of the dlR97 sample (though its name was incorrectly
IRAS19012-0742 in the published table). It appears in Pereyra \etal\
(2005) as a Li-rich K giant.	

{\bf 	IRAS19038-0026	}	$K_s$ for this object may not be on
the R-J side of the SED; it may be too cool to be a K giant. There
does not seem to be significant IR excess at 3.4 or 4.6 \mum, but the
MSX, AKARI, WISE, and IRAS points suggest an excess may be present
starting at $\sim$10 \mum. By  $>$20 \mum, assuming that the IRAS and
AKARI points are detecting flux density truly associated with this
source, there is a substantial IR excess. At $\sim$20 \mum, it has a
more subtle excess -- 3.4$-$[22]=1.45, $\chi_{[3.4],[22]}$=4.9. This
object appears in Castilho \etal\ (2000) and Pereyra \etal\ (2005) as
a Li rich giant, but it may be too cool to be a K giant.	The
\teff\ that appears in the literature for it is $\sim$3600 K.

{\bf 	Tyc9112-00430-1	}	This source does not have an
immediately obvious IR excess from the SED, but the $\chi$ calculation
supports there being a small but significant IR excess here:
[3.4]$-$[22]=0.86, $\chi_{[3.4],[22]}$=3.1.  It is identified as a
Li-rich K giant in Ruchti \etal\ (2011).

\clearpage

\section{Discussion: Characteristics of the Entire IR Excess Sample}
\label{sec:disc1}

We started with 316 sources. Out of those, 10 have too sparsely
populated SEDs for us to sensibly place any strong restrictions on
whether or not there is an excess. There are 36 sources that have
relatively sparse SEDs, and are often missing at least the [22] band.
For these, we can put some constraints on whether or not there is an
IR excess, and 2 out of those 36 sources could plausibly have an IR
excess. There are 24 sources that we suspect are subject to source
confusion, and we drop them from the sample. There are 218 sources
with well-populated SEDs and no evidence for IR excesses out to
$\sim$20 \mum. There are 28 sources that have well-populated SEDs that
do, in fact, have evidence for an IR excess.  Out of those 28, 5 are
probably giants but may not be K giants. We conclude that IR excesses
are rare among our sample of K giants, at best $\sim$10\%. Given the
biases in our sample (described in detail below), this fraction is
probably less in Li-rich RGs and substantially less in Li-poor RGs.

We now examine our ensemble population in several different ways.

\subsection{Comparison of IR Excesses to Literature}
\label{sec:irlit}

Not all of our IR excess sources are newly identified as having an IR
excess; after all, dlR97 and others in the literature identified these
sources based on their infrared properties. However, one important
goal of our paper was to reassess the IR excesses in these sources
given the higher spatial resolution data now available. Other recent
papers have identified IR excesses in some of our targets using
similar or the same data; McDonald \etal\ (2012) and Kumar \etal\
(2015) both identify sources with IR excesses.  Kumar \etal\ (2015)
has similar goals to our paper, and many targets overlap. We recover
all 7 of their IR excess K giants. 

\begin{figure}[h]
\epsscale{0.3}
\plotone{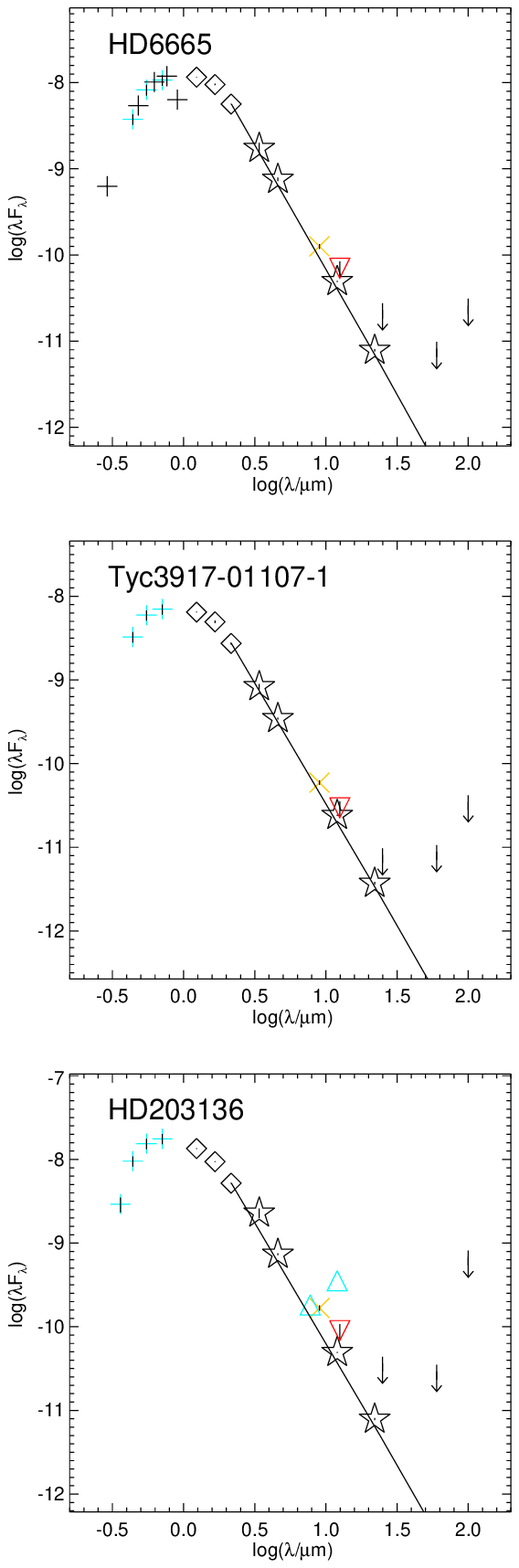}
\caption{SEDs for three sources identified in McDonald \etal\ (2012)
as potentially having IR excesses, but for which we do not identify an
IR excess. Notation is as in Fig.~\ref{fig:lgirxseds1}. See text for
discussion of individual objects.
\label{fig:noirxseds}}
\end{figure}

McDonald \etal\ (2012) did an analysis of more than 107,000 Hipparcos
stars, incorporating data from IRAS, SDSS, DENIS, 2MASS, MSX, AKARI,
and WISE, in order to identify stars with an IR excess. Out of our
targets, 118 are included in the catalog presented in McDonald \etal\
(2012).  Only some of the IR excess objects are explicitly discussed
in McDonald \etal\ (2012), so only one of our objects (HD65750) is
mentioned there as having an IR excess; we agree that it has an IR
excess.  Following the prescription laid out in their paper (in their
Fig.~7 and associated discussion), however, 5 more objects (out of the
118 we have in common) can be identified as having at least
potentially significant IR excesses: HD6665, HD96195, Tyc3917-01107-1,
HD203136, and HD219025. Two of those (HD96195, HD219025) are ones we
have already identified above as having IR excesses. The remaining
three (HD6665, Tyc3917-01107-1, and HD203136) do not appear to us to
have excesses (Figure~\ref{fig:noirxseds}); we investigated why the
McDonald \etal\ calculations might have identified them as excess
objects. HD6665 has the IRAS 12 \mum\ point slightly above the
photosphere, but all other detections are on the photosphere
([3.4]$-$[22]=0.09); the value of $E_{IR}$ calculated by McDonald
\etal\ is 2.16, likely a result of the IRAS point being slightly high.
Tyc3917-01107-1 is in a nearly identical situation, though in this
case, the IRAS 12 \mum\ point is closer to the photosphere; the
McDonald \etal\ $E_{IR}$ is 1.73, and [3.4]$-$[22]=0.07, so again, not
likely to have a real excess. HD203136 has some irregularities in its
SED, where there is a lot of scatter among the the MSX, AKARI, and
IRAS points near 8-12 \mum; they are inconsistent with each other and
the rest of the SED, and all of them are too high compared to the WISE
[12] and [22] points. The McDonald \etal\  $E_{IR}$ comes out to be
4.89 most likely because the MSX, AKARI, and IRAS photometric points
are high. We take WISE to be the most reliable, because it has the
highest spatial resolution; [3.4]$-$[22]=$-$0.21$\pm$0.14, so there is
no detectable IR excess in this object. 

Thus, we conclude that we have recovered all of the recent
literature-identified IR excess sources. We have identified 18 more
objects out of our aggregate data that have IR excesses, though not
all of them may be first ascent K giants.

Many of the sources in our sample had been identified as IR excess
sources from IRAS measurements. All 21 of the sources with the largest
IRAS IR excesses ([12]$-$[25]$>$0.5 mag) from detections -- not limits
-- in the PSC are recovered as IR excess sources here.  Nearly all, 11
of 14, of the sources with detections and [12]$-$[25]$>$0.5 in the FSC
are recovered. 

There are just three sources (HD76066, HD112859, HD203251)
for which the FSC detections at [12] and [25] result in
[12]$-$[25]$>$0.5, but the measured WISE flux densities are
substantially lower than the IRAS FSC flux densities, such that these
stars do not have detectable IR excesses. The [3.4]$-$[22] for these
objects are 0.05, 0.08, 0.08 mag, respectively, so these sources do
not have measurable IR excesses to 22 \mum. This is probably a direct
result of the higher spatial resolution of WISE better measuring the
flux density of the target.

However, many of the sources with smaller measured IRAS excesses are
not recovered as IR excess sources.  In the SEDs for these cases, one
can often see the IRAS PSC suggesting an IR excess, the IRAS FSC
suggesting less of an excess, and WISE (and sometimes AKARI)
suggesting a smaller or no excess. (Similarly, and more dramatically,
the upper limits are often pushed lower and lower in the SEDs.) In
these cases, what is most likely going on is that the increased
spatial resolution and the fainter sensitivity reached resolves out
extended emission and/or source multiplicity, lowering the overall
measured flux density. However, there are also some noticable
calibration offsets between IRAS and 2MASS+WISE; see
Sec.~\ref{sec:dlr97} below.

\subsection{SED interpretation}

IR excess around stars is commonly interpreted as due to circumstellar
dust in a shell or disk or ring. In Table~\ref{tab:irxsrcs}, we
include an approximate wavelength at which the excess appears. Sources
where the IR excess is already present between 2 and 5 \mum\ likely
have very large disks or envelopes of dust that reach nearly all the
way in to the star.  Sources where the IR excess does not start until
10 or 20 \mum\ likely have shells or rings of dust, where there is a
gap between the star and the dust. Obtaining total dust masses would
require detailed modeling of the star+dust SED (plus assumptions about
the composition of the dust) and is beyond the scope of this paper. On
the whole, larger excesses likely correspond to larger quantities of
dust.  For the 7 sources modeled by Kumar \etal\ (2015), under the
array of assumptions they made, they find dust temperatures between 75
and 260 K, but they do not estimate total dust mass. 

The circumstellar dust around these stars could plausibly be dust
ejected by the K giant, but it could also be residual debris disk dust
in the system heated afresh by the first ascent onto the RGB (e.g.,
Jura 1999). However, debris disks are typically relatively low-mass,
producing small IR excesses; (re-)illumination of an old debris disk
is not a particularly reasonable explanation for the very large IR
excesses. 

\clearpage

\subsection{Color-Magnitude Diagrams}
\label{sec:cmds}

\begin{figure}[ht]
\epsscale{0.8}
\plotone{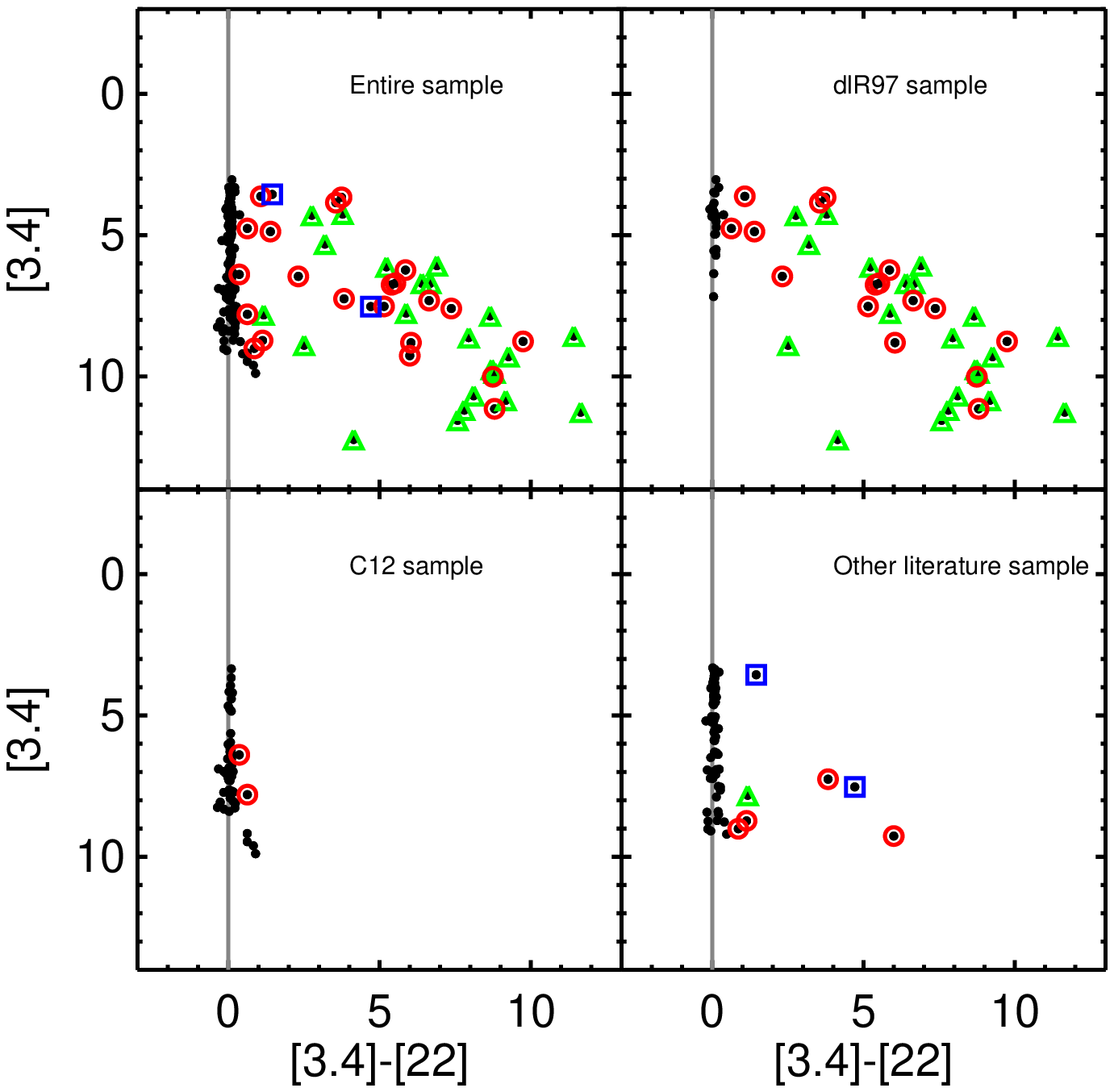}
\caption{[3.4] vs.\ [3.4]$-$[22] for the entire sample where [3.4] and
[22] are both detected (upper left), the dlR97 sample alone (upper
right), the C12 sample alone (lower left), and the remaining
literature  sample (lower right). All of the sources subject to source
confusion (\S\ref{sec:srcconf}) have additional green triangles
overplotted. The sources we identify as having an IR excess
(\S\ref{sec:irx}) are circled in red; the sources we called out as
having an IR excess but  potentially not K giants (\S\ref{sec:irx})
are overplotted in blue squares (some of these are not detected at
[3.4] and thus do not appear). The vertical line at [3.4]$-$[22]=0
indicates the value expected for photospheres. All of the sources
significantly redward of [3.4]$-$[22]$\sim$0 are either identified as
IR excess sources or are likely to be subject to source confusion. 
\label{fig:3422}}
\end{figure}

\begin{figure}[ht]
\epsscale{0.8}
\plotone{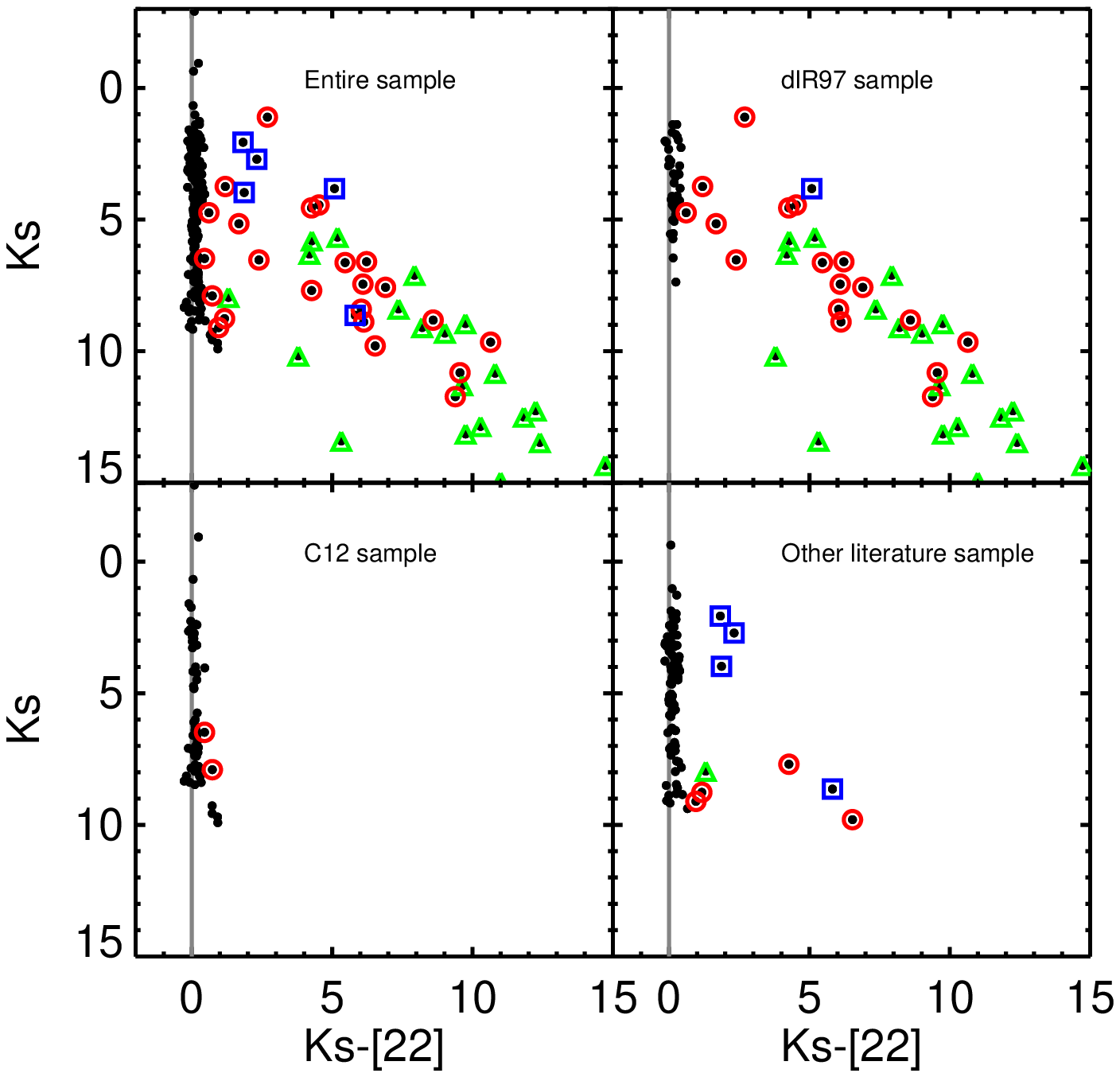}
\caption{$K_s$ vs.\ $K_s-$[22] for the entire sample where $K_s$ and
[22] are both detected (upper left), the dlR97 sample alone (upper
right), the C12 sample alone (lower left), and the remaining
literature  sample (lower right). Notation is as in
Fig~\ref{fig:3422}. There are more sources in this plot than in the
prior plot, but the same conclusions apply -- all of the sources
significantly redward of \ks$-$[22]$\sim$0
are either identified as IR excess sources or are subject to source
confusion. The scatter in the $K_s-$[22]=0 locus is larger in this
plot, reflecting larger $K_s$ uncertainties.
\label{fig:k22}}
\end{figure}

Another view of the IR excesses for the entire sample can be obtained
by plotting [3.4] vs.\ [3.4]$-$[22], as illustrated in Figure~\ref{fig:3422}.
Photospheres (that is, stars without any circumstellar dust) should
have [3.4]$-$[22]$\sim$0.  All of the sources significantly redward of
[3.4]$-$[22]$\sim$0 are either identified as IR excess sources or are
identified as subject to source confusion (\S\ref{sec:srcconf} above).
Even though [3.4]$-$[22] is a nearly ideal metric with which to assess
IR excesses, many of our objects are saturated at [3.4]. Therefore, it
is worthwhile to examine the $K_s-$[22] plot as well; see
Figure~\ref{fig:k22}. More sources are included in this plot, and
there are still very large excesses here, even among the objects 
thought to be K giants. There is more scatter in the $K_s-$[22]=0
photospheric locus, reflecting some of the larger uncertainties in the
measurements of the (bright) stars in $K_s$.

Many of the reddest sources (redward of [3.4]$-$[22]$\sim$5 or
\ks$-$[22]$\sim$5) in either of these Figures are the dropped sources;
their location in this diagram is not surprising given some of the
SEDs shown in Figs.~\ref{fig:srcconfseds1} and \ref{fig:srcconfseds2}.
Still, many of the large IR excess K giants we identified above are
very red in this diagram. For context, young stars in Taurus (which
are known to have substantial dusty disks and envelopes) have a
typical [3.4]$-$[22]$\sim$4, though some extend to
[3.4]$-$[22]$\sim$10 (Rebull \etal\ 2011). The range of IR excesses
for the objects we believe to be K giants is comparable to the range
of IR excesses found in young stars. Some of the IR excesses seen in
these K giants are very large indeed, with many well past 4 out to 10.
However, these large IR excesses are not distributed uniformly among
the subsamples. The dlR97 sample includes the large excesses; the
excesses in the literature Li-rich sample are much more moderate. The
C12 sample has very few excesses, and those are quite small.

\subsection{C12 sample}
\label{sec:c12}

Recall that the dlR97 sample is biased towards IR-bright sources, and
the literature sample is strongly biased towards high $A$(Li) stars.
The C12 sample of red giants (RGs) is unbiased with respect to $A$(Li)
and IR excess, though it does have a larger proportion of fast
rotators than a random RG field population.  Additionally, it has
$A$(Li), \vsini, and $^{12}$C/$^{13}$C measured for every star in the
sample, in contrast to the rest of the sources, for which only some of
these parameters are available.  We had hoped that we would detect IR
excesses in enough of these sources to look for correlations. However,
only two of the C12 sources (G0928+73.2600 and Tyc0276-00327-1) have
any excesses, and they are both small, $<$0.6 mag.  Out of the whole
86-star C12 sample, the fraction of stars with an IR excess is 2\%.
Considering just those with well-populated SEDs to 22 \mum, 2/79
(3$^{+4}_{-0.9}$\%) of the stars have an excess by [22] (using the
binomial statistics from the appendix in Burgasser \etal\ 2003 to
obtain uncertainties).  

The three best candidates for planet accretion listed in C12 are
G0928+73.2600, Tyc0647-00254-1, and Tyc3340-01195-1. If the IR excess
production process is directly related to planet accretion, one would
expect all three of these to have an IR excess, but only one of these
has a measurable IR excess. The SEDs for Tyc0647-00254-1 and
Tyc3340-01195-1 do not suggest excesses at 22 \mum. Admittedly, for
Tyc0647-00254-1, the sole non-WISE point beyond 3 \mum\ is an AKARI 9
\mum\ point that is slightly above the photosphere, though the WISE 12
and 22 \mum\ points are not consistent with AKARI and do not suggest
an IR excess. (In comparison, Tyc0276-00327-1 has a very similar SED,
but in that case, the 22 \mum\ point is enough above the photosphere
that a small IR excess is suggested.) Tyc3340-01195-1 has no points
other than WISE beyond 3 \mum, and neither does G0928+73.2600.  

The two stars with a significant IR excess are split in terms of
properties. Tyc0276-00327-1 seems to be a relatively unremarkable star
in the C12 data; it has a subsolar Li abundance ($A$(Li)$_{\rm
NLTE}$=$-$0.24 dex),  is a slow rotator (4.2 km s$^{-1}$), and has an
average $^{12}$C/$^{13}$C (17). However, G0928+73.2600 is a
particularly interesting star (C12, Carlberg \etal\ 2010) because it
has particularly high Li ($A$(Li)$_{\rm NLTE}$=3.30 dex), relatively
rapid rotation (8.4 km s$^{-1}$), and high $^{12}$C/$^{13}$C (28).

G0928+73.2600 was also included in the ensemble of Li-rich stars in Kumar
\etal\ (2011).  Its location on a Hertzsprung-Russell (H-R) diagram is
similar to many other Li-rich RGs in that sample. Kumar \etal\
(2011) noted that many Li-rich RGs have properties
consistent with red clump stars, and suggested the possibility that an
episode of Li regeneration may occur during the He flash for some red
giants.  G0928 is noteworthy among that sample of RGs for
having the largest $^{12}$C/$^{13}$C. (The importance of
$^{12}$C/$^{13}$C in interperting Li-rich stars is described in more
detail in Sec.~\ref{sec:abun} below.)  Carlberg \etal\ (2010) and C12
argued in favor of external replenishment as a source of the high Li
given given its relatively high $^{12}$C/$^{13}$C.

\clearpage

\subsection{The dlR97 Sample and the Original IRAS Color-Color Diagram}
\label{sec:dlr97}

The dlR97 sample is strongly biased towards sources that are bright in
the infrared, and so it is not surprising that there are many more
bright IR sources, and large IR excesses, found in the dlR97 sample
than in either the C12 or the `other literature' samples.  In de la
Reza \etal\ (1996), dlR97, and Siess \& Livio (1999), there is a plot
of IRAS colors for their targets, which they use to describe a
proposed evolutionary sequence of objects in the diagram. This plot is
described as a color-color diagram, where the following is their
definition of color: \begin{equation} [\lambda_1 - \lambda_2] = \log
(\lambda_2 F_1) - \log (\lambda_1 F_2) \end{equation} where $\lambda$
is wavelength and $F$ is flux density. In more recent papers, driven
at least in part by Spitzer and WISE work, the convention for color is
instead truly a difference of magnitudes: \begin{equation} M_1 - M_2 =
2.5 \times \log \left(\frac{F_2}{F_1}\right) \end{equation} And in the
infrared, where the band names are often the wavelength of the
bandpass, this difference in magnitudes would be written, e.g.,
$[\lambda_1] - [\lambda_2]$. In any case, the {\em shape} (if not the
specific {\em values}) of the distribution of points in the dlR97
color-color plot is recovered by using [12]$-$[25] and [25]$-$[60]
defined as in equation 3. 

\begin{figure}[h]
\epsscale{1.0}
\plotone{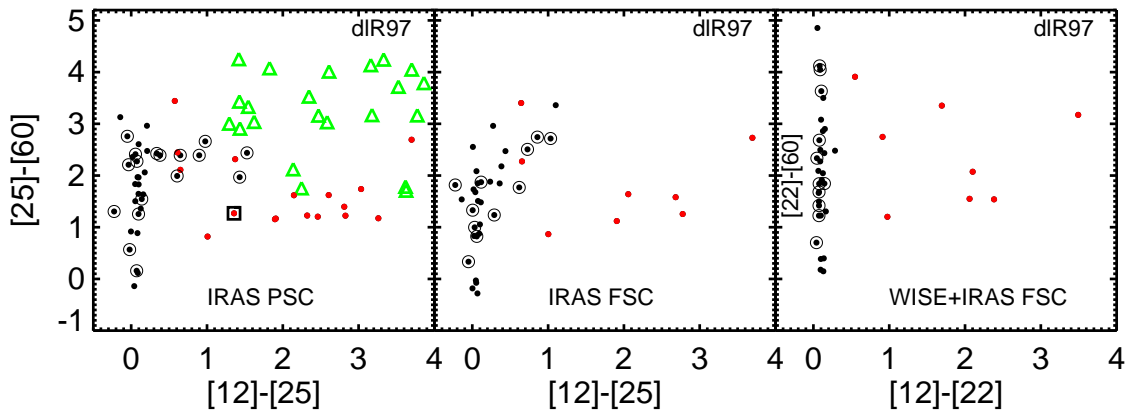}
\caption{Color-color plot of the dlR97 sample in the style of those
found in de la Reza \etal\ (1996), dlR97, and Siess \& Livio (1999),
where the IRAS [25]$-$[60] color is plotted against the IRAS
[12]$-$[25] color (see text for clarification of units). The left
panel uses IRAS PSC, the middle panel uses IRAS FSC, and the right
panel combines WISE [12] and [22] with IRAS FSC [60].  Green triangles
(first panel only) are objects we identify as likely subject to source
confusion. Red dots are  objects we identify as having an IR excess.
The square (first panel only) indicates the only object appearing in this plot
which we identify as having an IR excess, but that we suspect may not
be a K giant. Additional circles around sources are those for which
the IRAS PSC (or FSC) data quality flag in at least one of the
relevant bands is 1, indicating a limit. Objects may not be the same
between the first and second panels; all the objects in the third
panel also appear in the second panel. The most important features 
in these plots are: (1) the bulk of the distribution
delineating the reddest IR excesses is composed of objects likely
subject to confusion; (2) the envelope of the distribution shrinks
when going from the PSC to the FSC, and shrinks further when WISE is
used, suggesting that the apparent IR excesses measured in IRAS vanish
when higher spatial resolution observations are used. (3) All of the
largest excess K giants (those with detections and  [12]$-$[25]$>$0.5
mag) are recovered as IR excess sources, except for HD76066, discussed
in Sec.~\ref{sec:irlit}. 
\label{fig:vanishingirx}}
\end{figure}

In Figure~\ref{fig:vanishingirx}, we have made the plot analogous to
that from dlR97 and collaborators, namely [25]$-$[60] against
[12]$-$[25], just for the original dlR97 sample. The first panel uses
the values from the IRAS PSC, as in dlR97. The shape of the overall
distribution is similar to what they have obtained, in that there is a
locus near [12]$-$[25]$\sim$0 extending up to a range of [25]$-$[60]
values, and a broad `bubble' of points extending to the right. 
However, we find that the upper envelope of the distribution (near
[12]$-$[25]=1 to 4, and [25]$-$[60]$\sim$3 to 4) is defined entirely by
objects we suspect should be dropped from the sample because they are
subject to source confusion. Removing these points from the plot
signficantly reduces the range of colors found for K giants. We have
also indicated which of these sources are the ones for which we have
identified an IR excess. Most of the PSC points that are still thought
to be K giants with [12]$-$[25]$>$0.5 are also ones we identify as
having an IR excess, but there are several that we do not recover.
These sources are identified in at least one of [12], [25], or [60]
with data quality flag=1, e.g., a limit, not a detection. dlR97
started from the Pico dos Dias Survey (PDS; Gregorio-Hetem \etal\
1992), which reports having started from the IRAS PSC, only the
high-quality detections. However, several of the points in our version
of this diagram (which, again, is just the dlR97 sample) do not have
high quality detections at all 3 of the relevant IRAS bands.

The second panel of Figure~\ref{fig:vanishingirx} uses the IRAS FSC
instead of the PSC, again, just for the dlR97 sample. (Recall from
Sec.~\ref{sec:primarycatalogs} above that 36 of the dlR97 objects are
also detected in the FSC in any band, so not all of the objects from the first
panel appear in the second panel.) All of the objects we suspect
should be dropped do not have FSC measurements, so they do not appear
in the second panel. Again, most of the sources with detections (not
limits) in [12], [25], and [60], and large excesses with
[12]$-$[25]$>$0.5, are also ones we identify as having an IR excess,
but there is one that we do not recover. (It is HD76066, discussed
above as having significantly lower WISE flux densities than IRAS, and
not really having an excess.) Interestingly, in moving between the PSC
and FSC plots, the envelope delineating the red excursion of the
distribution of points shrinks dramatically in size, and there are
fewer points within the `bubble' -- more points are on the
[12]$-$[25]$\sim$0 locus where no IR excess is measured. Whereas
$\sim$45\% of the K giants in the first panel have [12]$-$[25]$>$0.5
mag, by the second panel, just $\sim$30\% have [12]$-$[25]$>$0.5 mag.
However, there are fewer objects overall in the second panel.

We can carry this further -- both IRAS and WISE had a 12 \mum\
channel, and IRAS had a 25 \mum\ channel, close to the WISE 22 \mum\
channel. Though the filter bandpasses are far from identical, they
ought to give similar estimates of the broadband IR excess near their
respective wavelengths. In the third panel of
Fig.~\ref{fig:vanishingirx}, we have used the WISE [12] in place of
the IRAS [12], and WISE [22] in place of the IRAS [25]; for 60 \mum,
we retain the IRAS FSC values. All of the objects from the middle
panel with WISE detections at [12] and [22] appear in the third panel.
The distribution continues to shrink towards the [12]$-$[22]=0 locus,
with now only $\sim$20\% of the sources having [12]$-$[22]$>$0.5 mag.
As higher spatial resolution and more sensitive observations are
available, the apparent IR excesses shrink.

However, in using WISE in place of the IRAS bands, we are implicitly
assuming that the two missions are calibrated in the same fashion, or
at least consistently with each other. It is possible that the change
in distributions (the shrinking of the `bubble') of points between the
panels of Fig.~\ref{fig:vanishingirx} can be accounted for, at least
in part, by different calibrations of the instruments.  To investigate
this, we compared the IRAS PSC to the FSC first, as a check, to make
sure that they were internally calibrated consistently with respect to
each other. Then, we compared the IRAS FSC to the WISE values. In all
cases, we used our entire sample, but only those with data quality
flags 3 or 2. The comparison of the 12 and 25 \mum\ channels
(calculating (PSC-FSC)/PSC with the values in magnitudes) shows that
they are indeed  well-matched to each other, with no significant
average offset between them. For [12], the center of the best-fit
Gaussian to the distribution is 0 mag, with a width of 0.04 mag. For
[25], the center is 0.02 mag, and the width is 0.04 mag.  However, in
comparing the WISE and IRAS FSC 12 \mum\ channels, there is a clear
offset of $\sim$30\% between WISE and IRAS, in the direction of the
FSC being brighter.  (At [12], calculating (FSC-WISE)/FSC in mag, the
center of the Gaussian is 0.26 mag and the width is 0.05 mag.) For 25
\mum, we expected larger differences between  WISE and IRAS because of
the different bandpasses. There is an offset between IRAS FSC and WISE
of $\sim$15\%. (The center of the Gaussian is 0.14 mag, and the width
is 0.06 mag.)  It is again in the direction of the FSC being
brighter.  Once one is made aware of this offset, one can see it
systematically in the SEDs of the ensemble where both IRAS and WISE
are available. This effect is in the same direction one would expect
if the higher spatial resolution of WISE was resolving out background
contributions to the IRAS flux, which may still be a part of what is
going on. (We note that we only did this comparison for the objects in
our sample, not the entire IRAS or WISE catalogs or over a controlled
range of backgrounds. Such a comparison is beyond the scope of our
study.) Nonetheless, the net effect in the last panel of
Fig.~\ref{fig:vanishingirx} is to move the envelope of points left and
up. Firstly, on the $x$-axis in this panel, [12] and [22] now both
come from WISE, and are internally well-calibrated, so stars without
excesses are closely clumped near 0 in [12]$-$[22]. They are more
tightly clumped than they were for IRAS, which collapses part of the
distribution. Stars with excesses have slightly smaller [12]$-$[22]
(from WISE) on average than [12]$-$[25] (from IRAS), because the
systematic calibration offset for [12] is slightly larger than for
[25] ([22]). Secondly, on the $y$-axis, [60] is still from IRAS,
presumably well-calibrated internally to the other IRAS bands. Since
both [12] and [25] are slightly systematically brighter compared to
WISE, if we assume [60] is also slightly brighter as a result of
calibration systematics, this will push the distribution of points
slightly up in the diagram -- which can be seen in the Figure.

So, we conclude that some (but not all) of the reduction in the range
of points in Fig.~\ref{fig:vanishingirx} can be accounted for in the
different calibrations of IRAS and WISE. The average calibration
effect for the ensemble is on the order of a few tenths of a
magnitude. However, the movement of individual objects between plots
is primarily a reflection of more accurate measurements of the IR flux
from the stars.

Kumar \etal\ (2015) also made plots like our
Fig.~\ref{fig:vanishingirx}, though in the same units as dlR97 (using
equation 2). Their plots have similar distributions of points as ours
do.

\section{Abundances and Rotation Rates}
\label{sec:abun}

One of our original goals of this paper was to seek a correlation
between lithium abundance (and rotation rate and the $^{12}$C/$^{13}$C
ratio) and IR excess in K giants, but with only $\sim$10\% of our
sample likely to have IR excesses, our ability to test correlations
with IR excess is somewhat limited. However, we can still infer some
things about the relationship among these parameters. In order to
better understand these relationships, however, we need to make sure
that our already biased sample is as clean and internally consistent
as possible.  

We now discuss how we limit the sample to identify Li-rich
stars, and likely first ascent K giants, and look at the relationship
between IR excess, Li abundance, rotation rate, and $^{12}$C/$^{13}$C.

\subsection{Definition of Li-Rich}

A substantial number of objects were added to our sample on the basis
of a paper in the literature asserting that the K giant was Li-rich.
However, everyone does not use the same definition of Li-rich.  For
example, dlR97 did not report Li abundances, but identified certain
stars as Li-rich based on equivalent widths. C12 included Li
abundances, and determined them under both LTE and NLTE assumptions.
Other literature sources sometimes report only equivalent widths, or
only LTE abundances. If we were to rely solely on literature
reporting, 183/316 sources are Li-rich. 

However, we wished to be a bit more restrictive, or at least
internally consistent. For those sources for which we have NLTE Li
abundances, we took those with $A$(Li)$_{\rm NLTE}$ $\geq$1.5 dex as
Li-rich.  If there was no NLTE abundance available, we took those with
$A$(Li)$_{\rm LTE}$ $\geq$1.5 dex as Li-rich.  If there was no
abundance in the literature (e.g., just equivalent widths), we did not
identify it as Li-rich. If we thought (based on the analysis above)
that it was not a K giant, we did not identify it as Li-rich (that
includes the 24 dropped sources, the carbon star, and the S-type
star). A total of 62 sources are missing $A$(Li), including the
dropped sources. Using this approach, 139 stars in our sample are
Li-rich.  Unfortunately, 10 of these sources are the ones with very
sparse SEDs, and 29 more have relatively sparse SEDs (though both of
the sources identified as having an IR excess from these SEDs are
Li-rich).  Just 9 of the remaining 100 sources have well-populated
SEDs and an IR excess.  There are 115 that are Li-poor, 7 of which
have sparse SEDs, and 6 of which have an IR excess.

\subsection{Restrictions on $\log g$ and \teff}

This study is aimed at understanding the Li-IR connection for K giants
(first ascent RGB stars and red clump stars); however, we have already
noted above that some of our sample are suspected to be more evolved
AGB stars or other non K-giant contaminants.  We can try to limit the
contamination by requiring that there be an estimate of $\log g$ and
\teff, and that these values fall within a certain range.  

We identify stars with $\log g >3.5$ as not likely giants. This
selection should weed out subgiant and dwarf stars for which high Li
may not be unusual. Out of the entire set of 316(-24 confused
sources), 46 have no $\log g$ estimate available to our knowledge. Of
the remaining sources, 16 have $\log g >3.5$ (with 2 more having
exactly 3.5 being left in the sample).  

We can put both upper and lower constraints on \teff. Temperatures
$<$3700 K are likely to be AGB stars, not first ascent K giants,
because the AGB reaches cooler temperatures. On the upper end,
temperatures $>$5200 K are likely to be dwarf stars or subgiants that
have not yet completed first dredge-up (the deepening of the
convection zone that reduces the surface Li abundances during the
post-MS phase).  Out of the entire set of 316(-24 confused sources),
20 have no \teff\ estimate at all. Of the ones with \teff, 6 are
cooler than 3700 K, several of which we identified above as `likely
too cool' for our sample. There are 3 stars with \teff=5200 K (left in
our sample), and there are 19 hotter than 5200 K.

The net loss of objects out of our sample by requiring that there be
an $A$(Li) and that \teff\ and $\log g$ are in the correct range is 97
objects (some objects counted in more than one omission category
above), leaving 219 in the sample. That sample includes the 10 very
sparse SEDs, and 33 of the relatively sparse SEDs. Of the 219, 119
(54$\pm$6\%)\footnote{The errors presented here are assumed to be the
larger of either Poisson errors or the binomial approximation found in
the appendix of  Burgasser \etal\ (2003).} are Li-rich as per our
definition above (with 10+26 of those being the very sparse/relatively
sparse SEDs). Just 13/219 (6$\pm$2\%) of this sample have an IR
excess. Bringing Li into it,  11/119 (9$\pm3$\%) of the Li-rich stars
have an IR excess, and 2/100 (2$^{+3}_{-0.6}$\%) of the Li-poor stars
have an IR excess (with 7 sparse SEDs included).

The sparse SEDs included in the above calculation likely miss more
subtle IR excesses, so we can repeat the analysis on the subset of 176 stars
with well-populated SEDS. There are 83 stars in this sub-sample that
are Li-rich, 93 that are Li-poor, and 11 that have an IR excess.  Of
the Li-rich stars, 9/83 (11$^{+4}_{-3}$\%) have an IR excess; of the
Li-poor stars, 2/93 (2$^{+3}_{-1}$) have an IR excess.  Although we
are well into the regime of small-number statistics, IR excesses
appear to be at least 2--3 times as common among Li-rich stars
compared to Li-poor stars.   Kumar \etal\ (2015) came to a similar
conclusion, that IR excesses are rare in the general K giant
population. They found that only $\sim$1\% of RGs (their sample is
dominated by Li-poor stars) have an IR excess. Of the 40 Li-rich stars
in their sample, they find 7 (18$^{+8}_{-4}$\%) with an IR excess.

\subsection{Relationships Among $A$(Li), \vsini, $^{12}$C/$^{13}$C, and
IR excess}

For the rest of this section, we will only use the cleanest possible
sample of 176 RGs with \teff\ and $\log g$ consistent with
K giant stars, and with well-populated SEDs, from the prior section.

\begin{figure}[h]
\epsscale{0.7}
\plotone{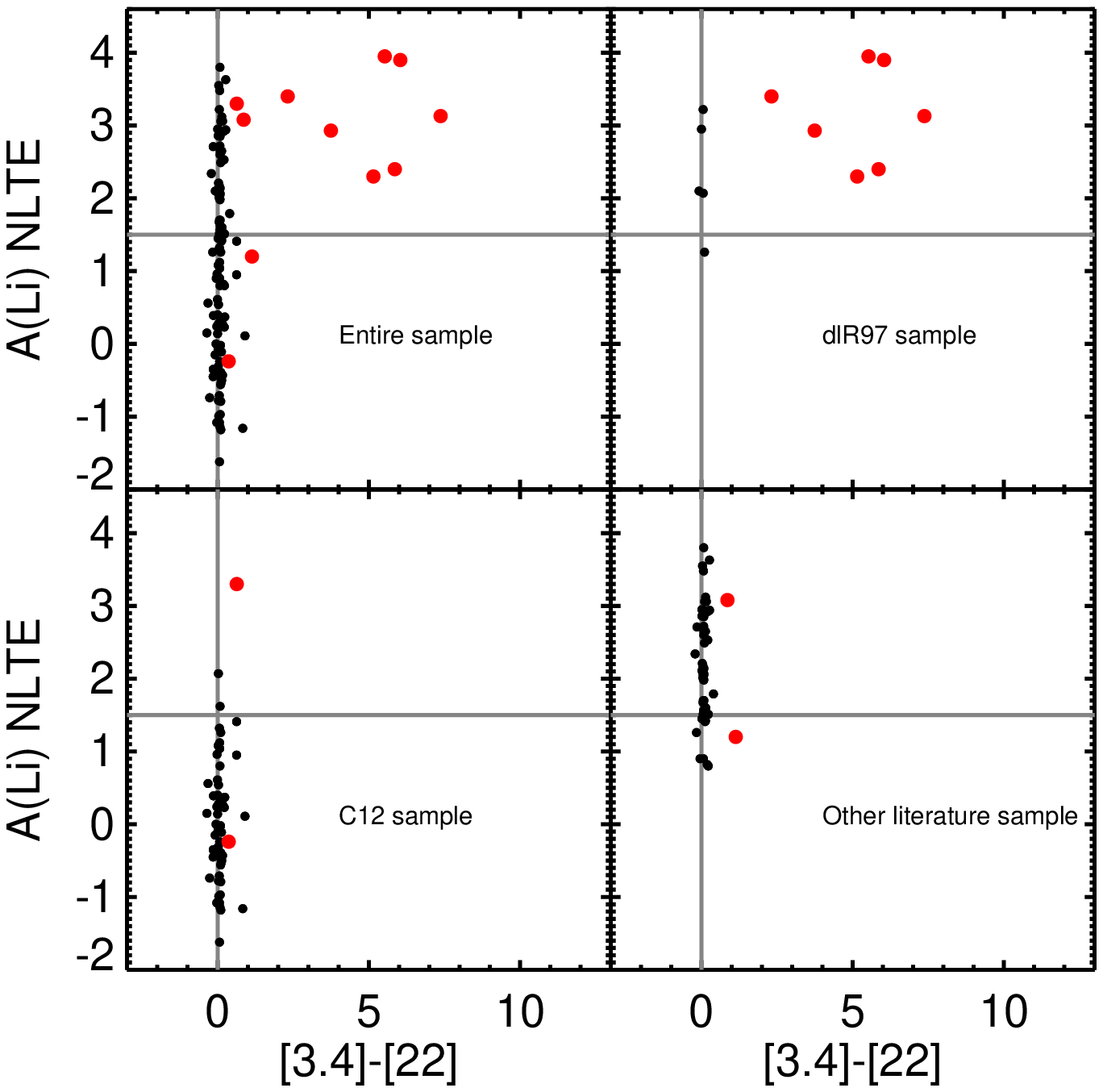}
\caption{$A$(Li)$_{\rm NLTE}$ vs.\ [3.4]$-$[22] for the 176 stars in the
cleanest possible sample. The vertical line at [3.4]$-$[22]=0
indicates the photospheric locus, and the red points are the IR excess
stars. The horizontal line at $A$(Li)=1.5 dex indicates our adopted
division between Li-rich and not Li-rich. The entire (available)
sample is plotted in the upper left, and the component samples (dlR97,
C12, and the other literature) are shown in separate panels. A very
similar plot is obtained  if $A$(Li)$_{\rm LTE}$  is used instead of
$A$(Li)$_{\rm NLTE}$, or $K_s-[22]$ rather than [3.4]$-$[22].  If a
star has a large IR excess, it probably has a large $A$(Li), but
having a large $A$(Li) does not necessarily indicate it has a large IR
excess. \label{fig:3422li}}
\end{figure}

Figure~\ref{fig:3422li} shows $A$(Li)$_{\rm NLTE}$ vs.\ [3.4]$-$[22] for the
176 stars in the cleanest possible sample. This plot suggests that if
a star has a large IR excess, it probably has a large $A$(Li), but
having a large $A$(Li) does not mean that it necessarily has a large
IR excess. Smaller excesses can be found at all abundance levels. Very
similar results are obtained if LTE rather than NLTE lithium
abundances are used, or if $K_s-[22]$ is used instead of [3.4]$-$[22].
Within the sample of Li-rich objects with IR excesses, there does not
seem to be a trend that, say, the largest Li abundances are always
found with the largest IR excesses.

\begin{figure}[h]
\epsscale{0.7}
\plotone{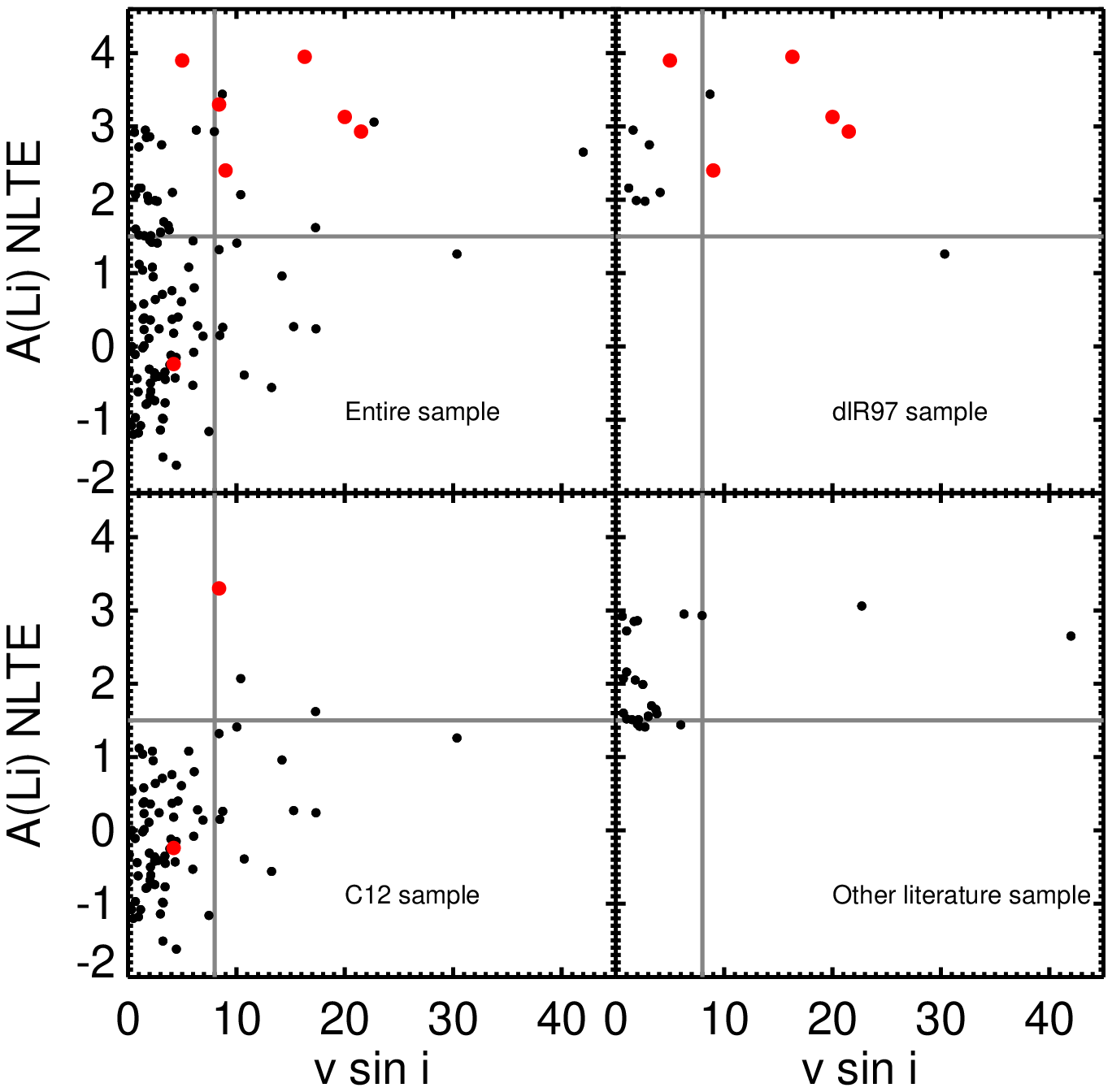}
\caption{$A$(Li)$_{\rm NLTE}$ vs.\ \vsini\ in km s$^{-1}$ for the
cleanest possible sample. The vertical line at 8 km s$^{-1}$ divides
the fast from the slow rotators. The red points are the IR excess
stars. The horizontal line at $A$(Li)=1.5 dex is the  division between
Li-rich and not Li-rich. The entire (available) sample is plotted in
the upper left, and the component samples (dlR97, C12, and the other
literature) are shown in separate panels. A very similar plot is
obtained  if $A$(Li)$_{\rm LTE}$ is used instead of $A$(Li)$_{\rm
NLTE}$.  Fast-rotating stars also often (but not exclusively) have
large $A$(Li), and often also have an IR excess.
\label{fig:vsinili}}
\end{figure}

Figure~\ref{fig:vsinili} shows $A$(Li)$_{\rm NLTE}$ vs.\ \vsini. 
Fast-rotating stars also often (but not exclusively) have large
$A$(Li), which has been previously noted (e.g., Drake \etal\ 2002).
What is further revealed by this plot is that many of the fast
rotating, Li-rich stars are also those with IR excesses. Half of the
Li-rich stars that also show fast rotation have an IR excess, whereas
only one Li-rich star among the more populated slow rotators has an IR
excess.  Additionally, only one RG with an IR excess shows neither
high rotation nor enriched Li. Thus, having both high Li and fast
rotation is a stronger predictor for an IR excess than high Li alone. 
This suggests that relatively enhanced angular momentum is necessary
for the ejection of circumstellar shells in Li-enriched stars.  We
note that there are several IR excess sources unable to be plotted in
this diagram because no \vsini\ is available, and that a fundamental
uncertainty in the use of projected rotational velocities is that
inclination effects can mask rapid rotation.

\begin{figure}[h]
\epsscale{0.7}
\plotone{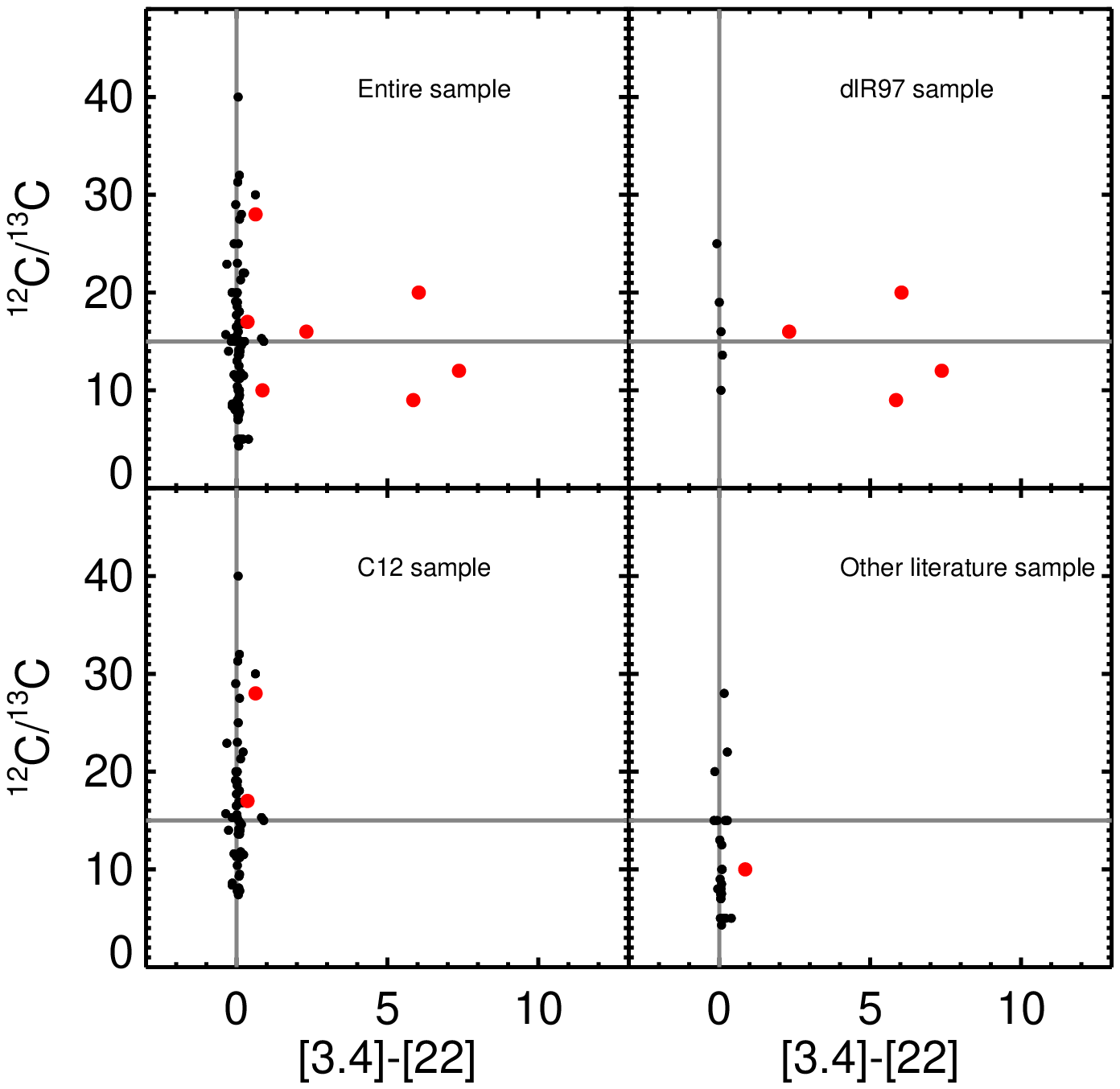}
\caption{$^{12}$C/$^{13}$C vs.\ [3.4]$-$[22] for the
cleanest possible sample. The vertical line at [3.4]$-$[22]=0
indicates the photospheric locus, and the red points are the IR excess
stars. The horizontal line at $^{12}$C/$^{13}$C=15 is the 
division between a high and low ratio. The entire (available)
sample is plotted in the upper left, and the component samples (dlR97,
C12, and the other literature) are shown in separate panels. A very
similar plot is obtained  if $A$(Li)$_{\rm LTE}$ is used instead of
$A$(Li)$_{\rm NLTE}$.  There is no discernible correlation of IR excess with
$^{12}$C/$^{13}$C.
\label{fig:3422cratiov3}}
\end{figure}

\begin{figure}[h]
\epsscale{0.7}
\plotone{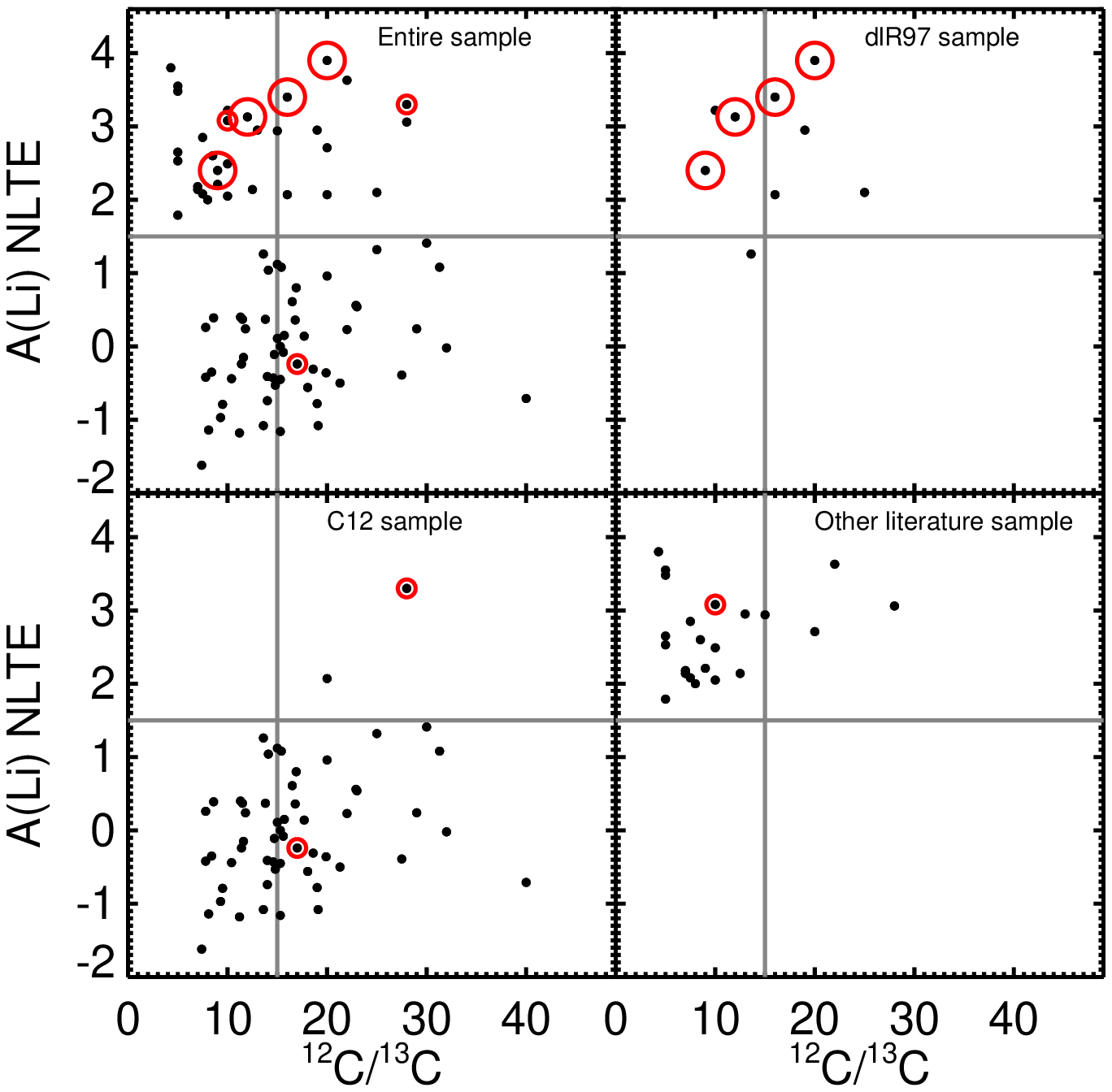}
\caption{$A$(Li) vs.\ $^{12}$C/$^{13}$C  for the cleanest possible
sample. The vertical line at $^{12}$C/$^{13}$C=15 is the  division
between a high and low ratio; the horizontal line at $A$(Li)=1.5 is
the division between Li-rich and Li-poor. The red points highlight the
IR excess stars; smaller circles are the smaller excesses, and larger
circles are larger excesses ([3.4]$-$[22]$>$1). The entire (available)
sample is plotted in the upper left, and the component samples (dlR97,
C12, and the other literature) are shown in separate panels. 
\label{fig:alicratioirx}}
\end{figure}

As seen in Figure~\ref{fig:3422cratiov3}, no correlations can be found
between IR excess and the carbon isotope ratio, $^{12}$C/$^{13}$C. Very low
$^{12}$C/$^{13}$C is thought to indicate substantial extra mixing. IR
excess sources (extreme and moderate) are found with both low and high
$^{12}$C/$^{13}$C, where we have taken 15 as the division between low
and high values. The largest IR excesses do not have unusual carbon
ratios.   However, we note that relatively few of our stars have a
measure of $^{12}$C/$^{13}$C, so this may introduce some additional
biases.

In Figure~\ref{fig:alicratioirx}, we plot the $A$(Li) versus
$^{12}$C/$^{13}$C, a plot that should give some insight into the Li
enrichment mechanism.  In the presence of extra mixing, the surface
values of both $^{12}$C/$^{13}$C and $A$(Li) will be reduced if the
mixing proceeds slowly (e.g., Denissenkov \& VandenBerg 2003).  
However, when the mixing proceeds rapidly, newly synthesized Li can be
brought into the convection zone, where it is long-lived, such that
the surface Li is enhanced while $^{12}$C/$^{13}$C decreases (e.g.,
Denissenkov \& Herwig 2004).  Once the source of the Li ($^3$He) is
considerably depleted, a net destruction of Li begins and $A$(Li) will
again be reduced.  The physical mechanism behind this very fast mixing
is still under investigation.  Denissenkov \& Herwig (2004) argued
that rotation-induced mixing was required, and since RGs are slow
rotators, an outside source of angular momentum such as binary
interactions or planet engulfment was required. However, Palacios
\etal\ (2006) argued that the shear turbulence caused by differential
rotation was not sufficiently fast to increase Li at the surface.
Magnetic buoyancy is another possible model (e.g., Guandalini \etal\
2009). 

Planet engulfment that does {\em not} trigger fast mixing could
be identified with higher than expected Li, but with a relatively high
$^{12}$C/$^{13}$C. In Figure~\ref{fig:alicratioirx}, the C12 sample
shows the most homogeneously measured $^{12}$C/$^{13}$C and the only
sample unbiased towards Li. The stars generally show the expected
linear trend of lower $^{12}$C/$^{13}$C and lower $A$(Li) for stars
that experience different degrees of mixing.  Other panels of the plot
show the addition of the large sample of all the literature Li-rich
stars, which span a large range of $^{12}$C/$^{13}$C.  Red circles
again indicate IR excess. Stars with the clearest sign of substantial
internal mixing (very low $^{12}$C/$^{13}$C and high Li) show no
evidence of an IR excess. However, this is not entirely unexpected. As
pointed out by Denissenkov \& Herwig (2004), the timescale over which
a shell is ejected and dissipates (e.g., the timescale for the IR
excess to appear/disappear) is of order 10$^4$-10$^5$ years, compared
to the timescale of Li regeneration, which is of order 10$^5$-10$^6$
yrs. Very low values of $^{12}$C/$^{13}$C are only reached at the
latest stages of Li regneration, well after the IR excess has
disappeared. Figure~\ref{fig:alicratioirx} does, however, exhibit a
tantalizing correlation between the Li and $^{12}$C/$^{13}$C of many
of the IR excess stars in this plot. We offer no explanation for this
trend but suggest that it may be informative on the conditions of the
star when the shell is ejected. Adding additional objects to this plot
(via determinations of $^{12}$C/$^{13}$C) will likely further
illuminate any relationship.

Given the relative shortness of the IR phase compared to the Li
synthesis phase, one would expect little change in $A$(Li) for any
particular star as it traverses the IR color-color diagram
(Figure~\ref{fig:vanishingirx}). This expectation is confirmed by the
fact that we do not see a correlation between level of Li-richness and
the strength of the IR-excess (Figure~\ref{fig:3422li}). Furthermore,
since IR excess is seen around a wide range of $A$(Li), we can speculate
that the Li regneration reaches different maximum $A$(Li) values in each
star.

\subsection{Biases and Future Work}

An additional concern in interpreting
Figs.~\ref{fig:3422li}-\ref{fig:3422cratiov3}  is that our sample has
many substantial biases within it. One very significant bias is that
many of the sources were discovered based on their IRAS colors, e.g.,
they are biased towards IR-bright sources. The subsamples are
separated out in Figs.~\ref{fig:3422li}-\ref{fig:3422cratiov3}
specifically because of this bias -- a substantial fraction of the
dlR97 sample and a smaller fraction of the literature sample is based
on IRAS colors. In Fig.~\ref{fig:3422li}, it can be seen that all of
the large IR excesses are from the dlR97 sample. (Some of the large
excesses seen in, e.g., the literature sample in Fig.~\ref{fig:3422}
do not appear in this plot because they do not have $\log g$ or \teff\
in the correct range, etc.) Those largest excesses draw the eye and
dominate the relationships found in the plots. The relationships are
not as obvious in just the C12 sample, which is the least biased with
respect to IR properties.

To first order, we hoped we could try to constrain the influence of
this bias by omitting K giants known primarily by their IRAS names.
Only 5 of the sources making it into our `cleanest possible' sample
have IRAS names, but 4 of them are among the largest IR
excesses ([3.4]$-$[22]$>$2). 

We also have biases in the sample due to the incomplete information
for stars in the sample. A number of the IR excess sources have no
\teff\ and/or $\log g$ and were removed from our detailed analysis of
IR excess among K giants. Furthermore, half of the Li-rich sample has
no measure of $^{12}$C/$^{13}$C and/or $v \sin i$.  This makes comparisons
between, e.g., Figures~\ref{fig:vsinili} and \ref{fig:3422cratiov3},
difficult because the IR excess stars appearing in each plot are not
all the same stars. Follow-up  high resolution optical spectra to
measure these missing stellar parameters would be extremely valuable.

It is also worth considering if the objects with the largest excesses
are not really old dusty giants ascending the giant branch, but young
dusty giants, still contracting along their Hayashi track. This could
explain both the very large excesses as well as the high lithium
abundances, since young stars are known to often have high $A$(Li).
Young, actively accreting stars with substantial disks would have
strong and variable H$\alpha$ profiles, and have significant
variability at essentially all wavelengths, both of which are
different than expectations for old giant stars. Detailed isotopic
ratios (such as $^{12}$C/$^{13}$C) that trace mixing and chemical
evolution would also be of help. Additional detailed spectroscopic
data and modeling is required to distinguish old dusty stars from
young dusty stars.

\clearpage

\section{Conclusions}
\label{sec:concl}

In the past, dlR97 and others have suggested a connection between
enhanced lithium and IR excesses in K giants. However, others (e.g.,
Fekel \& Watson 1998, Jasniewicz \etal\ 1999;  Lebzelter \etal\ 2012;
Kumar \etal\ 2015) have questioned the association between Li and IR
abundances.  We have assembled a set of 316 targets thought to be K
giants, two-thirds of which are thought to be Li-rich, in order to
test the association between IR excesses and lithium abundances. The
targets come from the dlR97 study (biased towards IR-bright sources),
the C12 study (assembled with limited biases to test correlations of
various stellar parameters), and a wide variety of literature
identifying Li-rich K giants.  For these targets, we assembled a
multiwavelength catalog spanning optical through 100 \mum\ data, using
SDSS, NOMAD, 2MASS, DENIS, WISE, IRAS, AKARI, MSX, and Spitzer
data.   

We inspected each source in as many different images as possible. In
24 cases, all identified first as IR-bright sources with IRAS, we
believe that source confusion is playing a role, in that either (a)
the source that is bright in the optical (and most likely the source
of which a spectrum was obtained to assess lithium) is not responsible
for the IR flux, or (b) there is more than one source responsible for
the IR flux as measured in IRAS. 

We looked for IR excesses by $\sim$20 \mum\ using two different
approaches : (a) simple SED construction and assessment, and (b) an
approach drawn from studies of young stars used to identify small but
significant IR excesses. We identify 19 stars with large IR excesses,
and 9 more stars that have small but significant IR excesses. However,
5 of these 28 may not be first ascent K giants.  (There are 2 more K
giants that may have IR excesses by $\sim$10 \mum, identified from
relatively sparse SEDs.) Ten of the 28 clear IR excess K giants were
already recently identified in the literature as having IR excesses,
but this is the first recent confirmation of IR excess for 18 of these
targets.  Some of these giants have IR excesses that start at or
before 5 \mum, but others have excesses that start near 20 \mum.

IR excesses by 20 \mum, though rare, are about twice as common among
Li-rich K giants (11$^{+4}_{-3}$\%) as in Li-poor K giants
(2$^{+3}_{-1}$\%).  Despite identifying very few IR excesses (by
number or fraction of sample), we find that if a RG has a large IR
excess, it probably has a large $A$(Li) and is a fast rotator, but
having a large $A$(Li) (or being a fast rotator) does not mean that it
necessarily has a large IR excess. Smaller excesses can be found at
all abundance levels.  This is consistent with the idea that the IR
excess lifetime of a single ejected shell is very short-lived compared
to the timescale of Li enrichment. It could also suggest that not all
Li-rich stars eject shells, and some other parameter (such as fast
rotation, or even rotation history) dictates whether shell ejection
occurs.  Stars with the clearest sign of substantial internal mixing
(very low $^{12}$C/$^{13}$C and high Li) show no evidence of an IR
excess. This could be also explained by a shorter timescale for the IR
excess than the Li regeneration, since the lowest $^{12}$C/$^{13}$C
are realized near the end of the Li-enrichment stage.  An external Li
regeneration mechanism identified in the literature is planet
injestion. However, we identify only one of the three best candidates
for planet accretion listed in C12 as having a measurable IR excess,
and it is a small excess. The largest IR excesses are all found in the
dlR97 sample, which is strongly biased towards IR-bright objects. 
There remains a possibility that at least some of these largest excess
objects may not be old dusty stars, but instead young dusty stars.

\acknowledgments

Support provided for this work by the NASA/IPAC Teacher Archive
Research Program (NITARP; http://nitarp.ipac.caltech.edu), which
partners small groups of high school educators with a mentor
astronomer for an authentic research project. It receives funding from
the NASA ADP program and the IPAC archives.  We acknowledge the
following students who helped out at various phases of this project:
Rosie Buhrley, Julie Herring, Kendall Jacoby, and Elena Mitchell,
from Walden School of Liberal Arts.

JKC was supported by an appointment to the NASA Postdoctoral Program
at the Goddard Space Flight Center, administered by Oak Ridge
Associated Universities through a contract with NASA.

This research has made etensive use of the NASA/ IPAC Infrared Science
Archive, which is operated by the Jet Propulsion Laboratory,
California Institute of Technology, under contract with the National
Aeronautics and Space Administration.

The Digitized Sky Survey was produced at the Space Telescope Science
Institute under U.S. Government grant NAG W-2166. The images of these
surveys are based on photographic data obtained using the Oschin
Schmidt Telescope on Palomar Mountain and the UK Schmidt Telescope.
The plates were processed into the present compressed digital form
with the permission of these institutions."

Funding for SDSS-III has been provided by the Alfred P. Sloan
Foundation, the Participating Institutions, the National Science
Foundation, and the U.S. Department of Energy Office of Science. The
SDSS-III web site is http://www.sdss3.org/. SDSS-III is managed by the
Astrophysical Research Consortium for the Participating Institutions
of the SDSS-III Collaboration including the University of Arizona, the
Brazilian Participation Group, Brookhaven National Laboratory,
University of Cambridge, Carnegie Mellon University, University of
Florida, the French Participation Group, the German Participation
Group, Harvard University, the Instituto de Astrofisica de Canarias,
the Michigan State/Notre Dame/JINA Participation Group, Johns Hopkins
University, Lawrence Berkeley National Laboratory, Max Planck
Institute for Astrophysics, Max Planck Institute for Extraterrestrial
Physics, New Mexico State University, New York University, Ohio State
University, Pennsylvania State University, University of Portsmouth,
Princeton University, the Spanish Participation Group, University of
Tokyo, University of Utah, Vanderbilt University, University of
Virginia, University of Washington, and Yale University.

This publication makes use of data products from the Two Micron All
Sky Survey, which is a joint project of the University of
Massachusetts and the Infrared Processing and Analysis
Center/California Institute of Technology, funded by the National
Aeronautics and Space Administration and the National Science
Foundation.

This publication makes use of data products from the Wide-field
Infrared Survey Explorer, which is a joint project of the University
of California, Los Angeles, and the Jet Propulsion
Laboratory/California Institute of Technology, funded by the National
Aeronautics and Space Administration.

This research is based on observations with AKARI, a JAXA project with
the participation of ESA.

This work is based in part on observations made with the Spitzer Space
Telescope, which is operated by the Jet Propulsion Laboratory,
California Institute of Technology under a contract with NASA. 

This research made use of data products from the Midcourse Space
Experiment. Processing of the data was funded by the Ballistic Missile
Defense Organization with additional support from NASA Office of Space
Science. This research has also made use of the NASA/ IPAC Infrared
Science Archive, which is operated by the Jet Propulsion Laboratory,
California Institute of Technology, under contract with the National
Aeronautics and Space Administration. 

This research has made use of NASA's Astrophysics Data System (ADS)
Abstract Service, and of the SIMBAD database, operated at CDS,
Strasbourg, France.

\clearpage
\appendix{Appendix: Notes on individual sources}

This Appendix contains human-readable notes on the entire set of
sources; position, photometric, and abundance data are in the
machine-readable table in Table~\ref{tab:bigdatatable}. Detailed notes
on the sources we suspect are subject to source confusion appear in
Table~\ref{tab:droppedsrcs}. Much more detailed notes on the sources
we believe have IR excesses are in Section~\ref{sec:irx}.

\begin{deluxetable}{lllp{3cm}lp{6cm}p{3cm}}
\tabletypesize{\scriptsize}
\rotate
\tablecaption{Special Notes on Targets \label{tab:specialnotes}}
\tablewidth{0pt}
\tablehead{\colhead{name} & \colhead{dlR97} & \colhead{C12} &
\colhead{Other lit}& \colhead{status}& \colhead{notes}&
\colhead{$A$(Li), $\log g$, \teff?}}
\startdata

HD787	& 	dlR97	& 	\nodata	& 	\nodata	& 	(no excess)	& 	\nodata	& 	\nodata			\\ 
Tyc3663-01966-1	&	\nodata	&	\nodata	&	Adamow \etal\ (2014)	& 	(no excess)	& 	\nodata	& 	 Not Li rich.			\\ 
HD4893	&	\nodata	&	\nodata	&	Castilho \etal\ (2000)	& 	(no excess)	& 	May be too cool to be a K giant.	& 	 Not Li rich.			\\ 
IRAS00483-7347	&	\nodata	&	\nodata	&	Castilho \etal\ (1998)	& 	IR excess (but maybe not K giant)	& 	May be too cool to be a K giant.	& 	 No $A$(Li).	 No \teff.	 No $\log g$.	\\ 
Scl 1004838	&	\nodata	&	\nodata	&	Kirby \etal\ (2012)	& 	sparse SED but probably no IR excess	& 	\nodata	& 	\nodata			\\ 
Scl 1004861	&	\nodata	&	\nodata	&	Kirby \etal\ (2012)	& 	sparse SED but probably no IR excess	& 	\nodata	& 	\nodata			\\ 
NGC 362 V2	&	\nodata	&	\nodata	&	Smith \etal\ (1999)	& 	IR excess (small)	& 	\nodata	& 	 Not Li rich.			\\ 
HD6665	& 	\nodata	& 	\nodata	& 	Kumar \etal\ (2011)	& 	(no excess)	& 	Identified in McDonald \etal\ (2012) as having an IR excess, but IR excess is not real.	& 	\nodata			\\ 
HD7087	&	\nodata	&	\nodata	&	Liu \etal\ (2014)	& 	(no excess)	& 	\nodata	& 	\nodata			\\ 
HD8676	& 	\nodata	& 	\nodata	& 	Kumar \etal\ (2011)	& 	(no excess)	& 	\nodata	& 	\nodata			\\ 
HD9746	& 	dlR97	& 	\nodata	& 	\nodata	& 	(no excess)	& 	\nodata	& 	\nodata			\\ 
HD10437	& 	\nodata	& 	\nodata	& 	Kumar \etal\ (2011)	& 	(no excess)	& 	\nodata	& 	\nodata			\\ 
HD12203	& 	\nodata	& 	\nodata	& 	Kumar \etal\ (2011)	& 	(no excess)	& 	\nodata	& 	\nodata			\\ 
CPD-55395	& 	dlR97	& 	\nodata	& 	\nodata	& 	(no excess)	& 	\nodata	& 	 Not Li rich.	 No \teff.	 No $\log g$.	\\ 
HD13189	& 	\nodata	& 	C12	& 	\nodata	& 	(no excess)	& 	\nodata	& 	 Not Li rich.			\\ 
HD15866	&	\nodata	&	\nodata	&	Liu \etal\ (2014), Luck \& Heiter (2007)	& 	(no excess)	& 	\nodata	& 	 	\teff\ too warm to be RG.	$\log g$ too large to be RG.	\\ 
For 55609	&	\nodata	&	\nodata	&	Kirby \etal\ (2012)	& 	sparse SED but probably no IR excess	& 	\nodata	& 	\nodata			\\ 
For 60521	&	\nodata	&	\nodata	&	Kirby \etal\ (2012)	& 	sparse SED but probably no IR excess	& 	\nodata	& 	\nodata			\\ 
Tyc3300-00133-1	&	\nodata	&	\nodata	&	Adamow \etal\ (2014)	& 	(no excess)	& 	\nodata	& 	\nodata			\\ 
For 90067	&	\nodata	&	\nodata	&	Kirby \etal\ (2012)	& 	sparse SED but possible IR excess	& 	\nodata	& 	\nodata			\\ 
For 100650	&	\nodata	&	\nodata	&	Kirby \etal\ (2012)	& 	sparse SED but probably no IR excess	& 	\nodata	& 	\nodata			\\ 
Tyc3304-00090-1	&	\nodata	&	\nodata	&	Adamow \etal\ (2014)	& 	(no excess)	& 	\nodata	& 	\nodata			\\ 
Tyc1780-00654-1	& 	\nodata	& 	C12	& 	\nodata	& 	(no excess)	& 	\nodata	& 	 Not Li rich.			\\ 
HD17144	&	\nodata	&	\nodata	&	Drake \etal\ (2002)	& 	(no excess)	& 	\nodata	& 	 Not Li rich.		 No $\log g$.	\\ 
SDSS J0245+7102	&	\nodata	&	\nodata	&	Martell \etal\ (2013)	& 	(no excess)	& 	\nodata	& 	\nodata			\\ 
Tyc0647-00254-1	& 	\nodata	& 	C12	& 	\nodata	& 	(no excess)	& 	\nodata	& 	\nodata			\\ 
SDSS J0301+7159	&	\nodata	&	\nodata	&	Martell \etal\ (2013)	& 	(no excess)	& 	\nodata	& 	\nodata			\\ 
G0300+00.29	& 	\nodata	& 	C12	& 	\nodata	& 	sparse SED but probably no IR excess	& 	\nodata	& 	 Not Li rich.			\\ 
Tyc3318-01333-1	&	\nodata	&	\nodata	&	Adamow \etal\ (2014)	& 	(no excess)	& 	\nodata	& 	\nodata			\\ 
SDSS J0304+3823	&	\nodata	&	\nodata	&	Martell \etal\ (2013)	& 	sparse SED but probably no IR excess	& 	\nodata	& 	\nodata			\\ 
Tyc5868-00337-1	& 	\nodata	& 	C12	& 	\nodata	& 	(no excess)	& 	\nodata	& 	 Not Li rich.			\\ 
HD19745	& 	dlR97	& 	\nodata	& 	\nodata	& 	IR excess	& 	Also identified in Kumar \etal\ (2015) as an IR excess star.	& 	\nodata			\\ 
Tyc3314-01371-1	&	\nodata	&	\nodata	&	Adamow \etal\ (2014)	& 	(no excess)	& 	\nodata	& 	 Not Li rich.			\\ 
HD21078	&	\nodata	&	\nodata	&	Fekel \& Watson (1998)	& 	(no excess)	& 	\nodata	& 	 Not Li rich.		 No $\log g$.	\\ 
G0319+56.5830	& 	\nodata	& 	C12	& 	\nodata	& 	(no excess)	& 	\nodata	& 	 Not Li rich.			\\ 
HD21018	& 	\nodata	& 	\nodata	& 	Kumar \etal\ (2011)	& 	(no excess)	& 	\nodata	& 	\nodata			\\ 
G0319+56.6888	& 	\nodata	& 	C12	& 	\nodata	& 	(no excess)	& 	\nodata	& 	 Not Li rich.			\\ 
Tyc5881-01156-1	& 	\nodata	& 	C12	& 	\nodata	& 	(no excess)	& 	A	& 	 Not Li rich.			\\ 
IRAS03520-3857	& 	dlR97	& 	\nodata	& 	\nodata	& 	IR excess	& 	Given position is offset $\sim$11$\arcsec$ from object taken as match (which is large in the context of the rest of this data set), but relatively isolated source and relatively clean field (e.g., unlikely to be confused). 	& 	 No $A$(Li).	 No \teff.	 No $\log g$.	\\ 
HD26162	& 	\nodata	& 	C12	& 	\nodata	& 	(no excess)	& 	\nodata	& 	 Not Li rich.			\\ 
Tyc3340-01195-1	& 	\nodata	& 	C12	& 	\nodata	& 	(no excess)	& 	\nodata	& 	 Not Li rich.			\\ 
HD27497	&	\nodata	&	\nodata	&	Jasniewicz \etal\  (1999)	& 	(no excess)	& 	\nodata	& 	 Not Li rich.			\\ 
RAVEJ043154.1-063210	&	\nodata	&	\nodata	&	Ruchti \etal\ (2011)	& 	sparse SED but probably no IR excess	& 	\nodata	& 	 	\teff\ too warm to be RG.		\\ 
IRASF04376-3238	&	\nodata	&	\nodata	&	Torres \etal\ (2000)	& 	IR excess	& 	Also CD-32 1919.	& 	 No $A$(Li).	 No \teff.	 No $\log g$.	\\ 
HD30238	& 	dlR97	& 	\nodata	& 	\nodata	& 	(no excess)	& 	\nodata	& 	 Not Li rich.			\\ 
HD30197	&	\nodata	&	\nodata	&	Liu \etal\ (2014), Luck \& Heiter (2007)	& 	(no excess)	& 	\nodata	& 	\nodata			\\ 
Tyc0684-00553-1	&	\nodata	&	\nodata	&	Adamow \etal\ (2014)	& 	(no excess)	& 	\nodata	& 	\nodata			\\ 
HD30834	& 	dlR97	& 	\nodata	& 	\nodata	& 	(no excess)	& 	\nodata	& 	\nodata			\\ 
Tyc5904-00513-1	& 	\nodata	& 	C12	& 	\nodata	& 	(no excess)	& 	\nodata	& 	 Not Li rich.			\\ 
G0453+00.90	& 	\nodata	& 	C12	& 	\nodata	& 	sparse SED but probably no IR excess	& 	\nodata	& 	 Not Li rich.			\\ 
HD31993	& 	dlR97	& 	C12	& 	\nodata	& 	(no excess)	& 	Only target in common between dlR97 and C12	& 	 Not Li rich.			\\ 
HD34198	& 	\nodata	& 	C12	& 	\nodata	& 	(no excess)	& 	Also UU Lep	& 	 Not Li rich.			\\ 
HD33798	&	\nodata	&	\nodata	&	Drake \etal\ (2002)	& 	(no excess)	& 	\nodata	& 	 		$\log g$ too large to be RG.	\\ 
HD33363	& 	\nodata	& 	C12	& 	\nodata	& 	(no excess)	& 	\nodata	& 	 Not Li rich.			\\ 
HD35984	&	\nodata	&	\nodata	&	Liu \etal\ (2014), Luck \& Heiter (2007)	& 	(no excess)	& 	\nodata	& 	 	\teff\ too warm to be RG.	$\log g$ too large to be RG.	\\ 
SDSS J0535+0514	&	\nodata	&	\nodata	&	Martell \etal\ (2013)	& 	sparse SED but probably no IR excess	& 	\nodata	& 	\nodata			\\ 
HD37719	& 	\nodata	& 	\nodata	& 	Kumar \etal\ (2011)	& 	(no excess)	& 	\nodata	& 	\nodata			\\ 
Be 21 T50	&	\nodata	&	\nodata	&	Hill \& Pasquini (1999)	& 	sparse SED but probably no IR excess	& 	Hill \& Pasquini call this T33 but it is not; based on photometry from Tosi \etal\ (1998), it is T50.	& 	\nodata			\\ 
HD39853	& 	dlR97	& 	\nodata	& 	\nodata	& 	(no excess)	& 	\nodata	& 	\nodata			\\ 
HD40359	&	\nodata	&	\nodata	&	Fekel \& Watson (1998)	& 	(no excess)	& 	\nodata	& 	 Not Li rich.		 No $\log g$.	\\ 
HD40168	& 	\nodata	& 	\nodata	& 	Kumar \etal\ (2011)	& 	(no excess)	& 	\nodata	& 	\nodata			\\ 
HD40827	& 	\nodata	& 	\nodata	& 	Kumar \etal\ (2011)	& 	(no excess)	& 	\nodata	& 	\nodata			\\ 
HD43827	&	\nodata	&	\nodata	&	Jasniewicz \etal\  (1999)	& 	(no excess)	& 	\nodata	& 	 Not Li rich.		 No $\log g$.	\\ 
Tyc1890-01314-1	& 	\nodata	& 	C12	& 	\nodata	& 	(no excess)	& 	\nodata	& 	 Not Li rich.			\\ 
HD44889	&	\nodata	&	\nodata	&	Castilho \etal\ (2000)	& 	(no excess)	& 	\nodata	& 	 Not Li rich.			\\ 
SDSS J0632+2604	&	\nodata	&	\nodata	&	Martell \etal\ (2013)	& 	sparse SED but probably IR excess	& 	\nodata	& 	\nodata			\\ 
Tr5 3416	&	\nodata	&	\nodata	&	Monaco \etal\ (2014)	& 	sparse SED but probably no IR excess	& 	\nodata	& 	\nodata			\\ 
HD47536	& 	\nodata	& 	C12	& 	\nodata	& 	(no excess)	& 	WISE measurements from AllWISE reject catalog	& 	 Not Li rich.			\\ 
IRAS06365+0223	& 	dlR97	& 	\nodata	& 	\nodata	& 	drop due to source confusion	& 	2MASS measurements from Extended Source Catalog; WISE has 2 similar sources at this location in catalog, but not in  image, so taking slightly closer. Extended source is origin of most of IR emission.	& 	 No $A$(Li).	 No \teff.	 No $\log g$.	\\ 
G0639+56.6179	& 	\nodata	& 	C12	& 	\nodata	& 	(no excess)	& 	\nodata	& 	 Not Li rich.			\\ 
SDSS J0652+4052	&	\nodata	&	\nodata	&	Martell \etal\ (2013)	& 	(no excess)	& 	\nodata	& 	\nodata			\\ 
Tyc3402-00280-1	& 	\nodata	& 	C12	& 	\nodata	& 	(no excess)	& 	\nodata	& 	 Not Li rich.			\\ 
SDSS J0654+4200	&	\nodata	&	\nodata	&	Martell \etal\ (2013)	& 	sparse SED but probably no IR excess	& 	\nodata	& 	\nodata			\\ 
HD51367	& 	\nodata	& 	\nodata	& 	Kumar \etal\ (2011)	& 	(no excess)	& 	\nodata	& 	\nodata			\\ 
G0653+16.552	& 	\nodata	& 	C12	& 	\nodata	& 	sparse SED but probably no IR excess	& 	\nodata	& 	 Not Li rich.			\\ 
G0654+16.235	& 	\nodata	& 	C12	& 	\nodata	& 	sparse SED but probably no IR excess	& 	\nodata	& 	 Not Li rich.			\\ 
HIP35253	& 	\nodata	& 	C12	& 	\nodata	& 	(no excess)	& 	\nodata	& 	 Not Li rich.			\\ 
SDSS J0720+3036	&	\nodata	&	\nodata	&	Martell \etal\ (2013)	& 	sparse SED but probably no IR excess	& 	\nodata	& 	 	\teff\ too warm to be RG.		\\ 
HD57669	&	\nodata	&	\nodata	&	Jasniewicz \etal\  (1999)	& 	(no excess)	& 	\nodata	& 	 		 No $\log g$.	\\ 
IRAS07227-1320(PDS132)	& 	dlR97	& 	\nodata	& 	\nodata	& 	IR excess	& 	Also GSC 05408-03215.	& 	 No $A$(Li).	 No \teff.	 No $\log g$.	\\ 
HD59686	& 	\nodata	& 	C12	& 	\nodata	& 	(no excess)	& 	\nodata	& 	 Not Li rich.			\\ 
HIP36896	& 	\nodata	& 	C12	& 	\nodata	& 	(no excess)	& 	\nodata	& 	 Not Li rich.			\\ 
NGC 2423 3	&	\nodata	&	\nodata	&	Carlberg in prep	& 	(no excess)	& 	\nodata	& 	\nodata			\\ 
IRAS07419-2514	&	\nodata	&	\nodata	&	Torres \etal\ (2000)	& 	drop due to source confusion	& 	Target position is photocenter of small group of objects. Torres \etal\ (2000) notes IRAS flux may come from CO cloud WB 1046.	& 	 No $A$(Li).	 No \teff.	 No $\log g$.	\\ 
HD62509(Pollux)	& 	\nodata	& 	C12	& 	\nodata	& 	(no excess)	& 	\nodata	& 	 Not Li rich.			\\ 
IRAS07456-4722(PDS135)	& 	dlR97	& 	\nodata	& 	\nodata	& 	IR excess	& 	\nodata	& 	 No $A$(Li).	 No \teff.	 No $\log g$.	\\ 
Tyc5981-00414-1	& 	\nodata	& 	C12	& 	\nodata	& 	(no excess)	& 	\nodata	& 	 Not Li rich.			\\ 
HD63798	& 	\nodata	& 	\nodata	& 	Kumar \etal\ (2011)	& 	(no excess)	& 	\nodata	& 	\nodata			\\ 
HD65750	& 	dlR97	& 	\nodata	& 	\nodata	& 	IR excess (small)	& 	POSS images have strong nebulosity. Also V341 Car -- a pulsating variable star. Identified in McDonald \etal\ (2012) as having an IR excess.	& 	 Not Li rich.	\teff\ too cool to be RG.		\\ 
HD65228	&	\nodata	&	\nodata	&	Liu \etal\ (2014)	& 	(no excess)	& 	\nodata	& 	 	\teff\ too warm to be RG.		\\ 
Tyc1938-00311-1	& 	\nodata	& 	C12	& 	\nodata	& 	(no excess)	& 	\nodata	& 	 Not Li rich.			\\ 
IRAS07577-2806(PDS260)	& 	dlR97	& 	\nodata	& 	\nodata	& 	IR excess	& 	\nodata	& 	 No $A$(Li).	 No \teff.	 No $\log g$.	\\ 
G0804+39.4755	& 	\nodata	& 	C12	& 	\nodata	& 	(no excess)	& 	\nodata	& 	 Not Li rich.			\\ 
SDSS J0808\_0815	&	\nodata	&	\nodata	&	Martell \etal\ (2013)	& 	sparse SED but probably no IR excess	& 	\nodata	& 	\nodata			\\ 
Tyc0195-02087-1	& 	\nodata	& 	C12	& 	\nodata	& 	(no excess)	& 	\nodata	& 	 Not Li rich.			\\ 
HD70522	&	\nodata	&	\nodata	&	Liu \etal\ (2014), Luck \& Heiter (2007)	& 	(no excess)	& 	\nodata	& 	 	\teff\ too warm to be RG.	$\log g$ too large to be RG.	\\ 
HD233517	& 	dlR97	& 	\nodata	& 	\nodata	& 	IR excess	& 	Also identified in Kumar \etal\ (2015) as an IR excess star, among many other references.	& 	\nodata			\\ 
Tyc0205-01287-1	& 	\nodata	& 	C12	& 	\nodata	& 	(no excess)	& 	\nodata	& 	 Not Li rich.			\\ 
G0827-16.3424	& 	\nodata	& 	C12	& 	\nodata	& 	(no excess)	& 	\nodata	& 	 Not Li rich.		$\log g$ too large to be RG.	\\ 
SDSS J0831+5402	&	\nodata	&	\nodata	&	Martell \etal\ (2013)	& 	sparse SED but probably no IR excess	& 	\nodata	& 	\nodata			\\ 
IRASF08359-1644	&	\nodata	&	\nodata	&	Torres \etal\ (2000)	& 	IR excess	& 	\nodata	& 	 No $A$(Li).	 No \teff.	 No $\log g$.	\\ 
HD73108	& 	\nodata	& 	C12	& 	\nodata	& 	(no excess)	& 	\nodata	& 	 Not Li rich.			\\ 
G0840+56.9122	& 	\nodata	& 	C12	& 	\nodata	& 	(no excess)	& 	\nodata	& 	 Not Li rich.			\\ 
G0840+56.5839	& 	\nodata	& 	C12	& 	\nodata	& 	(no excess)	& 	\nodata	& 	 Not Li rich.			\\ 
HD76066	& 	dlR97	& 	\nodata	& 	\nodata	& 	(no excess)	& 	IRAS FSC [12]-[25] $>$ 0.5 but WISE says no excess.	& 	 No $A$(Li).	 No \teff.	 No $\log g$.	\\ 
HD77361	& 	\nodata	& 	\nodata	& 	Kumar \etal\ (2011)	& 	(no excess)	& 	\nodata	& 	\nodata			\\ 
HD78668	&	\nodata	&	\nodata	&	Liu \etal\ (2014)	& 	(no excess)	& 	\nodata	& 	\nodata			\\ 
G0909-05.211	& 	\nodata	& 	C12	& 	\nodata	& 	(no excess)	& 	\nodata	& 	 Not Li rich.			\\ 
Tyc3809-01017-1	& 	\nodata	& 	C12	& 	\nodata	& 	(no excess)	& 	\nodata	& 	 Not Li rich.			\\ 
G0912-05.11	& 	\nodata	& 	C12	& 	\nodata	& 	(no excess)	& 	\nodata	& 	 Not Li rich.			\\ 
G0928+73.2600	& 	\nodata	& 	C12	& 	\nodata	& 	IR excess (small)	& 	\nodata	& 	\nodata			\\ 
HD82227	& 	dlR97	& 	\nodata	& 	\nodata	& 	(no excess)	& 	\nodata	& 	 No $A$(Li).		 No $\log g$.	\\ 
HD82421	& 	dlR97	& 	\nodata	& 	\nodata	& 	(no excess)	& 	\nodata	& 	 No $A$(Li).		 No $\log g$.	\\ 
HD82734	&	\nodata	&	\nodata	&	Liu \etal\ (2014)	& 	(no excess)	& 	\nodata	& 	\nodata			\\ 
SDSS J0936+2935	&	\nodata	&	\nodata	&	Martell \etal\ (2013)	& 	sparse SED but probably no IR excess	& 	\nodata	& 	\nodata			\\ 
G0935-05.152	& 	\nodata	& 	C12	& 	\nodata	& 	sparse SED but probably no IR excess	& 	\nodata	& 	 Not Li rich.			\\ 
G0946+00.48	& 	\nodata	& 	C12	& 	\nodata	& 	sparse SED but probably no IR excess	& 	\nodata	& 	 Not Li rich.			\\ 
HD85444	&	\nodata	&	\nodata	&	Liu \etal\ (2014)	& 	(no excess)	& 	\nodata	& 	\nodata			\\ 
IRAS09553-5621	& 	dlR97	& 	\nodata	& 	\nodata	& 	drop due to source confusion	& 	Target position is in between two sources	& 	 No $A$(Li).	 No \teff.	 No $\log g$.	\\ 
LeoI 71032	&	\nodata	&	\nodata	&	Kirby \etal\ (2012)	& 	sparse SED but probably no IR excess	& 	\nodata	& 	\nodata			\\ 
LeoI 60727	&	\nodata	&	\nodata	&	Kirby \etal\ (2012)	& 	sparse SED but probably no IR excess	& 	\nodata	& 	\nodata			\\ 
LeoI 32266	&	\nodata	&	\nodata	&	Kirby \etal\ (2012)	& 	sparse SED but probably no IR excess	& 	\nodata	& 	\nodata			\\ 
LeoI 21617	&	\nodata	&	\nodata	&	Kirby \etal\ (2012)	& 	sparse SED but probably no IR excess	& 	\nodata	& 	\nodata			\\ 
C1012254-203007	&	\nodata	&	\nodata	&	Ruchti \etal\ (2011)	& 	sparse SED but probably no IR excess	& 	\nodata	& 	\nodata			\\ 
HD88476	& 	\nodata	& 	\nodata	& 	Kumar \etal\ (2011)	& 	(no excess)	& 	\nodata	& 	\nodata			\\ 
Tyc3441-00140-1	& 	\nodata	& 	C12	& 	\nodata	& 	(no excess)	& 	\nodata	& 	 Not Li rich.			\\ 
BD+202457	&	\nodata	&	\nodata	&	Carlberg in prep	& 	(no excess)	& 	\nodata	& 	 Not Li rich.			\\ 
Tyc5496-00376-1=BD-12d3141	&	\nodata	&	\nodata	&	Ruchti \etal\ (2011)	& 	(no excess)	& 	\nodata	& 	\nodata			\\ 
HD90082	&	\nodata	&	\nodata	&	Castilho \etal\ (2000)	& 	(no excess)	& 	WISE measurements from AllWISE reject catalog	& 	 Not Li rich.	\teff\ too cool to be RG.		\\ 
Tyc3005-00827-1	& 	\nodata	& 	C12	& 	\nodata	& 	(no excess)	& 	\nodata	& 	 Not Li rich.			\\ 
HD90633	& 	\nodata	& 	\nodata	& 	Kumar \etal\ (2011)	& 	(no excess)	& 	\nodata	& 	\nodata			\\ 
HD92253	& 	dlR97	& 	\nodata	& 	\nodata	& 	(no excess)	& 	\nodata	& 	 No $A$(Li).		 No $\log g$.	\\ 
G1053+00.15	& 	\nodata	& 	C12	& 	\nodata	& 	(no excess)	& 	\nodata	& 	 Not Li rich.			\\ 
Tyc2521-01716-1	& 	\nodata	& 	C12	& 	\nodata	& 	(no excess)	& 	\nodata	& 	 Not Li rich.			\\ 
HD95799	& 	dlR97	& 	\nodata	& 	\nodata	& 	(no excess)	& 	\nodata	& 	\nodata			\\ 
HD96195	&	\nodata	&	\nodata	&	Castilho \etal\ (2000)	& 	IR excess (but maybe not K giant)	& 	May be too cool to be a K giant. Identified in McDonald \etal\ (2012) as having an IR excess.	& 	 Not Li rich.	\teff\ too cool to be RG.		\\ 
SDSS J1105+2850	&	\nodata	&	\nodata	&	Martell \etal\ (2013)	& 	sparse SED but probably no IR excess	& 	\nodata	& 	 	\teff\ too warm to be RG.	$\log g$ too large to be RG.	\\ 
IRAS11044-6127	& 	dlR97	& 	\nodata	& 	\nodata	& 	drop due to source confusion	& 	Optical counterpart hard to locate, steep SED.	& 	 No $A$(Li).	 No \teff.	 No $\log g$.	\\ 
HD96996	& 	dlR97	& 	\nodata	& 	\nodata	& 	(no excess)	& 	\nodata	& 	 No $A$(Li).	 No \teff.	 No $\log g$.	\\ 
HD97472	& 	dlR97	& 	\nodata	& 	\nodata	& 	(no excess)	& 	\nodata	& 	 No $A$(Li).		 No $\log g$.	\\ 
LeoII C-7-174	&	\nodata	&	\nodata	&	Kirby \etal\ (2012)	& 	sparse SED but probably no IR excess	& 	\nodata	& 	\nodata			\\ 
LeoII C-3-146	&	\nodata	&	\nodata	&	Kirby \etal\ (2012)	& 	sparse SED but probably no IR excess	& 	\nodata	& 	\nodata			\\ 
G1124-05.61	& 	\nodata	& 	C12	& 	\nodata	& 	(no excess)	& 	\nodata	& 	 Not Li rich.			\\ 
Tyc3013-01489-1	& 	\nodata	& 	C12	& 	\nodata	& 	(no excess)	& 	\nodata	& 	 Not Li rich.			\\ 
G1127-11.60	& 	\nodata	& 	C12	& 	\nodata	& 	sparse SED but probably no IR excess	& 	\nodata	& 	 Not Li rich.			\\ 
G1130+39.9414	& 	\nodata	& 	C12	& 	\nodata	& 	(no excess)	& 	G1130+37.9414 in C12 is a misprint for G1130+39.9414. It is also Tyc3013-01163-1.	& 	 Not Li rich.			\\ 
HD102845	&	\nodata	&	\nodata	&	Liu \etal\ (2014)	& 	(no excess)	& 	\nodata	& 	\nodata			\\ 
Tyc5523-00830-1	& 	\nodata	& 	C12	& 	\nodata	& 	(no excess)	& 	\nodata	& 	 Not Li rich.			\\ 
Tyc0276-00327-1	& 	\nodata	& 	C12	& 	\nodata	& IR excess (small)	& 	Also HD103915. In halo of bright galaxy(?) that appears by 12, 22 \mum. Likely high background, but probably ok. 	& 	 Not Li rich.			\\ 
G1200+67.3882	& 	\nodata	& 	C12	& 	\nodata	& 	(no excess)	& 	\nodata	& 	 Not Li rich.			\\ 
Tyc6094-01204-1	& 	\nodata	& 	C12	& 	\nodata	& 	(no excess)	& 	\nodata	& 	 Not Li rich.			\\ 
HD104985	& 	\nodata	& 	C12	& 	\nodata	& 	(no excess)	& 	\nodata	& 	 Not Li rich.			\\ 
Tyc2527-01442-1	& 	\nodata	& 	C12	& 	\nodata	& 	(no excess)	& 	\nodata	& 	 Not Li rich.			\\ 
G1213+33.15558	& 	\nodata	& 	C12	& 	\nodata	& 	(no excess)	& 	\nodata	& 	 Not Li rich.			\\ 
HD107484	& 	\nodata	& 	\nodata	& 	Kumar \etal\ (2011)	& 	(no excess)	& 	\nodata	& 	\nodata			\\ 
NGC 4349 127	&	\nodata	&	\nodata	&	Carlberg in prep	& 	(no excess)	& 	\nodata	& 	 Not Li rich.			\\ 
HD108225	& 	\nodata	& 	C12	& 	\nodata	& 	(no excess)	& 	\nodata	& 	 Not Li rich.			\\ 
IRAS12236-6302(PDS354)	& 	dlR97	& 	\nodata	& \nodata & 	drop due to source confusion	& 	Cluster of possible sources. Steep SED. Torres \etal\ (2000) mention that optical spectrum of source taken as counterpart has strong H$\alpha$ emission and could be an \ion{H}{2} region.	& 	 No $A$(Li).	 No \teff.	 No $\log g$.	\\ 
HD108471	& 	dlR97	& 	\nodata	& 	\nodata	& 	(no excess)	& 	\nodata	& 	\nodata			\\ 
IRAS12327-6523(PDS355)	& 	dlR97	& 	\nodata	& 	\nodata	& 	IR excess	& 	\nodata	& 	 No $A$(Li).			\\ 
HD109742	& 	\nodata	& 	C12	& 	\nodata	& 	(no excess)	& 	\nodata	& 	 Not Li rich.			\\ 
M68-A96=Cl* NGC 4590 HAR 1257	&	\nodata	&	\nodata	&	Ruchti \etal\ (2011)	& 	sparse SED but probably no IR excess	& 	The coordinates in the paper are incorrect; used finding chart, Fig 2, in Alcaino (1977) to ID by eye.	& 	\nodata			\\ 
G1240+56.8464	& 	\nodata	& 	C12	& 	\nodata	& 	(no excess)	& 	\nodata	& 	 Not Li rich.			\\ 
HD112127	& 	dlR97	& 	\nodata	& 	\nodata	& 	(no excess)	& 	\nodata	& 	\nodata			\\ 
HD111830	& 	dlR97	& 	\nodata	& 	\nodata	& 	IR excess (small)	& 	\nodata	& 	 No $A$(Li).		 No $\log g$.	\\ 
HD112859	& 	\nodata	& 	C12	& 	\nodata	& 	(no excess)	& 	IRAS FSC [12]-[25] $>$ 0.5 but WISE says no excess.	& 	\nodata			\\ 
\tablebreak
SDSS J1310\_0012	&	\nodata	&	\nodata	&	Martell \etal\ (2013)	& 	sparse SED but probably no IR excess	& 	\nodata	& 	\nodata			\\ 
HD115478	& 	\nodata	& 	C12	& 	\nodata	& 	(no excess)	& 	WISE measurements from AllWISE reject catalog	& 	 Not Li rich.			\\ 
HD115659	&	\nodata	&	\nodata	&	Liu \etal\ (2014)	& 	(no excess)	& 	\nodata	& 	\nodata			\\ 
HD116010	& 	\nodata	& 	C12	& 	\nodata	& 	(no excess)	& 	\nodata	& 	 Not Li rich.			\\ 
HD116292	& 	\nodata	& 	\nodata	& 	Liu \etal\ (2014), Kumar \etal\ (2011)	& 	(no excess)	& 	\nodata	& 	\nodata			\\ 
CVnI 195\_195	&	\nodata	&	\nodata	&	Kirby \etal\ (2012)	& 	sparse SED but probably no IR excess	& 	\nodata	& 	\nodata			\\ 
CVnI 196\_129	&	\nodata	&	\nodata	&	Kirby \etal\ (2012)	& 	sparse SED but probably no IR excess	& 	\nodata	& 	\nodata			\\ 
G1331+00.13	& 	\nodata	& 	C12	& 	\nodata	& 	(no excess)	& 	\nodata	& 	 Not Li rich.			\\ 
PDS365(IRAS13313-5838)	& 	dlR97	& 	\nodata	& 	\nodata	& 	IR excess	& 	\nodata	& 	\nodata			\\ 
HD118319	& 	\nodata	& 	\nodata	& 	Kumar \etal\ (2011)	& 	(no excess)	& 	\nodata	& 	\nodata			\\ 
HD118344	& 	dlR97	& 	\nodata	& 	\nodata	& 	(no excess)	& 	\nodata	& 	 No $A$(Li).		 No $\log g$.	\\ 
HD118839	& 	\nodata	& 	C12	& 	\nodata	& 	(no excess)	& 	\nodata	& 	 Not Li rich.			\\ 
M3-IV101=Cl* NGC 5272 SK 557	&	\nodata	&	\nodata	&	Kraft \etal\ (1999), Pilachowski \etal\ (2003), Ruchti \etal\ (2011)	& 	sparse SED but probably no IR excess	& 	\nodata	& 	\nodata			\\ 
HD119853	& 	dlR97	& 	\nodata	& 	\nodata	& 	(no excess)	& 	\nodata	& 	 No $A$(Li).		 No $\log g$.	\\ 
HD120048	&	\nodata	&	\nodata	&	Liu \etal\ (2014)	& 	(no excess)	& 	\nodata	& 	\nodata			\\ 
HD120602	& 	dlR97	& 	\nodata	& 	\nodata	& 	(no excess)	& 	\nodata	& 	\nodata			\\ 
HD121710(9Boo)	& 	dlR97	& 	\nodata	& 	\nodata	& 	(no excess)	& 	\nodata	& 	 Not Li rich.		 No $\log g$.	\\ 
PDS68(IRAS13539-4153)	& 	dlR97	& 	\nodata	& 	\nodata	& 	IR excess	& 	Also GSC 07798-00578.	& 	\nodata			\\ 
Tyc3027-01042-1	& 	\nodata	& 	C12	& 	\nodata	& 	(no excess)	& 	\nodata	& 	 Not Li rich.			\\ 
HD122430	& 	\nodata	& 	C12	& 	\nodata	& 	(no excess)	& 	\nodata	& 	 Not Li rich.			\\ 
Tyc0319-00231-1	& 	\nodata	& 	C12	& 	\nodata	& 	(no excess)	& 	\nodata	& 	 Not Li rich.			\\ 
HD124897(Arcturus)	& 	\nodata	& 	C12	& 	\nodata	& 	(no excess)	& 	high enough proper motions that automatic merging not possible; matches done via SIMBAD and by hand	& 	 Not Li rich.			\\ 
HD125618	& 	dlR97	& 	\nodata	& 	\nodata	& 	(no excess)	& 	\nodata	& 	 No $A$(Li).	 No \teff.	 No $\log g$.	\\ 
Tyc1469-01108-1	& 	\nodata	& 	C12	& 	\nodata	& 	(no excess)	& 	\nodata	& 	 Not Li rich.			\\ 
IRAS14198-6115	& 	dlR97	& 	\nodata	& 	\nodata	& 	drop due to source confusion	& 	Cluster of possible sources.	& 	 No $A$(Li).	 No \teff.	 No $\log g$.	\\ 
G1421+28.4625	& 	\nodata	& 	C12	& 	\nodata	& 	(no excess)	& 	\nodata	& 	 Not Li rich.			\\ 
RAVEJ142546.2-154629	&	\nodata	&	\nodata	&	Ruchti \etal\ (2011)	& 	(no excess)	& 	\nodata	& 	\nodata			\\ 
HD126868	&	\nodata	&	\nodata	&	Jasniewicz \etal\  (1999)	& 	(no excess)	& 	\nodata	& 	 	\teff\ too warm to be RG.	 No $\log g$.	\\ 
IRAS14257-6023	& 	dlR97	& 	\nodata	& 	\nodata	& 	drop due to source confusion	& 	Cluster of possible sources.	& 	 No $A$(Li).	 No \teff.	 No $\log g$.	\\ 
SDSS J1432+0814	&	\nodata	&	\nodata	&	Martell \etal\ (2013)	& 	sparse SED but probably no IR excess	& 	\nodata	& 	\nodata			\\ 
HD127740	&	\nodata	&	\nodata	&	Liu \etal\ (2014), Luck \& Heiter (2007)	& 	(no excess)	& 	\nodata	& 	 	\teff\ too warm to be RG.	$\log g$ too large to be RG.	\\ 
Tyc0913-01248-1	& 	\nodata	& 	C12	& 	\nodata	& 	(no excess)	& 	\nodata	& 	 Not Li rich.			\\ 
Tyc0914-00571-1	& 	\nodata	& 	C12	& 	\nodata	& 	(no excess)	& 	\nodata	& 	 Not Li rich.			\\ 
HD128309	& 	dlR97	& 	\nodata	& 	\nodata	& 	(no excess)	& 	\nodata	& 	 No $A$(Li).	 No \teff.	 No $\log g$.	\\ 
HD129955	& 	dlR97	& 	\nodata	& 	\nodata	& 	(no excess)	& 	\nodata	& 	 No $A$(Li).	 No \teff.	 No $\log g$.	\\ 
HD131530	& 	dlR97	& 	\nodata	& 	\nodata	& 	(no excess)	& 	\nodata	& 	 No $A$(Li).		 No $\log g$.	\\ 
HD133086	& 	\nodata	& 	\nodata	& 	Kumar \etal\ (2011)	& 	(no excess)	& 	\nodata	& 	\nodata			\\ 
Tyc0347-00762-1	& 	\nodata	& 	C12	& 	\nodata	& 	(no excess)	& 	\nodata	& 	 Not Li rich.			\\ 
M5 V42	&	\nodata	&	\nodata	&	Carney \etal\ (1998)	& 	(no excess)	& 	\nodata	& 	\nodata			\\ 
SDSS J1522+0655	&	\nodata	&	\nodata	&	Martell \etal\ (2013)	& 	sparse SED but probably no IR excess	& 	\nodata	& 	\nodata			\\ 
HD137759	& 	\nodata	& 	C12	& 	\nodata	& 	(no excess)	& 	\nodata	& 	 Not Li rich.			\\ 
\tablebreak
HD138525	&	\nodata	&	\nodata	&	Liu \etal\ (2014), Luck \& Heiter (2007)	& 	(no excess)	& 	\nodata	& 	 	\teff\ too warm to be RG.	$\log g$ too large to be RG.	\\ 
HD138688	&	\nodata	&	\nodata	&	Jasniewicz \etal\  (1999)	& 	(no excess)	& 	\nodata	& 	 Not Li rich.			\\ 
G1551+22.9456	& 	\nodata	& 	C12	& 	\nodata	& 	(no excess)	& 	\nodata	& 	 Not Li rich.			\\ 
SDSS J1607+0447	&	\nodata	&	\nodata	&	Martell \etal\ (2013)	& 	sparse SED but probably no IR excess	& 	\nodata	& 	\nodata			\\ 
HD145206	&	\nodata	&	\nodata	&	Jasniewicz \etal\  (1999)	& 	(no excess)	& 	\nodata	& 	 Not Li rich.			\\ 
HD145457	& 	\nodata	& 	\nodata	& 	Kumar \etal\ (2011)	& 	(no excess)	& 	\nodata	& 	\nodata			\\ 
IRAS16086-5255(PDS410)	& 	dlR97	& 	\nodata	& 	\nodata	& 	IR excess	& 	\nodata	& 	 No $A$(Li).	 No \teff.	 No $\log g$.	\\ 
IRAS16128-5109	& 	dlR97	& 	\nodata	& 	\nodata	& 	drop due to source confusion	& 	Appears in SIMBAD as an \ion{H}{2} region; the morphology of the image suggests a dense clump of sources from which emanate long streamers of extended emission. 	& 	 No $A$(Li).	 No \teff.	 No $\log g$.	\\ 
HD146850	& 	dlR97	& 	\nodata	& 	\nodata	& 	(no excess)	& 	\nodata	& 	\nodata			\\ 
HD146834	& 	dlR97	& 	\nodata	& 	\nodata	& 	IR excess (small)	& 	Also HR 6076. Also identified in McDonald \etal\ (2012) as having an IR excess.	& 	 No $A$(Li).		 No $\log g$.	\\ 
HD148293	& 	dlR97	& 	\nodata	& 	\nodata	& 	(no excess)	& 	\nodata	& 	\nodata			\\ 
Tyc2043-00747-1	& 	\nodata	& 	C12	& 	\nodata	& 	(no excess)	& 	\nodata	& 	 Not Li rich.			\\ 
IRAS16227-4839	& 	dlR97	& 	\nodata	& 	\nodata	& 	drop due to source confusion	& 	Match forced to be IR-bright source (not found automatically given this position). 	& 	 No $A$(Li).	 No \teff.	 No $\log g$.	\\ 
HD148317	&	\nodata	&	\nodata	&	Liu \etal\ (2014), Luck \& Heiter (2007)	& 	(no excess)	& 	\nodata	& 	 	\teff\ too warm to be RG.	$\log g$ too large to be RG.	\\ 
IRAS16252-5440	& 	dlR97	& 	\nodata	& 	\nodata	& 	drop due to source confusion	& 	Also PDS 146. Cluster of IR-bright sources. 	& 	 No $A$(Li).	 No \teff.	 No $\log g$.	\\ 
HD150902	& 	\nodata	& 	\nodata	& 	Kumar \etal\ (2011)	& 	(no excess)	& 	\nodata	& 	\nodata			\\ 
HIP81437	& 	\nodata	& 	C12	& 	\nodata	& 	(no excess)	& 	\nodata	& 	 Not Li rich.			\\ 
G1640+56.6327	& 	\nodata	& 	C12	& 	\nodata	& 	(no excess)	& 	\nodata	& 	 Not Li rich.			\\ 
IRAS16514-4625(PDS432)	& 	dlR97	& 	\nodata	& 	\nodata	& 	drop due to source confusion	& 	Torres \etal\ (2000) list it as a confirmed giant but the source that was measured may not be responsible for the IR flux. Cluster of soruces, steep SED.	& 	 No $A$(Li).	 No \teff.	 No $\log g$.	\\ 
HD153135	& 	dlR97	& 	\nodata	& 	\nodata	& 	(no excess)	& 	\nodata	& 	 No $A$(Li).		 No $\log g$.	\\ 
HD152786	&	\nodata	&	\nodata	&	Jasniewicz \etal\  (1999)	& 	(no excess)	& 	\nodata	& 	 Not Li rich.			\\ 
HD153687	&	\nodata	&	\nodata	&	Jasniewicz \etal\  (1999)	& 	(no excess)	& 	\nodata	& 	 Not Li rich.			\\ 
HD155646	&	\nodata	&	\nodata	&	Liu \etal\ (2014), Luck \& Heiter (2007)	& 	(no excess)	& 	\nodata	& 	 	\teff\ too warm to be RG.	$\log g$ too large to be RG.	\\ 
IRAS17102-3813	& 	dlR97	& 	\nodata	& 	\nodata	& 	drop due to source confusion	& 	Cluster of sources.	& 	 No $A$(Li).	 No \teff.	 No $\log g$.	\\ 
IRAS17120-4106	& 	dlR97	& 	\nodata	& 	\nodata	& 	drop due to source confusion	& 	Two possible sources.	& 	 No $A$(Li).	 No \teff.	 No $\log g$.	\\ 
HD156115	& 	dlR97	& 	\nodata	& 	\nodata	& 	(no excess)	& 	\nodata	& 	 No $A$(Li).	\teff\ too cool to be RG.	 No $\log g$.	\\ 
HD156061	& 	dlR97	& 	\nodata	& 	\nodata	& 	(no excess)	& 	\nodata	& 	 No $A$(Li).		 No $\log g$.	\\ 
IRAS17211-3458	& 	dlR97	& 	\nodata	& 	\nodata	& 	drop due to source confusion	& 	Two possible sources.	& 	 No $A$(Li).	 No \teff.	 No $\log g$.	\\ 
HD157457	&	\nodata	&	\nodata	&	Jasniewicz \etal\  (1999)	& 	(no excess)	& 	\nodata	& 	\nodata			\\ 
HD157919	&	\nodata	&	\nodata	&	Liu \etal\ (2014), Luck \& Heiter (2007)	& 	(no excess)	& 	\nodata	& 	 	\teff\ too warm to be RG.	$\log g$ too large to be RG.	\\ 
IRAS17442-2441	& 	dlR97	& 	\nodata	& 	\nodata	& 	drop due to source confusion	& 	Cluster of possible sources.	& 	 No $A$(Li).	 No \teff.	 No $\log g$.	\\ 
\tablebreak
PDS97(IRAS17554-3822)	& 	\nodata	& 	\nodata	& 	de la Reza, Drake, \& da Silva (1996)	& 	drop due to source confusion	& 	Target position in between two sources of comparable brightness.	& 	 No $A$(Li).	 No \teff.	 No $\log g$.	\\ 
IRAS17576-1845	& 	dlR97	& 	\nodata	& 	\nodata	& 	drop due to source confusion	& 	The multi-wavelength images suggest extinction in this field.  Coadella \etal\ (1995) list it as a candidate to be related to high-mass star forming regions with an ultracompact \ion{H}{2} region, though it remained undetected in their survey. 	& 	 No $A$(Li).	 No \teff.	 No $\log g$.	\\ 
IRAS17578-1700	& 	dlR97	& 	\nodata	& 	\nodata	& 	IR excess (but maybe not K giant)	& 	Also C* 2514,   CGCS 3922 - likely carbon star.	& 	 No $A$(Li).	 No \teff.	 No $\log g$.	\\ 
HD162298	& 	dlR97	& 	\nodata	& 	\nodata	& 	(no excess)	& 	\nodata	& 	 No $A$(Li).		 No $\log g$.	\\ 
G1800+61.12976	& 	\nodata	& 	C12	& 	\nodata	& 	(no excess)	& 	\nodata	& 	 Not Li rich.			\\ 
IRAS17582-2619	& 	dlR97	& 	\nodata	& 	\nodata	& 	drop due to source confusion	& 	Brightest source in IR has no optical counterpart. SIMBAD lists this as an OH/IR star. Yoon \etal\ (2014) and references therein identify it as a post-AGB star (OH4.02-1.68).	& 	 No $A$(Li).	 No \teff.	 No $\log g$.	\\ 
IRAS17590-2412	& 	dlR97	& 	\nodata	& 	\nodata	& 	drop due to source confusion	& 	Position shifted slightly to pick up WISE source. Diffuse emission can also be seen in the field in various bands. Messineo \etal\ (2004) identify a SiO emitter in this region but suggest that it may not be associated with the source from which an optical spectrum had been obtained by dlR97. They note that this IRAS source is the only mid-infrared source within their 86 GHz beam.	& 	 No $A$(Li).	 No \teff.	 No $\log g$.	\\ 
IRAS17596-3952(PDS485)	& 	dlR97	& 	\nodata	& 	\nodata	& 	IR excess	& 	Position shifted slightly to pick up WISE source. Also identified in Kumar \etal\ (2015) as IR excess star.	& 	\nodata			\\ 
Tyc0435-03332-1	&	\nodata	&	\nodata	&	Adamow \etal\ (2014)	& 	(no excess)	& 	\nodata	& 	\nodata			\\ 
HD164712	& 	dlR97	& 	\nodata	& 	\nodata	& 	(no excess)	& 	\nodata	& 	 No $A$(Li).	 No \teff.	 No $\log g$.	\\ 
HD167304	& 	\nodata	& 	\nodata	& 	Kumar \etal\ (2011)	& 	(no excess)	& 	\nodata	& 	\nodata			\\ 
HD170527	& 	\nodata	& 	\nodata	& 	Kumar \etal\ (2011)	& 	(no excess)	& 	\nodata	& 	\nodata			\\ 
HD169689	&	\nodata	&	\nodata	&	Jasniewicz \etal\  (1999)	& 	(no excess)	& 	\nodata	& 	 Not Li rich.			\\ 
V385 Sct	&	\nodata	&	\nodata	&	Castilho \etal\ (2000)	& 	IR excess (but maybe not K giant)	& 	Too cool to be a K giant.  S-type star.	& 	 Not Li rich.	\teff\ too cool to be RG.		\\ 
IRAS18334-0631(PDS524)	& 	dlR97	& 	\nodata	& 	\nodata	& 	drop due to source confusion	& 	No optical source at the target position, and cluster of IR sources.	& 	 No $A$(Li).	 No \teff.	 No $\log g$.	\\ 
Tyc3105-00152-1	&	\nodata	&	\nodata	&	Adamow \etal\ (2014)	& 	(no excess)	& 	\nodata	& 	\nodata			\\ 
Tyc3917-01107-1	&	\nodata	&	\nodata	&	Adamow \etal\ (2014)	& 	(no excess)	& 	Identified in McDonald \etal\ (2012) as having an IR excess, but IR excess is not real.	& 	\nodata			\\ 
IRAS18397-0400	& 	dlR97	& 	\nodata	& 	\nodata	& 	drop due to source confusion	& 	Brightest source in the IR has no optical counterpart.	& 	 No $A$(Li).	 No \teff.	 No $\log g$.	\\ 
Tyc3930-00681-1	&	\nodata	&	\nodata	&	Adamow \etal\ (2014)	& 	(no excess)	& 	\nodata	& 	 Not Li rich.			\\ 
HD175492	&	\nodata	&	\nodata	&	Jasniewicz \etal\  (1999)	& 	(no excess)	& 	\nodata	& 	 Not Li rich.	\teff\ too warm to be RG.		\\ 
IRAS18559+0140	& 	dlR97	& 	\nodata	& 	\nodata	& 	drop due to source confusion	& 	No optical source at the target position, and cluster of IR sources.	& 	 No $A$(Li).	 No \teff.	 No $\log g$.	\\ 
HD176588	& 	dlR97	& 	\nodata	& 	\nodata	& 	(no excess)	& 	\nodata	& 	\nodata			\\ 
SDSS J1901+3808	&	\nodata	&	\nodata	&	Martell \etal\ (2013)	& 	sparse SED but probably no IR excess	& 	\nodata	& 	\nodata			\\ 
HD176884	&	\nodata	&	\nodata	&	Jasniewicz \etal\  (1999)	& 	(no excess)	& 	\nodata	& 	 Not Li rich.			\\ 
IRAS19012-0747	& 	dlR97	& 	\nodata	& 	\nodata	& 	IR excess (small)	& 	Name as appearing in dlR97 had a typo; this is the correct name. 	& 	 No $A$(Li).			\\ 
HD177830	& 	\nodata	& 	C12	& 	\nodata	& 	(no excess)	& 	\nodata	& 	 Not Li rich.			\\ 
\tablebreak
IRAS19038-0026	&	\nodata	&	\nodata	&	Castilho \etal\ (2000)	& 	IR excess (small but maybe not K giant)	& 	May be too cool to be a K giant.	& 	 Not Li rich.	\teff\ too cool to be RG.		\\ 
HD177366	& 	dlR97	& 	\nodata	& 	\nodata	& 	(no excess)	& 	\nodata	& 	 No $A$(Li).		 No $\log g$.	\\ 
HD178168	&	\nodata	&	\nodata	&	Castilho \etal\ (2000)	& 	(no excess)	& 	\nodata	& 	 Not Li rich.			\\ 
SDSS J1909+3837	&	\nodata	&	\nodata	&	Martell \etal\ (2013)	& 	sparse SED but probably no IR excess	& 	\nodata	& 	\nodata			\\ 
IRAS19083+0119(PDS562)	& 	dlR97	& 	\nodata	& 	\nodata	& 	drop due to source confusion	& 	Brightest source in the IR has no optical counterpart. Steep SED.	& 	 No $A$(Li).	 No \teff.	 No $\log g$.	\\ 
KIC 5000307 	&	\nodata	&	\nodata	&	Silva Aguirre \etal\  (2014)	& 	(no excess)	& 	\nodata	& 	\nodata			\\ 
HD181154	& 	dlR97	& 	\nodata	& 	\nodata	& 	(no excess)	& 	\nodata	& 	 No $A$(Li).		 No $\log g$.	\\ 
IRAS19210+1715	& 	dlR97	& 	\nodata	& 	\nodata	& 	drop due to source confusion	& 	Brightest source in the IR has no optical counterpart. Steep SED.	& 	 No $A$(Li).	 No \teff.	 No $\log g$.	\\ 
HD182900	&	\nodata	&	\nodata	&	Liu \etal\ (2014), Luck \& Heiter (2007)	& 	(no excess)	& 	\nodata	& 	 	\teff\ too warm to be RG.	$\log g$ too large to be RG.	\\ 
HD182901	&	\nodata	&	\nodata	&	Liu \etal\ (2014), Luck \& Heiter (2007)	& 	(no excess)	& 	\nodata	& 	 	\teff\ too warm to be RG.	$\log g$ too large to be RG.	\\ 
HD183492	& 	\nodata	& 	\nodata	& 	Kumar \etal\ (2011)	& 	(no excess)	& 	\nodata	& 	\nodata			\\ 
HD183202	& 	dlR97	& 	\nodata	& 	\nodata	& 	(no excess)	& 	\nodata	& 	 No $A$(Li).	 No \teff.	 No $\log g$.	\\ 
PDS100	& 	dlR97	& 	\nodata	& 	\nodata	& 	IR excess	& 	Also V859 Aql and IRAS 19285+0517. Also identified in Kumar \etal\ (2015) as IR excess star.	& 	\nodata			\\ 
G1936+61.14369	& 	\nodata	& 	C12	& 	\nodata	& 	(no excess)	& 	\nodata	& 	 Not Li rich.			\\ 
HD185194	&	\nodata	&	\nodata	&	Liu \etal\ (2014)	& 	(no excess)	& 	\nodata	& 	\nodata			\\ 
KIC 4937011	&	\nodata	&	\nodata	&	Anthony-Twarog \etal\  (2013)	& 	sparse SED but probably no IR excess	& 	\nodata	& 	\nodata			\\ 
HD187114	& 	dlR97	& 	\nodata	& 	\nodata	& 	(no excess)	& 	\nodata	& 	 No $A$(Li).		 No $\log g$.	\\ 
RAVEJ195244.9-600813	&	\nodata	&	\nodata	&	Ruchti \etal\ (2011)	& 	(no excess)	& 	\nodata	& 	\nodata			\\ 
Tyc1058-02865-1	&	\nodata	&	\nodata	&	Adamow \etal\ (2014)	& 	(no excess)	& 	\nodata	& 	 Not Li rich.			\\ 
HD188376	&	\nodata	&	\nodata	&	Liu \etal\ (2014), Luck \& Heiter (2007)	& 	(no excess)	& 	\nodata	& 	 	\teff\ too warm to be RG.	$\log g$ too large to be RG.	\\ 
HD188993	&	\nodata	&	\nodata	&	Liu \etal\ (2014), Luck \& Heiter (2007)	& 	(no excess)	& 	\nodata	& 	 	\teff\ too warm to be RG.	$\log g$ too large to be RG.	\\ 
HD190299	& 	dlR97	& 	\nodata	& 	\nodata	& 	(no excess)	& 	\nodata	& 	 No $A$(Li).	 No \teff.	 No $\log g$.	\\ 
HD191277	& 	\nodata	& 	C12	& 	\nodata	& 	(no excess)	& 	\nodata	& 	 Not Li rich.			\\ 
SDSS J2019+6012	&	\nodata	&	\nodata	&	Martell \etal\ (2013)	& 	sparse SED but probably no IR excess	& 	\nodata	& 	\nodata			\\ 
HD194317(39Cyg)	& 	dlR97	& 	\nodata	& 	\nodata	& 	(no excess)	& 	\nodata	& 	 Not Li rich.		 No $\log g$.	\\ 
HD194937	& 	\nodata	& 	\nodata	& 	Liu \etal\ (2014), Luck \& Heiter (2007), Kumar \etal\ (2011)	& 	(no excess)	& 	\nodata	& 	\nodata			\\ 
Tyc9112-00430-1	&	\nodata	&	\nodata	&	Ruchti \etal\ (2011)	& 	IR excess (small)	& 	\nodata	& 	\nodata			\\ 
Tyc2185-00133-1	& 	\nodata	& 	C12	& 	\nodata	& 	(no excess)	& 	\nodata	& 	 Not Li rich.			\\ 
HD202261	&	\nodata	&	\nodata	&	Liu \etal\ (2014)	& 	(no excess)	& 	\nodata	& 	\nodata			\\ 
HD203136	& 	\nodata	& 	\nodata	& 	Kumar \etal\ (2011)	& 	(no excess)	& 	Identified in McDonald \etal\ (2012) as having an IR excess, but IR excess is not real.	& 	\nodata			\\ 
HD203251	& 	dlR97	& 	\nodata	& 	\nodata	& 	(no excess)	& 	IRAS FSC [12]-[25] $>$ 0.5 but WISE says no excess.	& 	 Not Li rich.	 No \teff.	 No $\log g$.	\\ 
HD204540	& 	dlR97	& 	\nodata	& 	\nodata	& 	(no excess)	& 	\nodata	& 	 No $A$(Li).		 No $\log g$.	\\ 
HD205349	& 	\nodata	& 	\nodata	& 	Kumar \etal\ (2011)	& 	(no excess)	& 	\nodata	& 	\nodata			\\ 
HD206445	& 	\nodata	& 	C12	& 	\nodata	& 	(no excess)	& 	\nodata	& 	 Not Li rich.			\\ 
Tyc6953-00510-1	&	\nodata	&	\nodata	&	Ruchti \etal\ (2011)	& 	(no excess)	& 	\nodata	& 	\nodata			\\ 
\tablebreak
G2200+56.3466	& 	\nodata	& 	C12	& 	\nodata	& 	(no excess)	& 	\nodata	& 	 Not Li rich.			\\ 
SDSS J2200+4559	&	\nodata	&	\nodata	&	Martell \etal\ (2013)	& 	sparse SED but probably no IR excess	& 	\nodata	& 	\nodata			\\ 
SDSS J2206+4531	&	\nodata	&	\nodata	&	Martell \etal\ (2013)	& 	sparse SED but probably no IR excess	& 	\nodata	& 	\nodata			\\ 
HD212271	&	\nodata	&	\nodata	&	Liu \etal\ (2014)	& 	(no excess)	& 	\nodata	& 	\nodata			\\ 
HD212430	&	\nodata	&	\nodata	&	Liu \etal\ (2014)	& 	(no excess)	& 	\nodata	& 	\nodata			\\ 
HD213619	&	\nodata	&	\nodata	&	Liu \etal\ (2014), Luck \& Heiter (2007)	& 	(no excess)	& 	\nodata	& 	 	\teff\ too warm to be RG.	$\log g$ too large to be RG.	\\ 
HD213930	&	\nodata	&	\nodata	&	Liu \etal\ (2014)	& 	(no excess)	& 	\nodata	& 	\nodata			\\ 
HD214995	& 	\nodata	& 	\nodata	& 	Liu \etal\ (2014), Luck \& Heiter (2007), Kumar \etal\ (2011)	& 	(no excess)	& 	\nodata	& 	\nodata			\\ 
HD217352	& 	\nodata	& 	\nodata	& 	Kumar \etal\ (2011)	& 	(no excess)	& 	\nodata	& 	\nodata			\\ 
HD218527	& 	dlR97	& 	\nodata	& 	\nodata	& 	(no excess)	& 	\nodata	& 	 No $A$(Li).	 No \teff.	 No $\log g$.	\\ 
Tyc8448-00121-1	&	\nodata	&	\nodata	&	Ruchti \etal\ (2011)	& 	(no excess)	& 	\nodata	& 	\nodata			\\ 
HD219025	& 	dlR97	& 	\nodata	& 	\nodata	& 	IR excess	& 	Also BI Ind. Also identified in Kumar \etal\ (2015) as an IR excess star.	& 	\nodata			\\ 
HD219449	& 	\nodata	& 	C12	& 	\nodata	& 	(no excess)	& 	\nodata	& 	 Not Li rich.			\\ 
HD221776	& 	dlR97	& 	\nodata	& 	\nodata	& 	(no excess)	& 	\nodata	& 	 No $A$(Li).		 No $\log g$.	\\ 
HD221862	& 	\nodata	& 	C12	& 	\nodata	& 	(no excess)	& 	\nodata	& 	 Not Li rich.			\\ 
SDSS J2353+5728	&	\nodata	&	\nodata	&	Martell \etal\ (2013)	& 	(no excess)	& 	\nodata	& 	\nodata			\\ 
SDSS J2356+5633	&	\nodata	&	\nodata	&	Martell \etal\ (2013)	& 	(no excess)	& 	\nodata	& 	\nodata			\\

\enddata
\end{deluxetable}

\end{document}